\documentclass{aastex631}

\graphicspath{{./}{figures/}}
\usepackage{amsmath}
\usepackage{bbm}
\usepackage{mathrsfs}
\usepackage{amsthm}
\theoremstyle{remark}
\newtheorem{remark}{Remark}
\usepackage{booktabs}
\usepackage{multirow}
\usepackage{calc}
\setcitestyle{aysep={,}}

\begin{document}

\title{Resampling-based confidence intervals and bands for the average treatment effect in observational studies with competing risks}

\correspondingauthor{Jasmin Rühl}
\email{jasmin.ruehl@math.uni-augsburg.de}

\author{Jasmin Rühl}
\affiliation{Department of Mathematical Statistics and Artificial Intelligence in Medicine \\
University of Augsburg \\
Augsburg, Germany}

\author{Sarah Friedrich}
\affiliation{Department of Mathematical Statistics and Artificial Intelligence in Medicine \\
University of Augsburg \\
Augsburg, Germany}
\affiliation{Centre for Advanced Analytics and Predictive Sciences (CAAPS) \\
University of Augsburg \\
Augsburg, Germany}

\begin{abstract}
The g-formula can be used to estimate the treatment effect while accounting for confounding bias in observational studies. With regard to time-to-event endpoints, possibly subject to competing risks, the construction of valid pointwise confidence intervals and time-simultaneous confidence bands for the causal risk difference is complicated, however. A convenient solution is to approximate the asymptotic distribution of the corresponding stochastic process by means of resampling approaches. In this paper, we consider three different resampling methods, namely the classical nonparametric bootstrap, the influence function equipped with a resampling approach as well as a martingale-based bootstrap version, the so-called wild bootstrap. For the latter, three sub-versions based on differing distributions of the underlying random multipliers are examined. We set up a simulation study to compare the accuracy of the different techniques, which reveals that the wild bootstrap should in general be preferred if the sample size is moderate and sufficient data on the event of interest have been accrued. For illustration, the resampling methods are further applied to data on the long-term survival in patients with early-stage Hodgkin's disease.
\end{abstract}


\section{Introduction}

Causal inference provides tools to compare treatment strategies in studies that do not permit random allocation of subjects to therapy groups, e.g.~for ethical reasons or simply because it is not feasible. Special analysis methods are necessary because in non-randomized trials, risk factors are likely to be distributed unequally across treatment groups and as a consequence, direct comparisons will lead to bias \citep{yang2010observational, norgaard2017confounding}. The idea of the counterfactual approach to causal inference is to eliminate this bias by modeling the mean outcome in a hypothetical world where all participants of the study are exposed to the same intervention - possibly `counter to the fact', i.e.~contrary to the treatment they actually received. Causal conclusions can then be drawn by contrasting the obtained estimates for the treatment levels of interest (\citealp{rubin1974estimating}; \citealp[section~I.1]{hernan2020causal}). \\
In case of time-to-event endpoints, statisticians need to take additional difficulties into account, however, as the analysis of right-censored data requires particular techniques. The hazard ratio, which is the common measure of the treatment effect for time-to-event data, comes along with several issues when the aim is to draw causal inferences: In the first place, it is non-collapsible. Thus, the causal effect estimate may differ from a conditional estimate that is adjusted for further variables even if these variables are no confounders \citep{martinussen2013on}. A related drawback is selection bias, which has e.g. been described by \citet{aalen2015does}. Apart from that, the hazard ratio - as a single value - fails to convey potentially time-varying effects and also depends on the duration of the study \citep{hernan2010the}. We therefore consider the risk difference in terms of the cumulative incidence function as effect measure instead. This way, a competing risks framework is accommodated on top, which covers the standard survival setting as a special case. Examples of observational studies that compare treatment effects using the cumulative incidence function include \citet{philipps2020pregnancy, butt2021vitamin, chauhan2022a}. \\
Besides the estimated average treatment effect, researchers are often also interested in further statistical inference. The stochastic process associated with the estimated cumulative incidence function is rather complex, making it difficult to derive exact confidence intervals and bands, though. A commonly applied remedy is the classical nonparametric bootstrap proposed by \citet{efron1981censored} \citep[cf.][]{stensrud2020separable, ryalen2020causal, neumann2016covariate}, even though this resampling method is not optimal in several situations, e.g. when the data lack independence \citep{singh1981on, friedrich2017permuting}. \citet{ozenne2020on} presented an alternative approach based on the influence function, and as counting processes are inherent to time-to-event analysis, resampling methods relying on martingale theory further suggest themselves.

In this paper, we illustrate that apart from the method proposed by \citet{ozenne2020on}, the classical bootstrap as well as the martingale-based wild bootstrap also accurately approximate the distribution of the stochastic process at hand. We compare the performance of these resampling approaches in terms of the resulting confidence intervals and bands by means of simulations as well as an applied data example recording the long-term outcomes of early-stage Hodgkin's disease patients. \\
The remainder of this manuscript is organized as follows. Section~\ref{sec:ATE} establishes the setting and notation as well as the causal estimator for the average treatment effect. In Section~\ref{sec:CIB}, we introduce the three mentioned resampling approaches. The simulation study and the analysis of the Hodgkin's disease data are presented in Sections~\ref{sec:simulations} and \ref{sec:hd_data}. Finally, the paper concludes with a discussion.

\section{Average treatment effect for right-censored data with competing risks \label{sec:ATE}}

We consider a competing risks setting with $K$ failure types. Let the absolutely continuous random variables $T$ and $C$ denote an individual's event and censoring time, respectively. The observed data include ${T \land C}$, the minimum of $T$ and $C$, as well as an indicator ${D \in \{0, 1, \dots, K\}}$, which represents the type of failure. W.l.o.g., let ${D = 1}$ imply that a subject experienced the event of interest. If $D = 0$, the event time is censored, i.e.~${C < T}$. Besides, we observe a binary treatment indicator $A$ and a bounded, $p$-dimensional vector $\boldsymbol{Z}$ of baseline covariates. Throughout this paper, suppose that the data sample $\smash{\{(T_i \!\land\! C_i, D_i, A_i, \boldsymbol{Z_i})\}_{i \in \{1, \dots, n\}}}$ is independent and identically distributed (i.i.d.), and does not include any tied event times. It is further assumed that $\smash{T_i}$ and $\smash{C_i}$ are conditionally independent given $\smash{(A_i, \boldsymbol{Z_i})}$.

For a fixed time point $t$ within the study time interval ${[0, \tau]}$, we define the average treatment effect of interest as $\smash{ATE(t) = \mathbb{E}\left(F_1^1(t) - F_1^0(t)\right)}$. The expression $\smash{F_1^a(t) = P(T^a \leq t, \, D^a = 1)}$ refers to the potential cumulative incidence function for cause $1$ under treatment $a$, applying the counterfactual notation as in \citet{hernan2020causal}. If the conditions of exchangeability, positivity and consistency along with a well-defined intervention and no interference are fulfilled (see e.g. \citealp[][section~I.3]{hernan2020causal} for a thorough description of these assumptions), the g-formula yields an estimate of the average treatment effect \citep{ozenne2020on}:
\[\widehat{ATE}(t) = \frac{1}{n} \sum_{i=1}^n \left(\hat{F}_1(t \mid A = 1, \boldsymbol{Z_i}) - \hat{F}_1(t \mid A = 0, \boldsymbol{Z_i})\right).\]
Despite the issues pointed out by \citet{aalen2015does}, it is reasonable to derive the cumulative incidence function - and hence $\smash{\widehat{ATE}}$ - from hazard rates; the key point is that the causal interpretation of the effect estimate relies on $\smash{\hat{F}_1}$. Let therefore $\smash{\hat{\Lambda}_k(t \mid a, \boldsymbol{z})}$, ${k \in \{1, \dots, K\}}$ be the estimator of the cause-specific, conditional cumulative hazard, and define
\[\hat{F}_1(t \mid a, \boldsymbol{z}) = \int_0^t \exp\left(-\sum_{k=1}^K \hat{\Lambda}_k(s \mid a, \boldsymbol{z})\right) \, \text{d}\hat{\Lambda}_1(s \mid a, \boldsymbol{z}),\]
in line with the characterization proposed by \citet{benichou1990estimates}. One possibility to obtain $\smash{\hat{\Lambda}_k(t \mid a, \boldsymbol{z})}$ is to fit a cause-$k$~specific Cox model with covariates $A$ and $\boldsymbol{Z}$, i.e.
\[\hat{\Lambda}_k(t \mid a, \boldsymbol{z}) = \hat{\Lambda}_{0k}(t) \exp(\hat{\beta}_{kA} a + \skew{3.5}\hat{\boldsymbol{\beta}}_{\boldsymbol{kZ}}^T \boldsymbol{z}),\]
with $\smash{\skew{3.5}\hat{\boldsymbol{\beta}}_{\boldsymbol{k}} = (\hat{\beta}_{kA}, \skew{3.5}\hat{\boldsymbol{\beta}}_{\boldsymbol{kZ}}^T)^T}$ representing the estimated vector of regression coefficients. In fact, the covariates may vary for different causes as long as $A$ is included in the model for the cause of interest. The Breslow estimator eventually yields the approximation
\[\hat{\Lambda}_{0k}(t) = \int_0^t \frac{\text{d}N_k(s)}{\sum_{i=1}^n Y_i(s) \exp(\hat{\beta}_{kA} A_i + \skew{3.5}\hat{\boldsymbol{\beta}}_{\boldsymbol{kZ}}^T \boldsymbol{Z_i})}\]
of the cumulative baseline hazard \citep{breslow1972contribution}. Letting ${\mathbbm{1}\{\cdot\}}$ denote the indicator function, the counting process $\smash{N_k(t)}$ is defined as $\smash{\sum_{i=1}^n N_{ki}(t)}$ with $\smash{N_{ki}(t) = \mathbbm{1}\{T_i \!\land\! C_i \leq t, \, D_i = k\}}$, such that $\smash{\text{d}N_k(t)}$ represents the increment of $\smash{N_k(t)}$ over the infinitesimal time interval ${[t, t+dt)}$. The at-risk indicator $\smash{Y_i(t) = \mathbbm{1}\{T_i \!\land\! C_i \geq t\}}$ further specifies whether subject $i$ is part of the risk set just prior to time $t$.

\section{Confidence intervals and bands \label{sec:CIB}}

Pointwise confidence intervals and time-simultaneous confidence bands are routinely reported in clinical trials as they help to assess the (un)certainty of an estimate. It is not straightforward to define such intervals for the average treatment effect, however, due to the complexity of the stochastic process $\smash{U_n(t) = \sqrt{n} \, (\widehat{ATE}(t) - ATE(t))}$. As a workaround, we aim to approximate the limiting distribution of $\smash{U_n}$ by means of different resampling approaches.

\subsection{Efron's bootstrap}

The most common way to derive confidence intervals for $ATE$ is the use of the classical nonparametric bootstrap \citep{efron1981censored}, which does not require knowledge of the true underlying distribution. By repeatedly drawing with replacement from the data and calculating a statistical functional of interest in each of the drawn samples, one tries to approach the distribution of the functional in the target population. In the given context, we obtain the estimates $\smash{\{\smash{\widehat{ATE}}^*_b(t)\}_{b \in \{1, \dots, B\}}}$ from $B$ bootstrap samples of the original data, each having size $n$. A confidence interval at level ${(1-\alpha)}$ can, for instance, be determined by setting the empirical $\smash{\tfrac{\alpha}{2}}$ and $\smash{(1 - \tfrac{\alpha}{2})}$ quantiles of the bootstrap estimates as limits. Furthermore, we construct a simultaneous confidence band over the time interval $\smash{[t_1, t_2]}$ as
\[\left[\widehat{ATE}(t) - q_{1-\alpha}^{E\!B} \sqrt{\hat{\nu}^{E\!B}(t)}, \ \widehat{ATE}(t) + q_{1-\alpha}^{E\!B} \sqrt{\hat{\nu}^{E\!B}(t)}\right],\] 
with $\smash{\hat{\nu}^{E\!B}(t)}$ referring to the empirical variance of the bootstrap estimates and $\smash{q_{1-\alpha}^{E\!B}}$ denoting the ${(1-\alpha)}$ quantile of 
\[\left\{\sup_{t \in [t_1, t_2]} \left|\frac{\smash{\widehat{ATE}}^*_b(t) - \tfrac{1}{B} \sum_{\tilde{b}=1}^B \smash{\widehat{ATE}}^*_{\tilde{b}}(t)}{\sqrt{\hat{\nu}^{E\!B}(t)}}\right|\right\}_{b \in \{1, \dots, B\}}.\]
This approach yields asymptotically correct results in many less intricate settings (as long as the considered data are i.i.d.), and its theoretical validity is proven in \citet{ruehl2023asymptotic} based on martingale arguments. \\
While the implementation of \citeauthor{efron1981censored}'s bootstrap is rather simple, the computation time can become excessive with large sample sizes and multiple bootstrap iterations.

\subsection{Influence function \label{subsec:asymptotics_IF}}

Another method to obtain confidence intervals for $ATE$ has been described by \citet{ozenne2020on}. Provided that the underlying model is correct, the functional delta method yields an approximation of the asymptotic distribution of $\smash{U_n}$ at a given time point w.r.t. the influence function of the average treatment effect. More specifically,
\[U_n(t) = \frac{1}{\sqrt{n}} \sum_{i=1}^n I\!F(t; \, T_i \!\land\! C_i, \! D_i, \! A_i, \! \boldsymbol{Z_i}) + o_P(1) \  \overset{\mathscr{D}}{\longrightarrow} \  \mathcal{N}\left(0, \int \left(I\!F(t; \, s, \! d, \! a, \! z)\right)^2 \, \text{d}P(s, \! d, \! a, \! z)\right),\]
as $n$ tends to infinity. The definition of the influence function ${I\!F}$ according to \citet{ozenne2020on, ozenne2017riskRegression} can be found in the supplementary material. Besides, we use $\mathcal{N}$ throughout this paper to symbolize the normal distribution. It follows that the plug-in estimator $\smash{\hat{\nu}^{I\!F}(t) = \tfrac{1}{n} \sum_{i=1}^n \left(\smash{\widehat{I\!F}(t; \, T_i \!\land\! C_i, \! D_i, \! A_i, \! \boldsymbol{Z_i}})\right)^2}$ is consistent for the asymptotic variance of $\smash{U_n(t)}$ and thus, confidence intervals are easy to calculate. The construction of confidence bands, on the other hand, is more involved. This is because the dependence between the increments of the process $\smash{U_n}$ must be taken into account when making inferences concerning multiple time points. It can be shown that $\smash{U_n}$ converges weakly to a zero-mean Gaussian process on the Skorokhod space ${\mathcal{D}[0, \tau]}$ \citep{ruehl2023asymptotic}, and thus, we can derive a ${(1-\alpha)}$ confidence band for ${ATE}$ over the interval $\smash{[t_1, t_2]}$ in line with the resampling approach described by \citet{scheike2008flexible}: 
\[\left[\widehat{ATE}(t) - q_{1-\alpha}^{I\!F} \sqrt{\hat{\nu}^{I\!F}(t)}, \ \widehat{ATE}(t) + q_{1-\alpha}^{I\!F} \sqrt{\hat{\nu}^{I\!F}(t)}\right].\]
Here, $\smash{q_{1-\alpha}^{I\!F}}$ denotes the ${(1-\alpha)}$ quantile of 
\[\left\{\sup_{t \in [t_1, t_2]} \left|\sum_{i=1}^n \frac{\widehat{I\!F}(t; \, T_i \!\land\! C_i, \! D_i, \! A_i, \! \boldsymbol{Z_i})}{\sqrt{\hat{\nu}^{I\!F}(t)}} \cdot G_i^{I\!F; (b)} \right|\right\}_{b \in \{1, \dots, B\}}\] 
for $B$ independent standard normal vectors $\smash{\{(G_1^{I\!F; (b)}, \dots, G_n^{I\!F; (b)})^T\}_{b \in \{1, \dots, B\}}}$. \\
As compared to the classical bootstrap, the influence function approach significantly reduces the computation time, considering that the resampling step builds upon repeated generation of random variables rather than the recalculation of functionals based on various individual data sets.

\subsection{Wild bootstrap \label{subsec:asymptotics_wild_bs}}

A third resampling method arises from the fact that the limiting distribution of $\smash{U_n}$ may be represented in terms of martingales: It can be shown that 
\[U_n(t) = \sum_{k=1}^K \sum_{i=1}^n \left(\int_0^t H_{k1i}(s,t) \, \text{d}M_{ki}(s) + \int_0^\tau H_{k2i}(s,t) \, \text{d}M_{ki}(s)\right) + o_p(1)\]
for functions $\smash{H_{k1i}}$ and $\smash{H_{k2i}}$ as defined in the supplementary material and $\smash{M_{ki}(t) = N_{ki}(t) - \int_0^t Y_i(s) \, \text{d}\Lambda_k(s \mid A_i, \boldsymbol{Z_i})}$, ${k \in \{1, \dots, K\}}$, ${i \in \{1, \dots, n\}}$ \citep{ruehl2023asymptotic}. Note that $\smash{M_{ki}}$ is a martingale relative to the history $\smash{\left(\mathscr{F}_t\right)_{t \geq 0}}$ that is generated by the data observed until a given time, i.e.~$\smash{\mathbb{E}\left(\text{d}M_{ki}(t) \mid \mathscr{F}_{t-}\right) = 0}$ and
\[\text{Var}\left(\text{d}M_{ki}(t) \mid \mathscr{F}_{t-}\right) = Y_i(t) \, \text{d}\Lambda_k(t \mid A_i, \boldsymbol{Z_i}).\]
Provided that Aalen's multiplicative intensity model \citep{aalen1978nonparametric} applies, the characterization of the variance equals the conditional expectation of $\smash{\text{d}N_{ki}(t)}$ given the past $\smash{\mathscr{F}_{t-}}$. This motivates the general idea of the wild bootstrap: By replacing $\smash{\text{d}M_{ki}(t)}$ with the product of $\smash{\text{d}N_{ki}(t)}$ and suitable random multipliers $\smash{G_i^{W\!B}}$, $k \in \{1, \dots, K\}$, $i \in \{1, \dots, n\}$, we can approximate the asymptotic distribution of $\smash{U_n}$. The initial method described by \citet{lin1993checking} only covered standard normal multipliers, but was later extended to more general resampling schemes (\citealp[cf.][]{beyersmann2013weak, dobler2017non}). In \citet{ruehl2023asymptotic}, we followed ideas of \citet{cheng1998prediction, beyersmann2013weak} and \citet{dobler2017non} to formally prove that, conditional on the data,
\[\hat{U}_n(t) = \sum_{k=1}^K \sum_{i=1}^n \left(\hat{H}_{k1i}(T_i \!\land\! C_i, t) \, N_{ki}(t) G_i^{W\!B} + \hat{H}_{k2i}(T_i \!\land\! C_i, t) \, N_{ki}(\tau) G_i^{W\!B} \right)\] 
converges weakly to the same process as $\smash{U_n}$ on ${\mathcal{D}[0, \tau]}$. (Here, the estimates $\smash{\hat{H}_{k1i}}$ and $\smash{\hat{H}_{k2i}}$ are calculated by plugging appropriate sample estimates into the definition of $\smash{H_{k1i}}$ and $\smash{H_{k2i}}$.)
\begin{remark} \label{rem:multipliers}
The following choices of multipliers $\smash{G_i^{W\!B}}$ fulfill the necessary conditions for the wild bootstrap \citep[cf.][]{dobler2017non}:
\begin{itemize}
\item $\smash{G_i^{W\!B} \overset{\text{i.i.d.}}{\sim} \mathcal{N}(0,1)}$, i.e.~independent standard normal multipliers (according to the original resampling approach by \citealp{lin1993checking});
\item $\smash{G_i^{W\!B} \overset{\text{i.i.d.}}{\sim} \mathcal{P}ois(1) - 1}$, that is, independent and centered unit Poisson multipliers (in line with the proposition of \citealp{beyersmann2013weak});
\item $\smash{G_i^{W\!B} \sim \mathcal{B}\left(Y(T_i \!\land\! C_i), \frac{1}{Y(T_i \land C_i)}\right) - 1}$ with $\smash{Y(t) = \sum_{i=1}^n Y_i(t)}$ and $\smash{(G_{i_1}^{W\!B} \perp \!\!\! \perp G_{i_2}^{W\!B}) \mid \mathscr{F}_{\tau}}$ for $\smash{i_1 \neq i_2}$, \newline
i.e.~conditionally independent, centered binomial multipliers. This version of the wild bootstrap is equivalent to the so-called weird bootstrap described in \citet[subsection~IV.1.4]{andersen1993statistical}, as \citet{dobler2017non} illustrate.
\end{itemize}
\end{remark}
For multiple multiplier realizations $\smash{\{(G_1^{W\!B; (b)}, \dots, G_n^{W\!B; (b)})^T\}_{b \in \{1, \dots, B\}}}$, one obtains the ${(1 - \alpha)}$ confidence interval
\[\left[\widehat{ATE}(t) - \frac{1}{\sqrt{n}} \, q_{1 - \alpha}^{W\!B}(t), \  \widehat{ATE}(t) + \frac{1}{\sqrt{n}} \, q_{1 - \alpha}^{W\!B}(t)\right]\]
with ${(1 - \alpha)}$ quantile $\smash{q_{1 - \alpha}^{W\!B}(t)}$ of $\smash{\{\left|\smash{\hat{U}_n^{(b)}(t)}\right|\}_{b \in \{1, \dots, B\}}}$. Similarly,
\[\left[\widehat{ATE}(t) - \frac{1}{\sqrt{n}} \, q_{1 - \alpha}^{W\!B} \sqrt{\hat{\nu}^{W\!B}(t)}, \  \widehat{ATE}(t) + \frac{1}{\sqrt{n}} \, q_{1 - \alpha}^{W\!B} \sqrt{\hat{\nu}^{W\!B}(t)} \right]\]
specifies a simultaneous ${(1 - \alpha)}$ confidence band over the interval $\smash{[t_1, t_2]}$, considering the empirical variance estimator $\smash{\hat{\nu}^{W\!B}(t)}$ of $\smash{\{\hat{U}_n^{(b)}(t)\}_{b \in \{1, \dots, B\}}}$ and the ${(1-\alpha)}$ quantile $\smash{q_{1-\alpha}^{W\!B}}$ of 
\[\left\{\sup_{t \in [t_1, t_2]} \left|\frac{\hat{U}_n^{(b)}(t)}{\sqrt{\hat{\nu}^{W\!B}(t)}}\right|\right\}_{b \in \{1, \dots, B\}}.\]
The described bootstrap, just like the influence function, takes only a fraction of the time required by the classical bootstrap. In addition, martingale-based resampling approaches are built upon the condition of independent right-censoring (\citealp[][subsection~III.2.2]{andersen1993statistical}, \citealp[cf.][]{ruehl2022general}) and do not rely on a strict i.i.d. setup. Therefore, they are less sensitive to deviations from standard assumptions where \citeauthor{efron1981censored}'s approach is known to fail, including dependencies inherent to the data \citep{singh1981on, friedrich2017permuting}.

\section{Simulation study \label{sec:simulations}}

In order to compare the performance of the resampling approaches described in Section~\ref{sec:CIB}, we simulated competing risks data following the same scheme as in \citet{ozenne2020on}, and constructed confidence intervals and bands using the proposed methods.

\subsection{Data generation}

The generated data comprised twelve independent covariates, namely ${Z_1, \dots, Z_6}$ following a mean-zero normal distribution and ${Z_7, \dots, Z_{12}}$ being Bernoulli distributed with parameter 0.5. Each covariate affected the treatment probability, the event time distributions of two competing failure causes and a conditionally independent censoring time in an individual manner (see Table~\ref{tab:covariate_effects}). The treatment indicator $A$ was for instance derived from a logistic regression model with linear predictor ${\alpha_0 + \log(2) \cdot \left(Z_1 - Z_2 + Z_6 + Z_7 - Z_8 + Z_{12}\right)}$. Here, the intercept $\alpha_0$ controls the overall frequency of treatment. Apart from that, we simulated censoring and event times according to a Weibull distribution with hazard ${\lambda(t) = 0.02 \, t \exp\left(\smash{\beta_{dA} A + \boldsymbol{\beta}^T_{\boldsymbol{dZ}} \boldsymbol{Z}}\right)}$ for corresponding parameters $\beta_{dA}$ and $\boldsymbol{\beta_{dZ}}$, ${d \in \{0, 1, 2\}}$, respectively. The minimum of the three resulting (latent) times determined the type of observation (i.e.~censored, type~1 or type~2 event).
\begin{table}[h!tbp]
\centering
\caption{Effects of the covariates on the treatment probability, event and censoring times. \label{tab:covariate_effects}}
\begin{tabular}{ccccc}
\toprule
\multirow{2}{*}{Covariate} & \multirow{2}{*}{\parbox{3.1cm}{\centering Odds Ratio w.r.t. \\treatment probability}} & \multirow{2}{*}{\parbox{3.1cm}{\centering Hazard ratio w.r.t. \\event of interest}} & \multirow{2}{*}{\parbox{3.1cm}{\centering Hazard ratio w.r.t. \\competing event}} & \multirow{2}{*}{\parbox{3.1cm}{\centering Hazard ratio w.r.t. \\censoring}} \\
& & & & \\
\midrule
\multirow{2}{*}{$A$} & \multirow{2}{*}{-} & $\exp\left(\beta_{1A}\right)$, & \multirow{2}{*}{1.0} & \multirow{2}{*}{1.0} \\[-2pt]
& & $\beta_{1A} \in \{-2, 0, 2\}$ & & \\
$Z_1$ / $Z_7$ & 2.0 & 2.0 & 0.5 & 0.5 \\
$Z_2$ / $Z_8$ & 0.5 & 1.0 & 1.0 & 1.0 \\
$Z_3$ / $Z_9$ & 1.0 & 2.0 & 1.0 & 1.0 \\
$Z_4$ / $Z_{10}$ & 1.0 & 1.0 & 1.0 & 2.0 \\
$Z_5$ / $Z_{11}$ & 1.0 & 1.0 & 2.0 & 1.0 \\
$Z_6$ / $Z_{12}$ & 2.0 & 2.0 & 2.0 & 0.5 \\
\bottomrule
\end{tabular}
\end{table}

This general simulation scheme served as a basis for a variety of scenarios, each implemented with sample sizes of ${n \in \{50, 75, 100, 200, 300\}}$ and treatment effects according to parameter ${\beta_{1A} \in \{-2,0,2\}}$. By default, about half of the observations were assigned to be treated, and the event of interest was observed in a third, half or two thirds of the subjects until time ${t = 9}$, corresponding to the case where $\smash{\beta_{1A} = -2,0,2}$, respectively. The frequency of censoring amounted to 17\%, 14\% or 11\% by ${t = 9}$, whereas the competing event affected 41\%, 31\% or 21\% of the subjects. (If necessary, the data generation step was repeated until at least 10 events of both causes were observed, such that meaningless regression outcomes could be averted.) Among the examined scenarios were settings with varying degrees of censoring (namely 0\%, 14\% and 30\% in the case without treatment effect, i.e.~$\smash{\beta_{1A}=0}$), treatment frequencies of 22\% as well as 86\% and non-unit variances (0.25 and 4) of the normally distributed covariates. Besides, we considered a standard survival scenario without competing events that involved type~II censoring with staggered entry in order to investigate a setting with independent, but not random censoring \citep{ruehl2022general}. For an overview of the different scenarios, see Table~\ref{tab:simu_scenarios}.

\begin{table}[h!tbp]
\centering
\caption{Overview of the simulation scenarios. \label{tab:simu_scenarios}}
\begin{tabular}{lcccccccc}
\toprule
\multirow{2}{*}{Scenario} & \multicolumn{3}{c}{\% censored at $t = 9$\textsuperscript{*}} & \multicolumn{3}{c}{\% type~1 events at $t = 9$\textsuperscript{*}} & \multirow{2}{*}{\% treated} & \multirow{2}{*}{$Var(Z_1)$} \\
\cmidrule(lr{.75em}){2-4} \cmidrule(lr{.75em}){5-7}
& $\beta_{1A} \!=\! -2$ & $\beta_{1A} \!=\! 0$ & $\beta_{1A} \!=\! 2$ & $\beta_{1A} \!=\! -2$ & $\beta_{1A} \!=\! 0$ & $\beta_{1A} \!=\! 2$ & & \\
\midrule
No censoring & \hphantom{0}0.0 & \hphantom{0}0.0 & \hphantom{0}0.0 & 35.7 & 56.1 & 70.3 & 56.4 & 1.00 \\
Light censoring & 16.7 & 14.0 & 11.0 & 32.2 & 51.5 & 66.2 & 56.4 & 1.00 \\
Heavy censoring & 35.3 & 29.7 & 23.0 & 27.0 & 44.5 & 60.1 & 56.4 & 1.00 \\
Low treatment probability & 14.9 & 14.0 & 13.1 & 43.7 & 51.5 & 56.6 & 22.3 & 1.00 \\
High treatment probability & 18.2 & 14.0 & \hphantom{0}8.3 & 23.5 & 51.5 & 75.6 & 85.8 & 1.00 \\
Low variance of the covariates & 13.7 & 10.7 & \hphantom{0}7.3 & 32.4 & 55.2 & 72.0 & 57.4 & 0.25 \\
High variance of the covariates & 22.0 & 20.2 & 17.9 & 32.6 & 45.6 & 56.4 & 54.6 & 4.00 \\
Type~II censoring & 49.7 & 39.2 & 25.0 & 50.0 & 49.5 & 48.4 & 56.4 & 1.00 \\
\bottomrule
\end{tabular}
\parbox{0.93\textwidth}{\footnotesize *: For the scenario with type~II censoring, the percentages of censoring and type~1 events are determined at ${t = 10}$, ${t = 5}$, and \\
\hphantom{*: } ${t = 2.5}$, for $\smash{\beta_{1A} \in \{-2, 0, 2\}}$, respectively.}
\end{table}

Confidence intervals (at time points ${t \in \{1,3,5,7,9\}}$) and bands (over the time interval ${[0,9]}$) for the average treatment effect were derived by applying \citeauthor{efron1981censored}'s bootstrap (EBS), the influence function approach (IF) and the wild bootstrap (WBS) to each generated data set. While the EBS was implemented with $1,000$ repetitions to maintain reasonable runtimes, we considered $10,000$ random multipliers for the two remaining methods. The WBS was realized with standard normal, Poisson and binomial multipliers according to Remark~\ref{rem:multipliers}. We then assessed the performance of the distinct methods by means of the associated 95\% coverage probabilities and the widths of the confidence ranges. The maximum Monte Carlo standard errors amounted to 0.77\% and 0.013 for the coverage and the width, respectively, as all simulation scenarios were repeated $5,000$ times.

\begin{figure}[h!tbp]
\centering
\caption{True average treatment effect. \label{fig:true_ATE}}
\includegraphics[width = 0.55\textwidth]{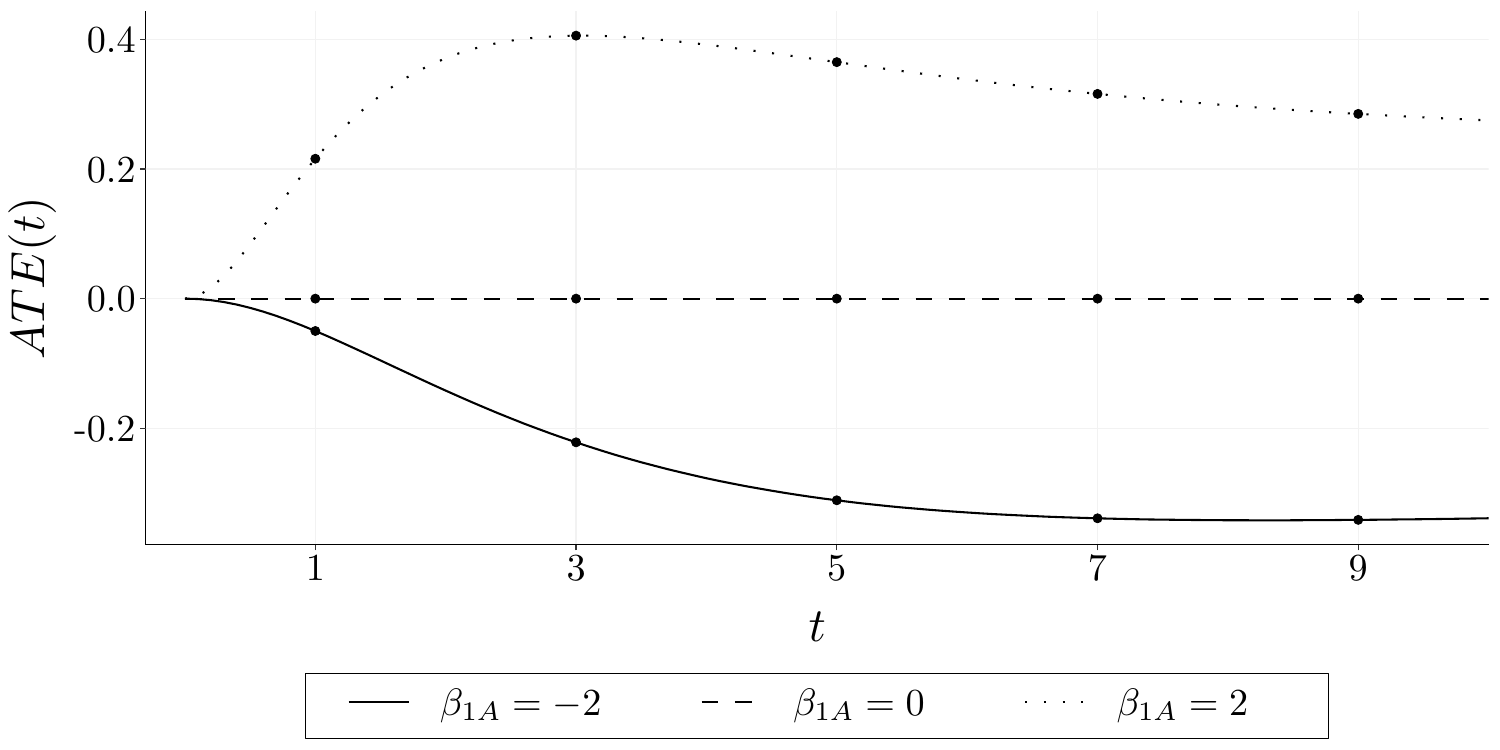}
\end{figure}

In order to determine the true average treatment effect for each of the mentioned scenarios, we considered $1,000$ simulated data sets with sample size ${n = 100,000}$, random treatment assignment independent of the covariates and no censoring. Figure~\ref{fig:true_ATE} depicts $ATE(t)$ except for the scenarios with non-unit variance of the covariates ${Z_1, \dots, Z_6}$ or type~II censoring.

\subsection{Results \label{subsec:sim_results}}

\begin{figure}[h!tbp]
\centering
\caption{Coverage of the confidence intervals in the scenario with light censoring and a positive average treatment effect. \label{fig:CI_coverage_ATE_lowCens}}
\includegraphics[width = \textwidth]{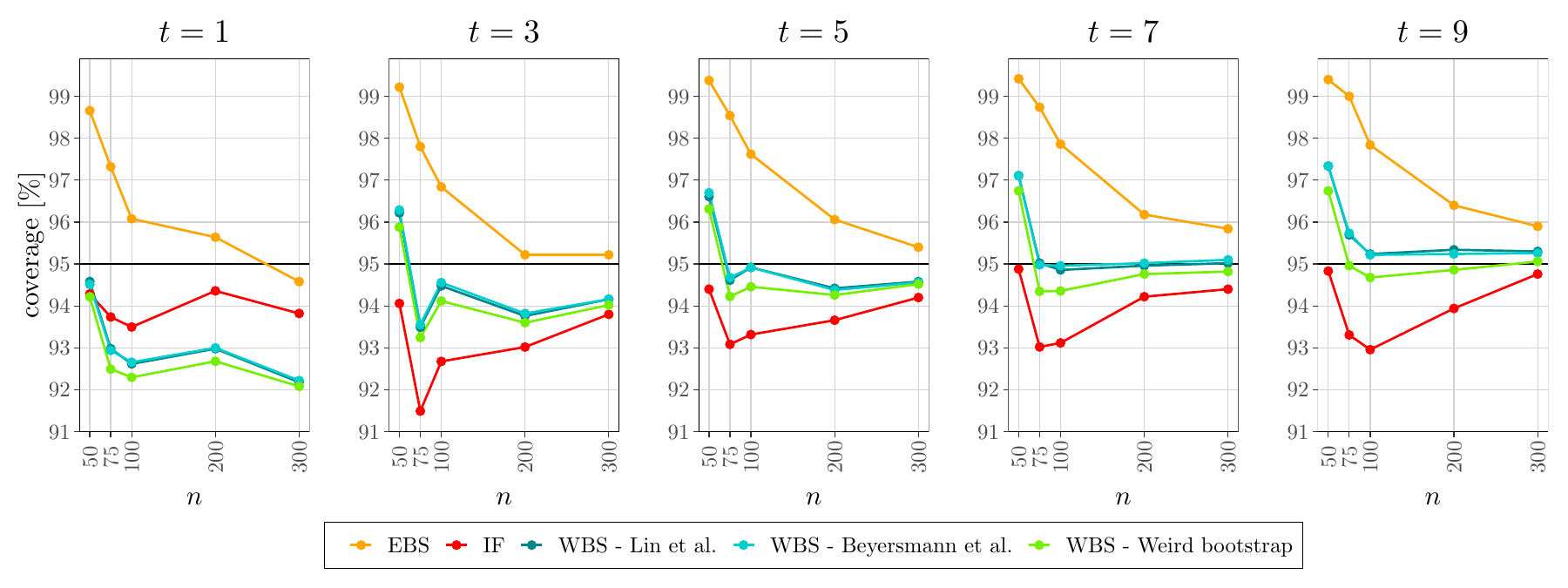}
\end{figure}

\begin{figure}[h!tbp]
\centering
\caption{Coverage of the confidence intervals in the scenario without censoring and a negative average treatment effect. \label{fig:CI_coverage_advATE_noCens}}
\includegraphics[width = \textwidth]{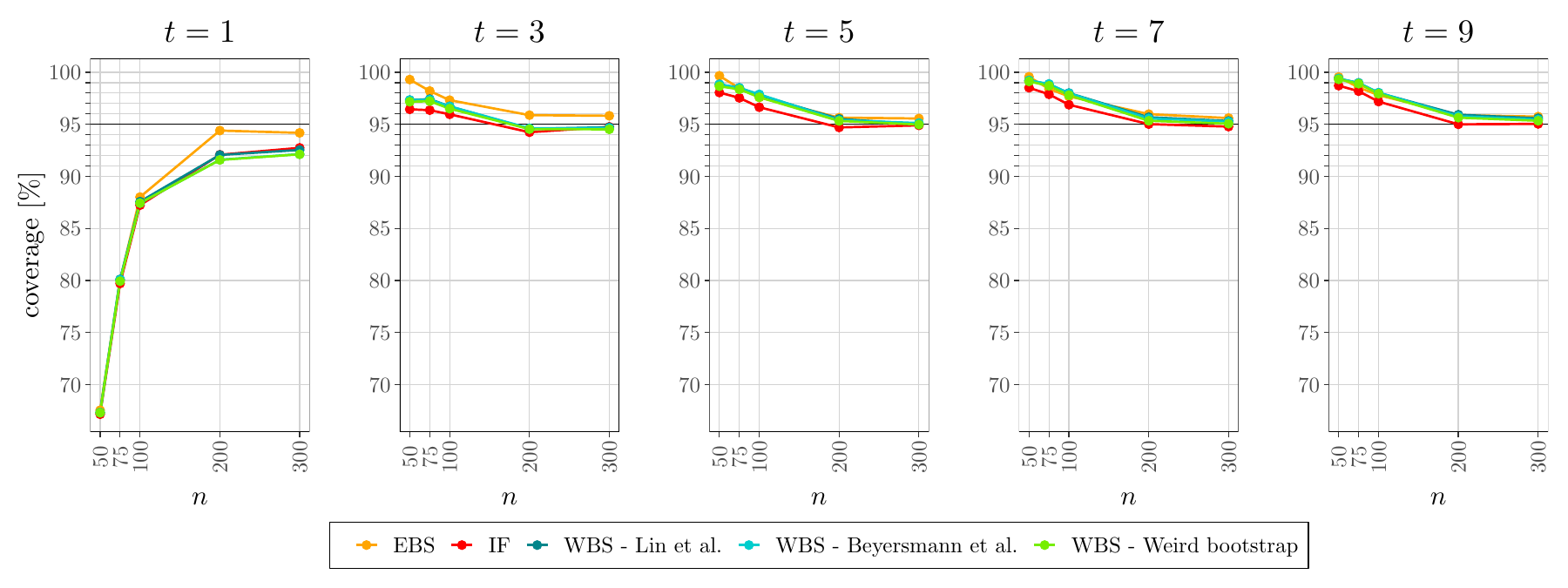}
\end{figure}

The WBS attained coverage probabilities that were, in total, the closest to the target level of 95\%. The mean absolute deviation across all scenarios, sample sizes and time points was 2.40\% for the WBS vs. 2.43\% and 2.63\% for the IF and the EBS, respectively. Throughout nearly all settings, the confidence intervals obtained by the EBS yielded coverages above those derived from the different WBS versions, whereas the IF intervals included the true average treatment effect the least frequently. Figure~\ref{fig:CI_coverage_ATE_lowCens} illustrates this ranking in the case with low-level censoring and a positive average treatment effect (i.e.~$\smash{\beta_{1A}=2}$, referring here and in the following to the sign of the causal risk difference, i.e.~a positive average treatment effect indicates that the potential cumulative incidence under treatment is higher than that under no treatment). We observed similar outcomes in the other scenarios that involved treatment effects according to $\smash{\beta_{1A} \in \{0,2\}}$ (see supplementary material), even though the performance of the resampling methods varied for small sample sizes up to 75 (e.g. with treatment probabilities larger than 0.5).
An exception was the setting with widely dispersed covariates: Here, all methods provided rather conservative confidence intervals, and as a consequence, the IF approach achieved the most accurate coverages. The same effect was also encountered in the scenarios with negative treatment effect. It should be noted, however, that the differences between the distinct resampling techniques were negligible in most cases with $\smash{\beta_{1A} = -2}$, apart from very early time points or sample sizes below 100 (see Figure~\ref{fig:CI_coverage_advATE_noCens}). Greater dissimilarities were only present in the setting with high covariance of the covariates (where the EBS performed best for larger sample sizes).
A common feature of all the schemes that yielded coverages along the lines of Figure~\ref{fig:CI_coverage_advATE_noCens} is that the proportion of observed type~1 events was lower than in the scenarios with $\smash{\beta_{1A} \in \{0,2\}}$. This is due to the prevalence of the competing event, and the IF approach seems to be slightly more suitable to cope with that condition than the bootstrap methods. \\
The WBS, in contrast, generally reached its full potential towards later time points, when a sufficient amount of data was available. This became apparent in the setting with type~II censoring and a positive average treatment effect: Because of the absence of any competing events, we evaluated the confidence intervals at earlier times ${t \in \{0.5, 1, 1.5, 2, 2.5\}}$, and the WBS did not attain coverages as close to 95\% as those obtained by the IF and the EBS until $t = 2$. \\
Against our expectations, the simulations revealed no considerable superiority of the martingale-based methods in case of type~II censoring with staggered entry, despite non-random censoring. It appears as if the dependence within the data was too weak for the sample sizes considered and the lack of additional pressure (e.g.~by internal left-truncation, \citealp[cf.][]{ruehl2022general}).
 
\begin{figure}[h!tbp]
\centering
\caption{Coverage of the confidence bands in the scenario with high treatment probability and a positive average treatment effect. \label{fig:CB_coverage_ATE_highTreatProb}}
\includegraphics[width = 0.65\textwidth]{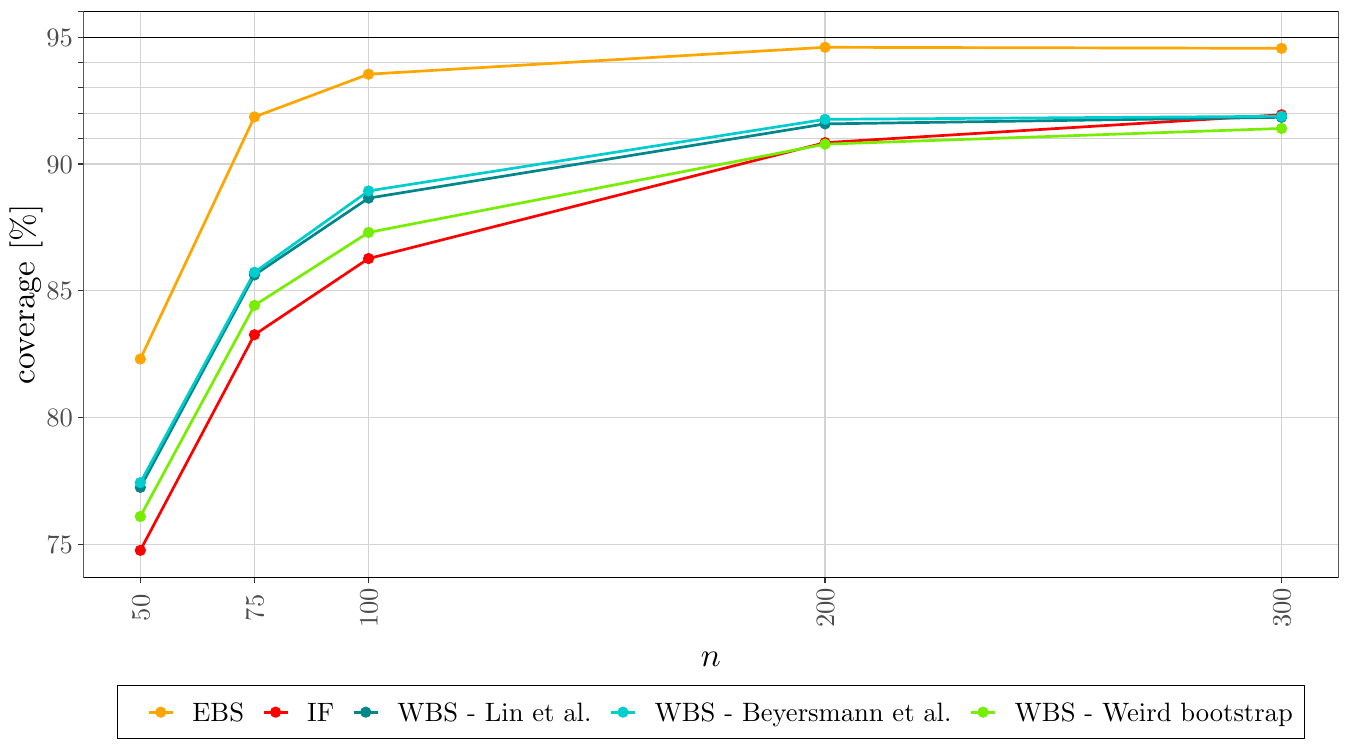}
\end{figure}

The coverage probabilities of the time-simultaneous confidence bands followed a similar trend as was observed for the pointwise intervals: While almost all scenarios with positive or no average treatment effect had the highest and lowest coverages for the EBS and IF, respectively, there were virtually no differences in most of the settings with $\smash{\beta_{1A} = -2}$ (see supplementary material). However, the EBS bands were especially accurate given positive average treatment effects (see Figure~\ref{fig:CB_coverage_ATE_highTreatProb}). On average, the mean absolute discrepancy between the simulated coverages and the nominal level of 95\% was 4.44\% in comparison to 5.31\% and 5.46\% for the WBS and the IF approach, respectively.

\begin{figure}[h!tbp]
\centering
\caption{Width of the confidence intervals at time ${t = 5}$ in the scenario with light censoring and a positive average treatment effect. \label{fig:CI_width_ATE_lowCens}}
\includegraphics[width = 0.65\textwidth]{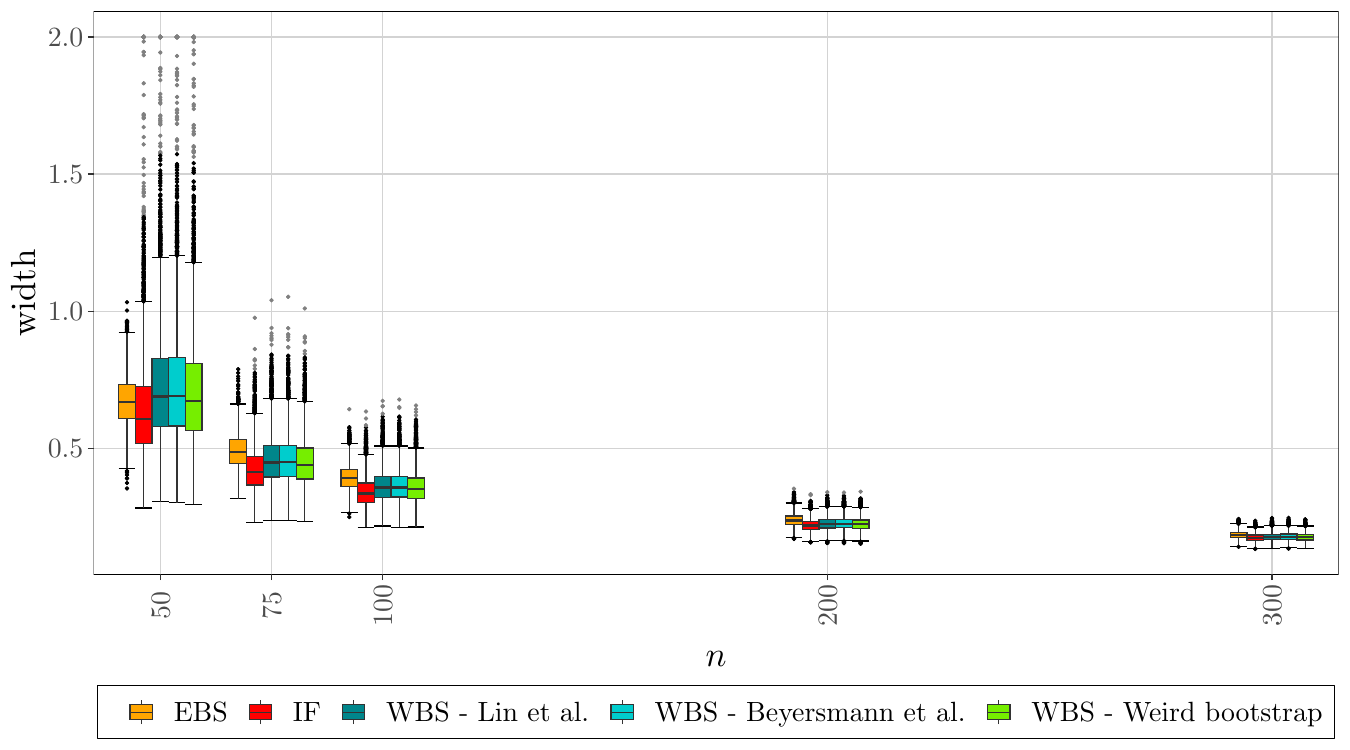}
\end{figure}

Our results imply further that the choice of the multiplier for the WBS does not have any significant impact. Since the confidence intervals derived using the approaches of \citet{lin1993checking} and \citet{beyersmann2013weak} were occasionally wider than those resulting from the weird bootstrap, the latter method attained lower coverages. Which of the multipliers provided the most accurate outcomes varied depending on the situation, however. \\
Other than that, the IF produced narrower intervals than any of the WBS versions, and in case of a negative average treatment effect, either approach lead to considerably greater variation in the interval width by comparison with the situations where $\smash{\beta_{1A} \in \{0, 2\}}$. Interestingly, this effect did not apply to the EBS. The extent of the EBS-based intervals ranged between or above the remaining widths, apart from the settings with $\smash{\beta_{1A} = -2}$. As the sample sizes increased, however, all resampling methods lead to nearly equally wide confidence intervals (cf. Figure~\ref{fig:CI_width_ATE_lowCens}).
The widths of the confidence bands furthermore related to one another in the same way as their pointwise counterparts.

\begin{figure}[h!tbp]
\centering
\caption{Computation times in the scenario with light censoring and a positive average treatment effect. \label{fig:comp_time} (The height of the bars illustrates the mean computation time.)}
\includegraphics[width = 0.65\textwidth]{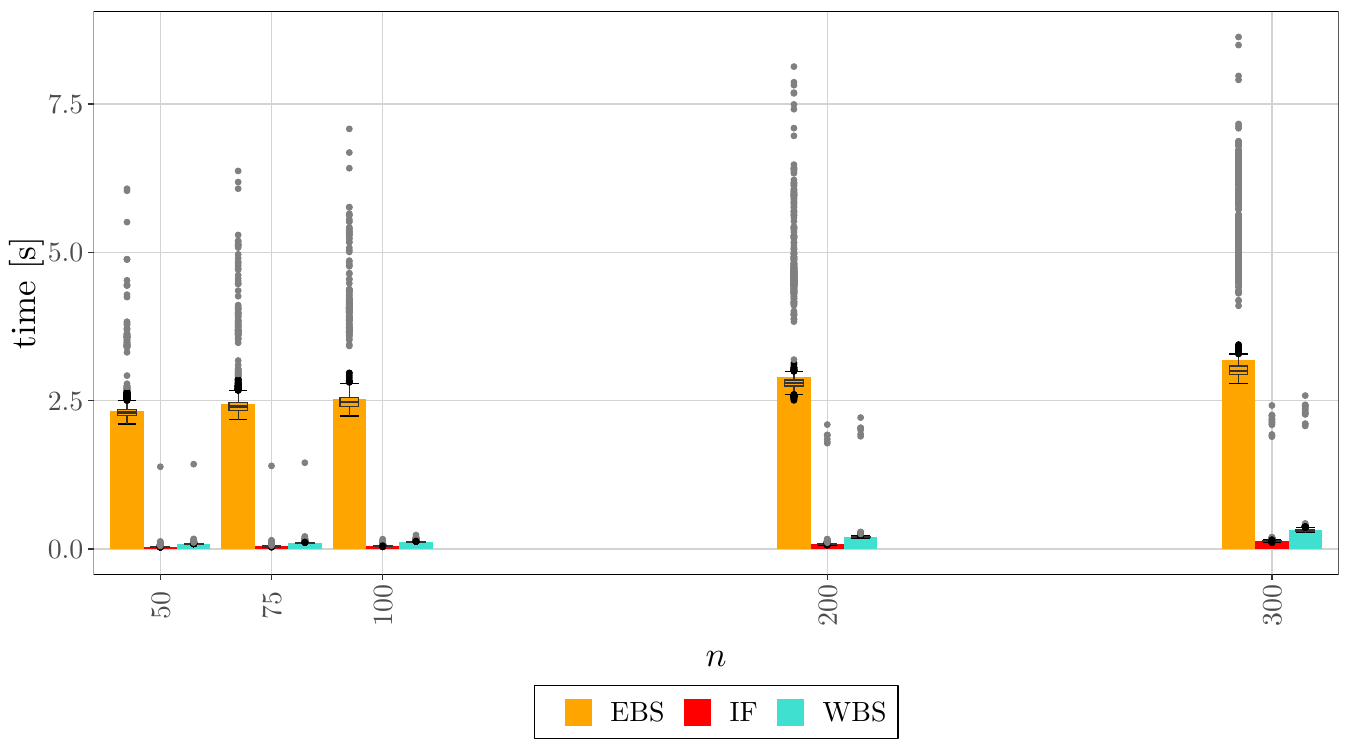}
\end{figure}

Eventually, a last note is in order about the computation times of the distinct methods: The IF and EBS approaches have been implemented in the function `\textit{ate}' of the \texttt{R} \citep{R} package \texttt{riskRegression} by \citet{riskRegression}. The calculations are sped up significantly by interfacing \texttt{C++} code for the IF method and parallelizing the computation of the bootstrap replicates for the EBS. We extracted and adapted the parts of the code that were relevant for our simulations. In addition, \texttt{C++} was also integrated to implement the WBS. The resulting computation times are summarized in Figure~\ref{fig:comp_time}.
Clearly, the EBS is several times slower than the multiplier-based methods. However, while the computation time of the IF approach is somewhat lower than that of the WBS, one should note that three different multipliers have been regarded for the latter.

\section{Real data application \label{sec:hd_data}}

To illustrate the performance of the resampling approaches when applied to real-world study data, we considered records of the long-term disease progression among patients with early-stage Hodgkin's lymphoma (i.e.~stage~I or II) \citep{pintilie2006competing}. These data are available within the \texttt{R}~package \texttt{randomForestSRC} (data '\textit{hd}',  \citealp{randomForestSRC}) and comprise information on 865 subjects who were treated at the Princess Margaret Hospital in Toronto between 1968 and 1986, either with radiation alone (${n = 616}$) or a combination of radiation and chemotherapy (${n = 249}$). We considered the time from diagnosis until death (in years), with prior relapse regarded as competing event. Covariates recorded include age, sex, clinical stage of the lymphoma, size of mediastinum involvement and whether the disease is extranodal (see Table~\ref{tab:hd_data} for a summary of the data). 
\begin{table}[h!tbp]
\centering
\caption{Summary of the Hodgkin's disease data. \label{tab:hd_data}}
\begin{tabular}{lcc}
\toprule
\multirow{3}{*}{Covariate} & \multicolumn{2}{c}{Treatment} \\
\cmidrule(lr{.75em}){2-3}
& \multirow{2}{*}{\parbox{\widthof{Radiation \& chemotherapy}}{\centering Radiation alone \\ ($n = 616$)}} & \multirow{2}{*}{\parbox{\widthof{Radiation \& chemotherapy}}{\centering Radiation \& chemotherapy \\ ($n = 249$)}} \\
& & \\
\midrule
Age, mean (sd) & \parbox{\widthof{331 (53.73\%)}}{\raggedleft 35.93 (16.37)} & \parbox{\widthof{331 (53.73\%)}}{\raggedleft 33.77 (12.86)} \\
Sex: male & \parbox{\widthof{331 (53.73\%)}}{\raggedleft 331 (53.73\%)} & \parbox{\widthof{331 (53.73\%)}}{\raggedleft 132 (53.01\%)} \\
Lymphoma stage: I & \parbox{\widthof{331 (53.73\%)}}{\raggedleft 266 (43.18\%)} & \parbox{\widthof{331 (53.73\%)}}{\raggedleft 30 (12.05\%)} \\
Mediastinum involvement \\
\hspace{3mm} none & \parbox{\widthof{331 (53.73\%)}}{\raggedleft 382 (62.01\%)} & \parbox{\widthof{331 (53.73\%)}}{\raggedleft 82 (32.93\%)} \\
\hspace{3mm} small & \parbox{\widthof{331 (53.73\%)}}{\raggedleft 211 (34.25\%)} & \parbox{\widthof{331 (53.73\%)}}{\raggedleft 77 (30.92\%)} \\
\hspace{3mm} large & \parbox{\widthof{331 (53.73\%)}}{\raggedleft 23 (3.73\%)} & \parbox{\widthof{331 (53.73\%)}}{\raggedleft 90 (36.14\%)} \\
Extranodal disease & \parbox{\widthof{331 (53.73\%)}}{\raggedleft 29 (4.70\%)} & \parbox{\widthof{331 (53.73\%)}}{\raggedleft 50 (20.08\%)} \\
\bottomrule
\end{tabular}
\end{table}

\begin{figure}[h!tbp]
\centering
\caption{Confidence intervals (left) and bands (rights) for the average treatment effect on the risk of death. \label{fig:hd_analysis}}
\includegraphics[width = \textwidth]{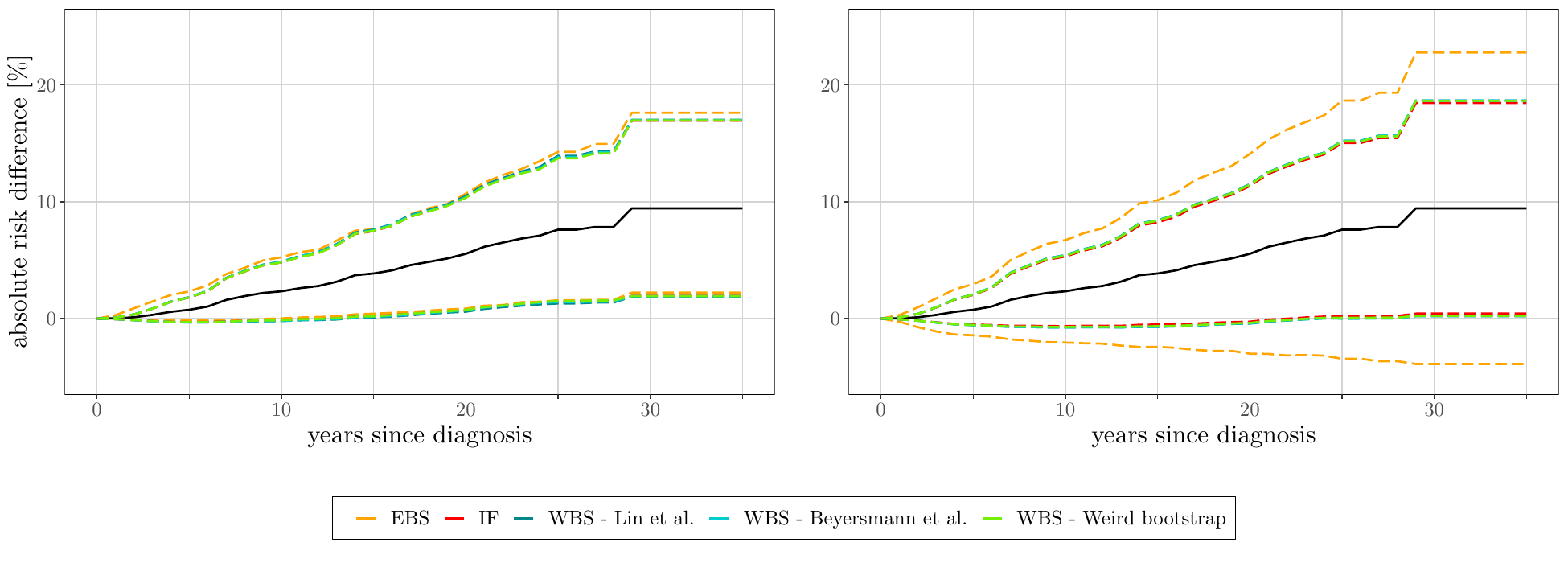}
\end{figure}

The resulting estimate of the average treatment effect (evaluating the combination of radiation and chemotherapy as opposed to radiation alone) on the risk of death as well as the corresponding confidence intervals and bands are depicted in Figure~\ref{fig:hd_analysis}.
As it can be seen, the EBS confidence bands are notably wider than those derived from the remaining resampling methods. For the confidence intervals, the difference is less pronounced, but still visible.

\section{Discussion \label{sec:discussion}}

The article at hand compares three resampling methods for the derivation of confidence intervals and bands for the average treatment effect in competing risks settings. As our simulations show, the wild bootstrap yields correct coverage levels for pointwise confidence intervals in the presence of rather small data sets, provided that sufficient events have been observed until the considered time point. This applies regardless of the type of multiplier that is implemented (i.e.~standard normal, centered Poisson, or weird bootstrap multipliers). The theory behind the wild bootstrap relies on martingales and therefore accommodates counting processes, which are naturally used to represent time-to-event data. As a consequence, it is straightforward to tackle common issues in survival analysis, such as e.g.~left-truncation. (Note the controversy about left-truncation in causal contexts, though, cf.~\citealp{vandenbroucke2015point, hernan2015counterpoint}.) In case that competing events prevail, one may prefer the influence function approach, however, and if earlier time points are examined, the classical bootstrap seems to be a reasonable choice. The latter also achieves very accurate coverages with respect to time-simultaneous confidence bands. As the amount of available data increases, the differences between the distinct resampling approaches fade away. Efron's simple bootstrap, which is most commonly used in practice, requires considerable computation time, however. What is more, dependencies might cause issues with this resampling method \citep{singh1981on, friedrich2017permuting, ruehl2022general}, even though our simulations did not disclose any major bias in this context. \\
The three covered approaches were additionally compared given real data on the long-term survival among patients with early-stage Hodgkin's disease \citep{pintilie2006competing}. While the outcomes resulting from the influence function approach and the wild bootstrap are fairly similar, Efron's bootstrap generated wider intervals and, in particular, bands.  

It should be noted that for the estimated average treatment effect to be consistent, the model for the cumulative incidence function must be correctly specified. Instead of the cause-specific Cox model used here, one might employ alternatives such as the nonparametric additive hazards model proposed by \citet{aalen1980a} (cf.~\citealp{ryalen2018transforming}), or the Fine-Gray regression model for $\smash{F_1(t \mid a, z)}$ adopting the subdistribution approach (see \citealp{rudolph2020causal} or the more technical discourse by \citealp{young2020a} for a discussion on cause-specific vs. subdistribution measures in causal frameworks). In the latter case, however, additional considerations on the associated stochastic process are necessary to make inferences on $\smash{\widehat{ATE}}$. For the same reason, we did not address estimators based on inverse probability of treatment weighting (IPTW, which requires correct specification of a treatment model rather than the outcome model) or the doubly-robust version combining both the g-formula and IPTW. More details on these estimators and appropriate resampling techniques based on the influence function are given by \citet{ozenne2020on}. \\
In order to handle complex conditions that are often observed in real-world trials with time-varying treatments, a possible subject of future work is the extension of the investigated resampling methods to settings that involve time-dependent confounding. The standard time-dependent Cox analysis has been shown to yield incorrect results in such settings \citep{hernan2000marginal}, which is why it is important to incorporate appropriate models \citep[see e.g.][]{keogh2023causal}.

\begin{acknowledgments}
Support by the DFG (Grant FR 4121/2-1) is gratefully acknowledged.
The authors also appreciate the helpful feedback from B.~Ozenne.
\end{acknowledgments}

\section*{Supporting information}

The supplementary material referenced in Subsections~\ref{subsec:asymptotics_IF}, \ref{subsec:asymptotics_wild_bs}, and \ref{subsec:sim_results} includes definitions and additional simulation results not shown here. \\
An \texttt{R} package for the computation of the average treatment together with the proposed confidence intervals and bands, as well as the code to reproduce the results of the simulation study and the real data analysis is available on github (\url{https://github.com/jruehl/ATESurvival}).

\section*{Data availability statement}

The data that support the findings of this study are openly available in the \texttt{R} package \texttt{randomForestSRC} at \url{https://cran.r-project.org/web/packages/randomForestSRC} \citep{randomForestSRC}.

\bibliography{bib_Simulations}{}
\bibliographystyle{apalike}

\end{document}


\title{Supporting information for \\
`Resampling-based confidence regions for the average treatment effect in competing risks data'}

\correspondingauthor{Jasmin Rühl}
\email{jasmin.ruehl@math.uni-augsburg.de}

\author{Jasmin Rühl}
\affiliation{Department of Mathematical Statistics and Artificial Intelligence in Medicine \\
University of Augsburg \\
Augsburg, Germany}

\author{Sarah Friedrich}
\affiliation{Department of Mathematical Statistics and Artificial Intelligence in Medicine \\
University of Augsburg \\
Augsburg, Germany}
\affiliation{Centre for Advanced Analytics and Predictive Sciences (CAAPS) \\
University of Augsburg \\
Augsburg, Germany}


\begin{abstract}
\noindent The supporting material presented hereafter includes definitions and additional simulation results that are not shown in the main paper. \\
\begin{adjustwidth}{2cm}{}
\small
\hyperref[sec:1]{1.\hphantom{..} Definitions} \\[0.05cm]
\hyperref[subsec:11]{\hphantom{1...} 1.1.\hphantom{.} Influence function of the average treatment effect} \\[0.02cm]
\hyperref[subsec:12]{\hphantom{1...} 1.2.\hphantom{.} Functions appearing in the martingale representation of $U_n$} \\[0.02cm]
\hyperref[sec:2]{2.\hphantom{..} Coverage of the confidence intervals} \\[0.05cm]
\hyperref[subsec:21]{\hphantom{2...} 2.1.\hphantom{.} No censoring} \\[0.02cm]
\hyperref[subsec:22]{\hphantom{2...} 2.2.\hphantom{.} Light censoring} \\[0.02cm]
\hyperref[subsec:23]{\hphantom{2...} 2.3.\hphantom{.} Heavy censoring} \\[0.02cm]
\hyperref[subsec:24]{\hphantom{2...} 2.4.\hphantom{.} Low treatment probability} \\[0.02cm]
\hyperref[subsec:25]{\hphantom{2...} 2.5.\hphantom{.} High treatment probability} \\[0.02cm]
\hyperref[subsec:26]{\hphantom{2...} 2.6.\hphantom{.} Low covariance of the covariates} \\[0.02cm]
\hyperref[subsec:27]{\hphantom{2...} 2.7.\hphantom{.} High covariance of the covariates} \\[0.02cm]
\hyperref[subsec:28]{\hphantom{2...} 2.8.\hphantom{.} Type II censoring} \\[0.1cm]
\hyperref[sec:3]{3.\hphantom{..} Coverage of the confidence bands} \\[0.05cm]
\hyperref[subsec:31]{\hphantom{3...} 3.1.\hphantom{.} No censoring} \\[0.02cm]
\hyperref[subsec:32]{\hphantom{3...} 3.2.\hphantom{.} Light censoring} \\[0.02cm]
\hyperref[subsec:33]{\hphantom{3...} 3.3.\hphantom{.} Heavy censoring} \\[0.02cm]
\hyperref[subsec:34]{\hphantom{3...} 3.4.\hphantom{.} Low treatment probability} \\[0.02cm]
\hyperref[subsec:35]{\hphantom{3...} 3.5.\hphantom{.} High treatment probability} \\[0.02cm]
\hyperref[subsec:36]{\hphantom{3...} 3.6.\hphantom{.} Low covariance of the covariates} \\[0.02cm]
\hyperref[subsec:37]{\hphantom{3...} 3.7.\hphantom{.} High covariance of the covariates} \\[0.02cm]
\hyperref[subsec:38]{\hphantom{3...} 3.8.\hphantom{.} Type II censoring} 
\end{adjustwidth}
\end{abstract}


\section{Definitions \label{sec:1}}

\subsection{Influence function of the average treatment effect \label{subsec:11}}

With $O_i$ denoting $\smash{\{T_i \land C_i, D_i, A_i, \boldsymbol{Z_i}\}}$, ${i \in \{1, \dots, n\}}$, the influence function $\smash{I\!F_{ATE}(t; \, O_i)}$ of the average treatment effect, as described in \citep{ozenne2020on} and \citep{ozenne2017riskRegression}, is defined as
\begin{align*}
I\!F_{ATE}(t; \, O_i) &= \hat{F}_{1}(t \mid A = 1, \boldsymbol{Z_i}) - \hat{F}_1(t \mid A = 0, \boldsymbol{Z_i}) \\
&\hphantom{=} \hphantom{n} - \mathbb{E}\left(F_1(t \mid A = 1, \boldsymbol{Z}) - F_1(t \mid A = 0, \boldsymbol{Z})\right) \\
&\hphantom{=} \hphantom{n} + \mathbb{E}\left(I\!F_{F_1}(t, A = 1, \boldsymbol{Z}; \, O_i) \mid O_i \right),
\end{align*}
with 
\begin{align*}
I\!F_{F_1}(t, a, \boldsymbol{z}; \, O_i) &= \int_0^t \exp\left( - \sum_{k=1}^K \Lambda_k(u \mid a, \boldsymbol{z})\right) \, \text{d}I\!F_{\Lambda_1}(u, a, \boldsymbol{z}; \, O_i) \\[-0.1cm]
&\hphantom{=} \hphantom{n} - \int_0^t  \exp\left( - \sum_{k=1}^K \Lambda_k(u \mid a, \boldsymbol{z})\right) \left(\sum_{k=1}^K I\!F_{\Lambda_k}(u, a, \boldsymbol{z}; \, O_i)\right) \, \text{d}\Lambda_1(u \mid a, \boldsymbol{z}), \\[0.5cm]
I\!F_{\Lambda_k}(t, a, \boldsymbol{z}; \, O_i) &= \exp\left(\beta_{kA} a + \boldsymbol{\beta}_{\boldsymbol{kZ}}^T \boldsymbol{z}\right) \left(I\!F_{\Lambda_{0k}}(t; \, O_i) +  \Lambda_{0k}(t) \, (a, \boldsymbol{z})^T \boldsymbol{I\!F}_{\boldsymbol{\beta_k}}(O_i)\right), \\[0.5cm]
I\!F_{\Lambda_{0k}}(t; \, O_i) &= - \boldsymbol{I\!F}_{\boldsymbol{\beta_k}}(O_i)^T \int_0^t \frac{\boldsymbol{s^{(1)}}(\boldsymbol{\beta_k}, u)}{s^{(0)}(\boldsymbol{\beta_k}, u)} \, \text{d}\Lambda_{0k}(u) - \exp\left(\beta_{kA} A_i + \boldsymbol{\beta}_{\boldsymbol{kZ}}^T \boldsymbol{Z}_{\boldsymbol{i}}\right) \int_0^{t \land (T_i \land C_i)} \hspace{-0.6cm} \frac{1}{s^{(0)}(\boldsymbol{\beta_k}, u)} \, \text{d}\Lambda_{0k}(u) \\
&\hphantom{=} \hphantom{n} + \frac{\mathbbm{1}\{T_i\!\land\!C_i \leq t, \, D_i = k\}}{s^{(0)}(\boldsymbol{\beta_k}, T_i\!\land\!C_i)}, \\[0.5cm]
\boldsymbol{I\!F}_{\boldsymbol{\beta_k}}(O_i) &= \boldsymbol{\mathcal{I}}(\boldsymbol{\beta_k})^{-1} \left(\mathbbm{1}\{D_i = k\} \left((A_i, \boldsymbol{Z_i}^T)^T - \frac{\boldsymbol{s^{(1)}}(\boldsymbol{\beta_k}, T_i\!\land\!C_i)}{s^{(0)}(\boldsymbol{\beta_k}, T_i\!\land\!C_i)}\right) \right. \\
&\hphantom{=} \hphantom{n} \left. - \exp\left(\beta_{kA} A_i + \boldsymbol{\beta}_{\boldsymbol{kZ}}^T \boldsymbol{Z}_{\boldsymbol{i}}\right) \mathbb{E}\left(\frac{\mathbbm{1}\{T\!\land\!C \leq T_i \!\land\! C_i, D = k\}}{s^{(0)}(\boldsymbol{\beta_k}, T\!\land\!C)} \left((A_i, \boldsymbol{Z}_{\boldsymbol{i}}^T)^T - \frac{\boldsymbol{s^{(1)}}(\boldsymbol{\beta_k}, T\!\land\!C)}{s^{(0)}(\boldsymbol{\beta_k}, T\!\land\!C)}\right)\right) \right), \\[0.5cm]
\boldsymbol{\mathcal{I}}(\boldsymbol{\beta_k}) &= \mathbb{E}\left(\mathbbm{1}\{D = k\} \left(\frac{\boldsymbol{s^{(2)}}(\boldsymbol{\beta_k}, T\!\land\!C)}{s^{(0)}(\boldsymbol{\beta_k}, T\!\land\!C)} - \frac{\boldsymbol{s^{(1)}}(\boldsymbol{\beta_k}, T\!\land\!C)}{s^{(0)}(\boldsymbol{\beta_k}, T\!\land\!C)} \left(\frac{\boldsymbol{s^{(1)}}(\boldsymbol{\beta_k}, T\!\land\!C)}{s^{(0)}(\boldsymbol{\beta_k}, T\!\land\!C)}\right)^T\right)\right), \\[0.5cm]
s^{(0)}(\boldsymbol{\beta_k},t) &= \mathbb{E}\left(Y(t) \exp\left(\beta_{kA} A + \boldsymbol{\beta}_{\boldsymbol{kZ}}^T \boldsymbol{Z}\right)\right), \\
\boldsymbol{s^{(1)}}(\boldsymbol{\beta_k},t) &= \mathbb{E}\left(Y(t) \exp\left(\beta_{kA} A + \boldsymbol{\beta}_{\boldsymbol{kZ}}^T \boldsymbol{Z}\right) \left(A, \boldsymbol{Z}^T\right)^T\right), \\[-0.1cm]
\boldsymbol{s^{(2)}}(\boldsymbol{\beta_k},t) &= \mathbb{E}\left(Y(t) \exp\left(\beta_{kA} A + \boldsymbol{\beta}_{\boldsymbol{kZ}}^T \boldsymbol{Z}\right) \! \left(A, \boldsymbol{Z}^T\right)^T \left(A, \boldsymbol{Z}^T\right)\right).
\end{align*}

\clearpage

\subsection{Functions appearing in the martingale representation of $U_n$ \label{subsec:12}}

Furthermore, the functions $\smash{H_{k1i}}$ and $\smash{H_{k2i}}$, ${k \in \{1, \dots, K\}}, {i \in \{1, \dots, n\}}$, are given by
\[H_{k1i}(s,t) = \frac{\tilde{H}_{k1}(s,t)}{\sqrt{n} \cdot S^{(0)}(\boldsymbol{\beta_k}, s)}\]
with 
\begin{align*}
\tilde{H}_{k1}(s,t) &= \frac{1}{n} \sum_{j=1}^n \left(\left(\mathbbm{1}\{k\!=\!1\} \exp\left(-\sum_{l=1}^K \Lambda_l(s \mid A\!=\!1, \boldsymbol{Z_j})\right) - F_1(t \mid A\!=\!1, \boldsymbol{Z_j}) + F_1(s \mid A\!=\!1, \boldsymbol{Z_j})\right) \exp\left(\beta_{kA} + \boldsymbol{\beta}_{\boldsymbol{kZ}}^T \boldsymbol{Z_j}\right)\right. \\
&\hphantom{=} \hphantom{\frac{1}{n} \sum_{j=1}^n \left( \right.} \left. \hphantom{n} - \left(\mathbbm{1}\{k\!=\!1\} \exp\left(-\sum_{l=1}^K \Lambda_l(s \mid A\!=\!0, \boldsymbol{Z_j})\right) - F_1(t \mid A\!=\!0, \boldsymbol{Z_j}) + F_1(s \mid A\!=\!0, \boldsymbol{Z_j})\right) \exp\left(\boldsymbol{\beta}_{\boldsymbol{kZ}}^T \boldsymbol{Z_j}\right) \right)
\end{align*}
and
\[H_{k2i}(s,t) = \frac{1}{\sqrt{n}} \left(\tilde{H}_{k2}(t)\right)^T \boldsymbol{\Sigma}_{\boldsymbol{k}}^{-1} \left((A_i, \boldsymbol{Z_i}^T)^T - \frac{\boldsymbol{S^{(1)}}(\boldsymbol{\beta_k}, s)}{S^{(0)}(\boldsymbol{\beta_k}, s)}\right)\]
with 
\begin{align*}
\tilde{H}_{k2}(t) &= \frac{1}{n} \sum_{j=1}^n \left(\left(\mathbbm{1}\{k\!=\!1\} \varphi_1(t \mid A\!=\!1, \boldsymbol{Z_j}) - \psi_k(t \mid A\!=\!1, \boldsymbol{Z_j})\right) - \left(\mathbbm{1}\{k\!=\!1\} \varphi_1(t \mid A\!=\!0, \boldsymbol{Z_j}) - \psi_{1k}(t \mid A\!=\!0, \boldsymbol{Z_j})\right)\right), \\[0.5cm]
\boldsymbol{\Sigma_k} &= \int_0^\tau \left(\frac{\boldsymbol{s^{(2)}}(\boldsymbol{\beta_k}, u)}{s^{(0)}(\boldsymbol{\beta_k}, u)} - \frac{\boldsymbol{s^{(1)}}(\boldsymbol{\beta_k}, u)}{s^{(0)}(\boldsymbol{\beta_k}, u)} \left(\frac{\boldsymbol{s^{(1)}}(\boldsymbol{\beta_k}, u)}{s^{(0)}(\boldsymbol{\beta_k}, u)}\right)^T\right) s^{(0)}(\boldsymbol{\beta_k}, u) \, \text{d}\Lambda_k(u), \\[0.5cm]
\varphi_1(t \mid a, \boldsymbol{z}) &= \int_0^t \exp\left(-\sum_{l=1}^K \Lambda_l(s \mid A\!=\!a, \boldsymbol{z})\right) \left(\left(a, \boldsymbol{z}^T\right)^T - \frac{\boldsymbol{s^{(1)}}(\boldsymbol{\beta_k}, u)}{s^{(0)}(\boldsymbol{\beta_k}, u)}\right) \, \text{d}\Lambda_k(u \mid a, \boldsymbol{z}), \\
\psi_{1k}(t \mid a, \boldsymbol{z}) &= \int_0^t \left(F_1(t \mid a, \boldsymbol{z}) - F_1(u \mid a, \boldsymbol{z})\right) \left(\left(a, \boldsymbol{z}^T\right)^T - \frac{\boldsymbol{s^{(1)}}(\boldsymbol{\beta_k}, u)}{s^{(0)}(\boldsymbol{\beta_k}, u)}\right) \, \text{d}\Lambda_k(u \mid a, \boldsymbol{z}), \\[0.5cm]
S^{(0)}(\boldsymbol{\beta_k},t) &= \frac{1}{n} \sum_{j=1}^n Y_j(t) \exp\left(\beta_{kA} A_j + \boldsymbol{\beta}_{\boldsymbol{kZ}}^T \boldsymbol{Z_j}\right), \\
\boldsymbol{S^{(1)}}(\boldsymbol{\beta_k},t) &= \frac{1}{n} \sum_{j=1}^n Y_j(t) \exp\left(\beta_{kA} A_j + \boldsymbol{\beta}_{\boldsymbol{kZ}}^T \boldsymbol{Z_j}\right) \left(A_j, \boldsymbol{Z}_{\boldsymbol{j}}^T\right)^T, \\
\boldsymbol{S^{(2)}}(\boldsymbol{\beta_k},t) &= \frac{1}{n} \sum_{j=1}^n Y_j(t) \exp\left(\beta_{kA} A_j + \boldsymbol{\beta}_{\boldsymbol{kZ}}^T \boldsymbol{Z_j}\right) \! \left(A_j, \boldsymbol{Z}_{\boldsymbol{j}}^T\right)^T \left(A_j, \boldsymbol{Z}_{\boldsymbol{j}}^T\right).
\end{align*}
\clearpage

\section{Coverage of the confidence intervals \label{sec:2}}

\subsection{No censoring \label{subsec:21}}

\begin{figure}[h!tbp]
\centering \label{fig:21}
\includegraphics[width = \textwidth]{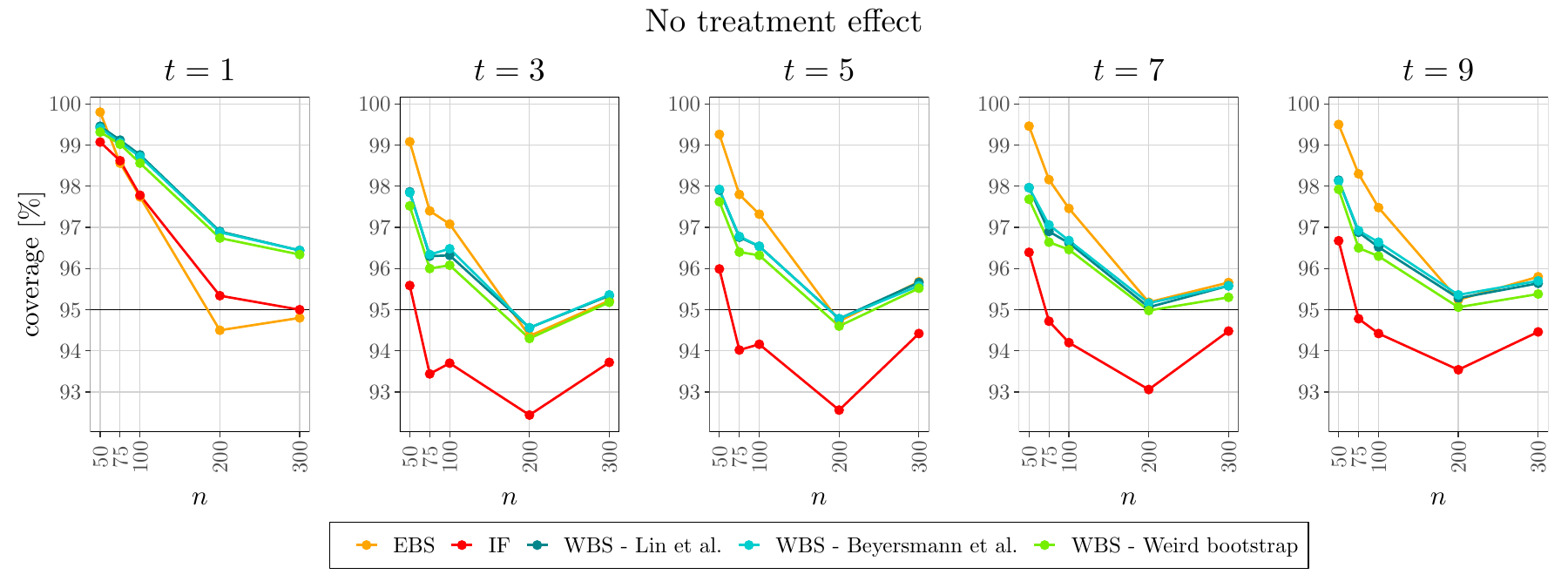}
\end{figure}

\begin{figure}[h!tbp]
\centering \label{fig:22}
\includegraphics[width = \textwidth]{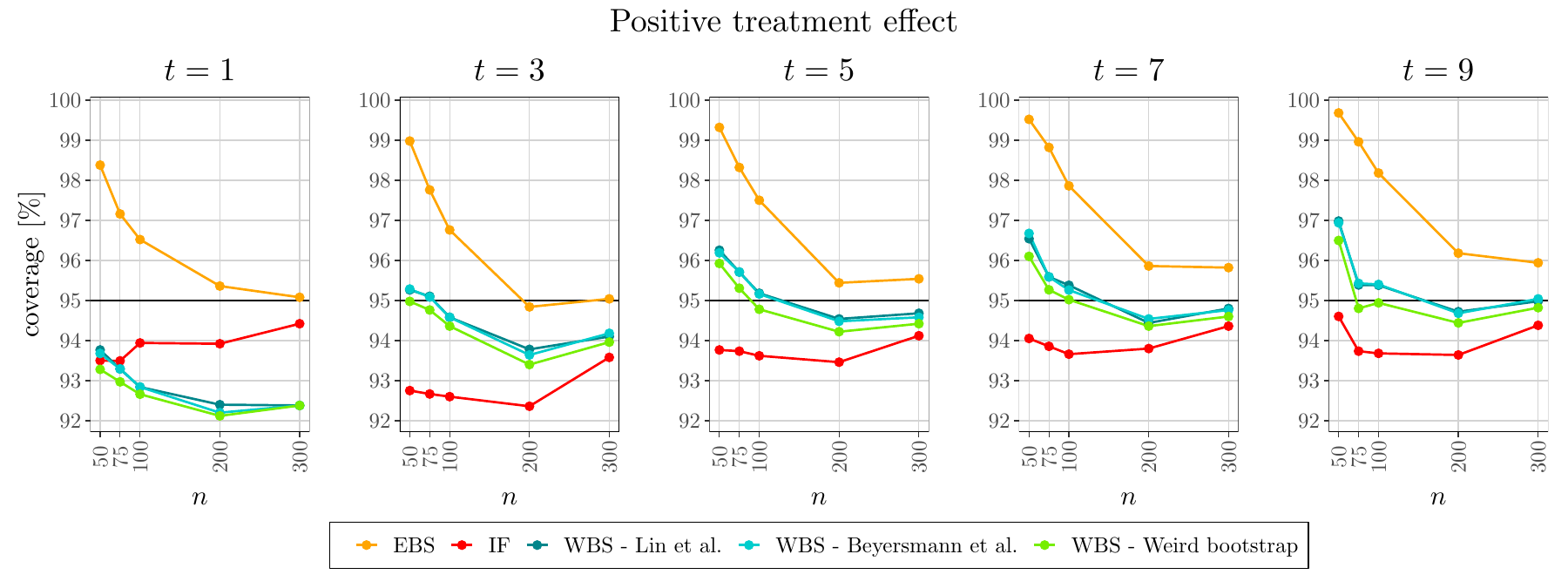}
\end{figure}

\clearpage

\subsection{Light censoring \label{subsec:22}}

\begin{figure}[h!tbp]
\centering \label{fig:23}
\includegraphics[width = \textwidth]{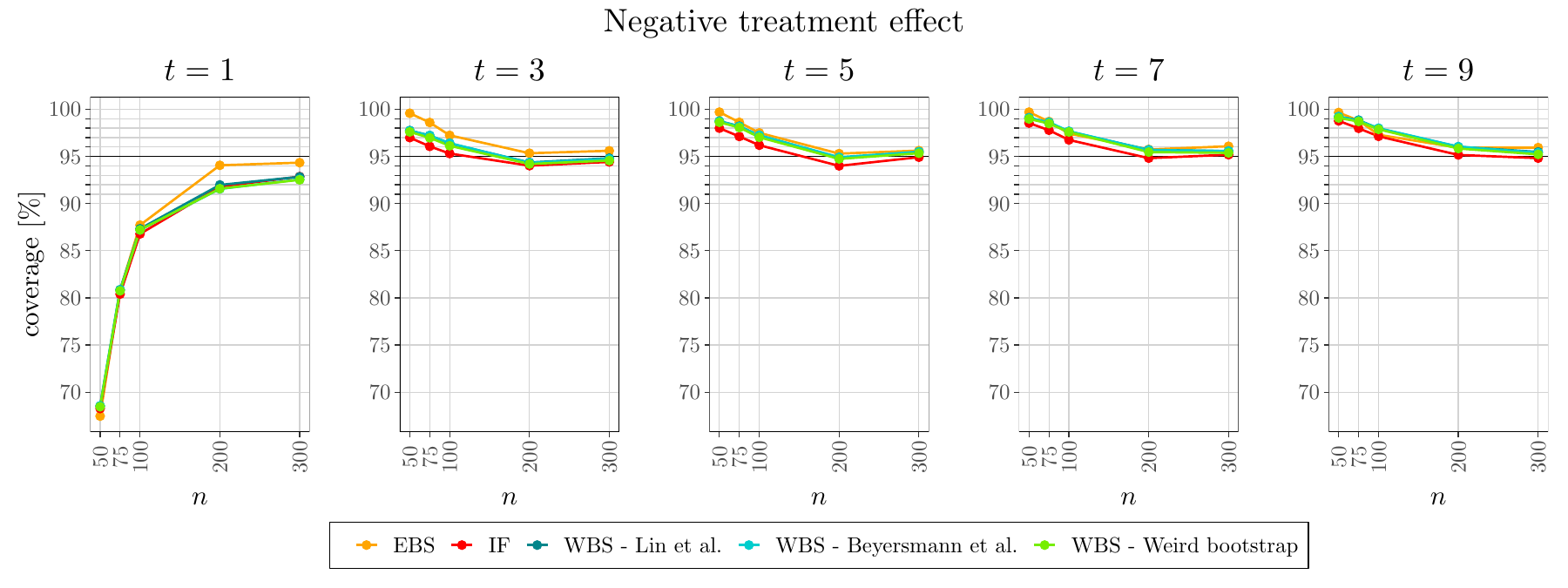}
\end{figure}

\begin{figure}[h!tbp]
\centering \label{fig:24}
\includegraphics[width = \textwidth]{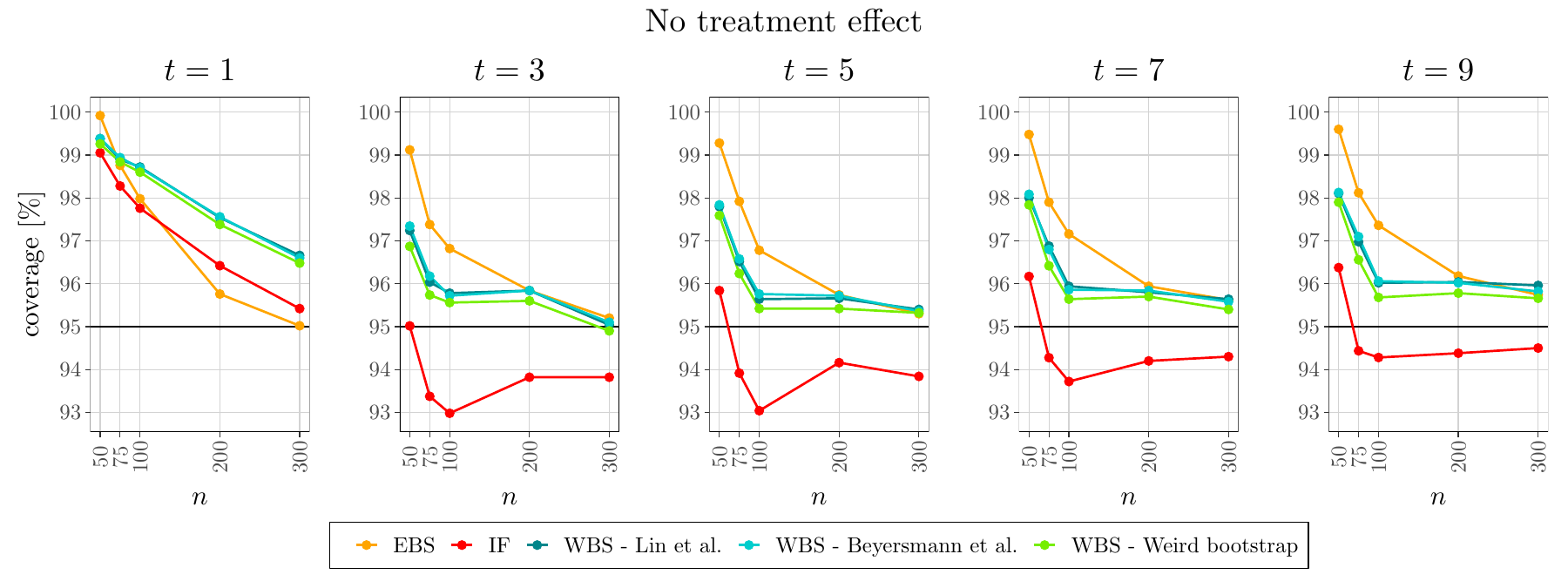}
\end{figure}

\clearpage

\subsection{Heavy censoring \label{subsec:23}}
\enlargethispage{1cm}

\begin{figure}[h!tbp]
\centering \label{fig:25}
\includegraphics[width = \textwidth]{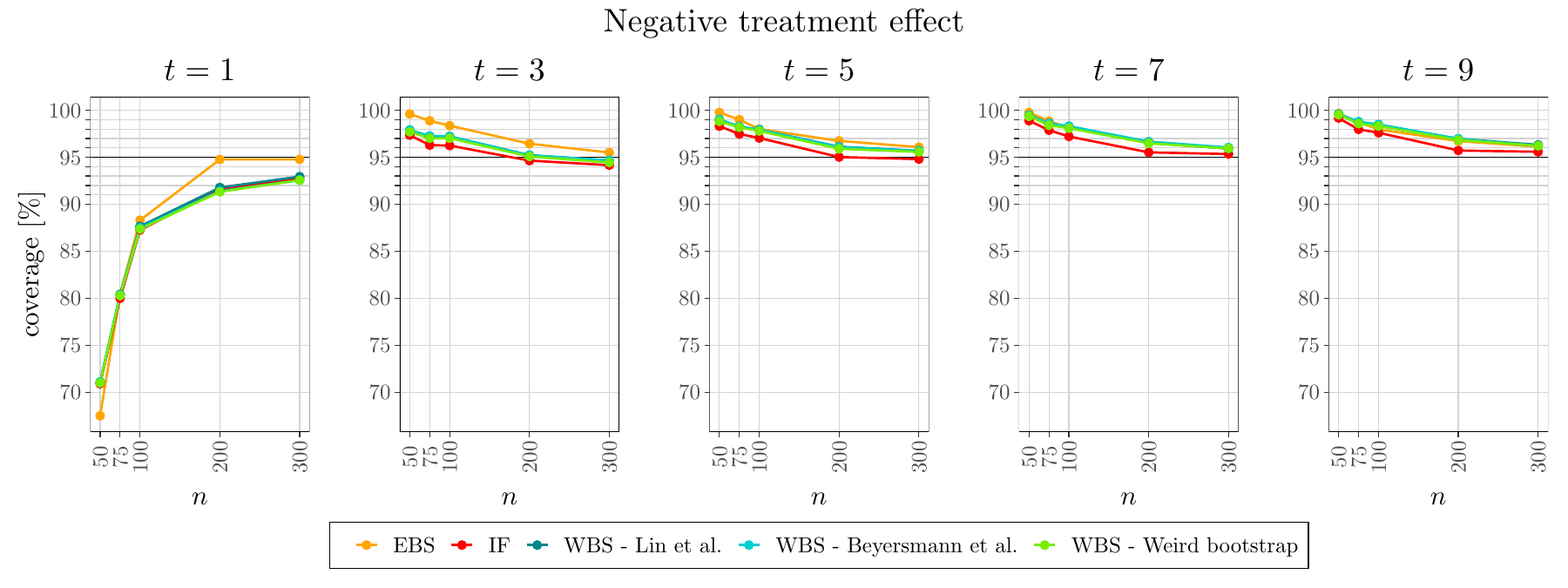}
\end{figure}

\begin{figure}[h!tbp]
\centering \label{fig:26}
\includegraphics[width = \textwidth]{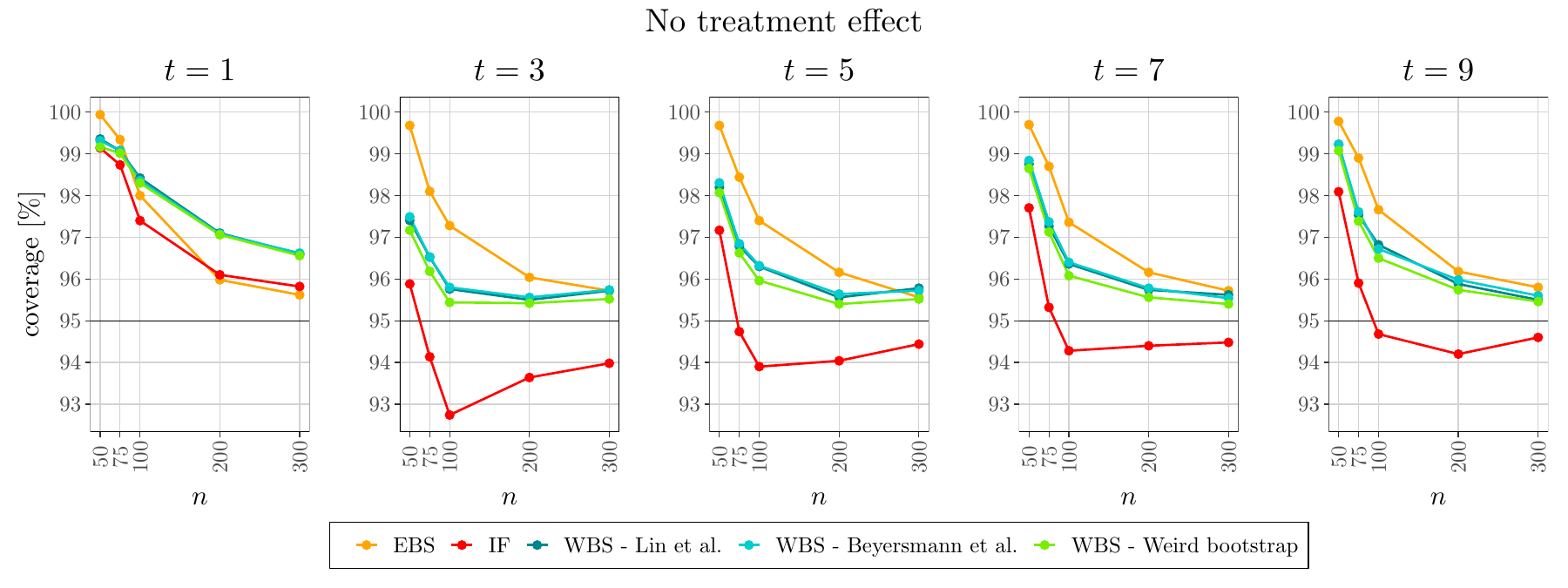}
\end{figure}

\begin{figure}[h!tbp]
\centering \label{fig:27}
\includegraphics[width = \textwidth]{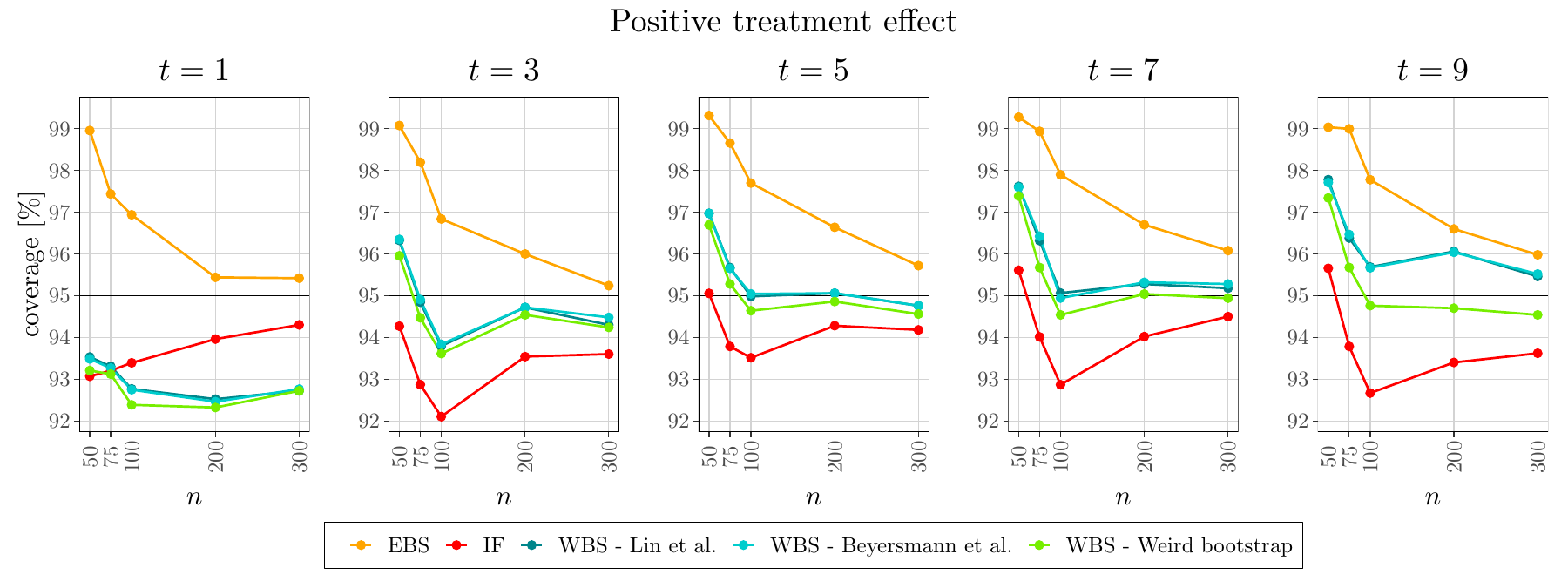}
\end{figure}

\clearpage

\subsection{Low treatment probability \label{subsec:24}}
\enlargethispage{1cm}

\begin{figure}[h!tbp]
\centering \label{fig:28}
\includegraphics[width = \textwidth]{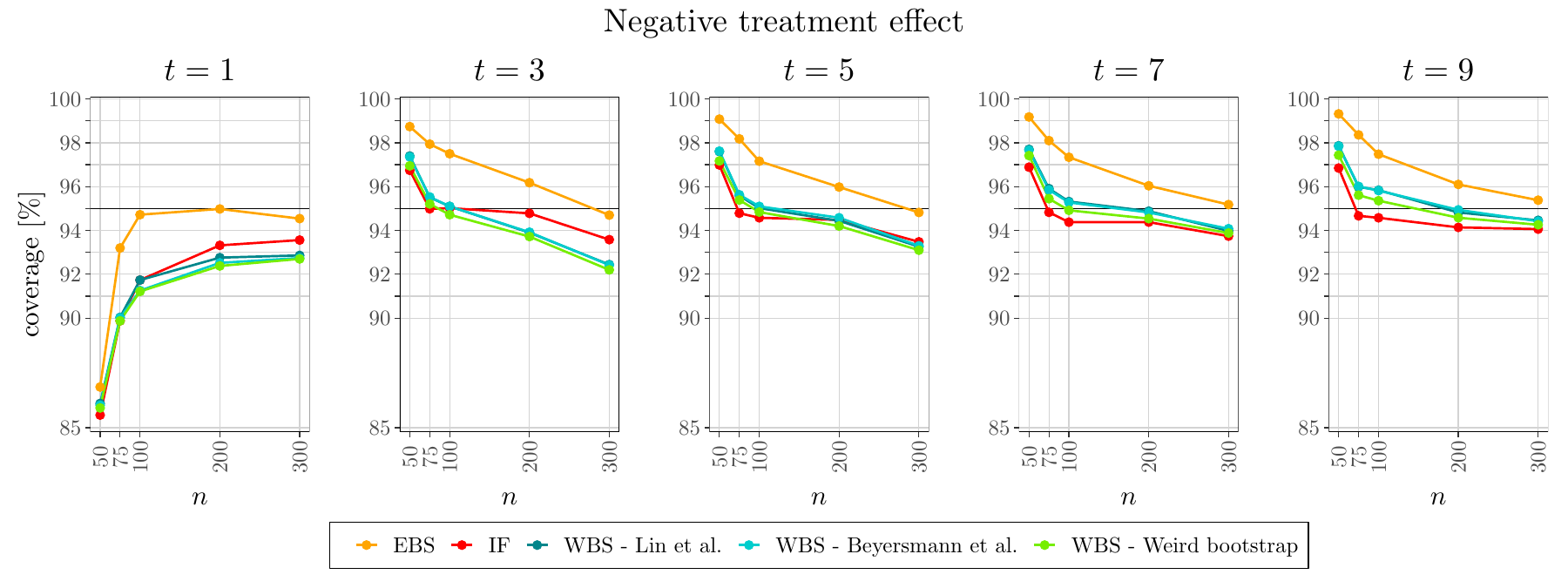}
\end{figure}

\begin{figure}[h!tbp]
\centering \label{fig:29}
\includegraphics[width = \textwidth]{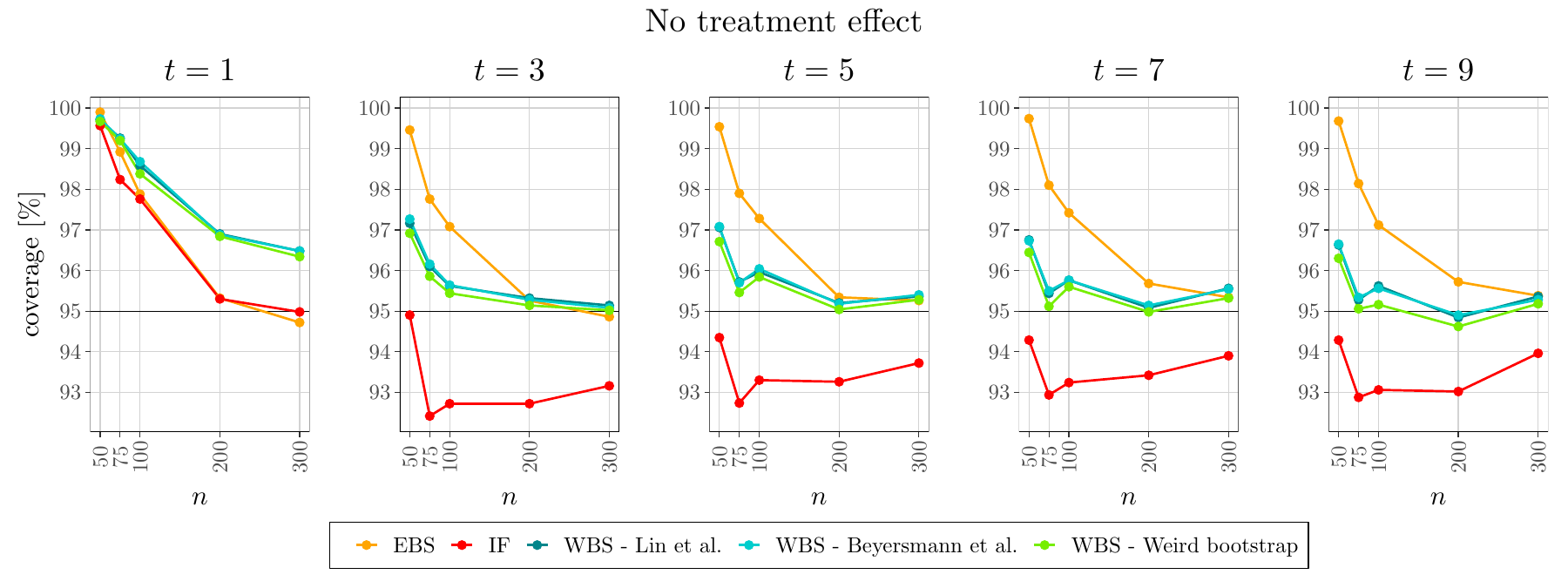}
\end{figure}

\begin{figure}[h!tbp]
\centering \label{fig:210}
\includegraphics[width = \textwidth]{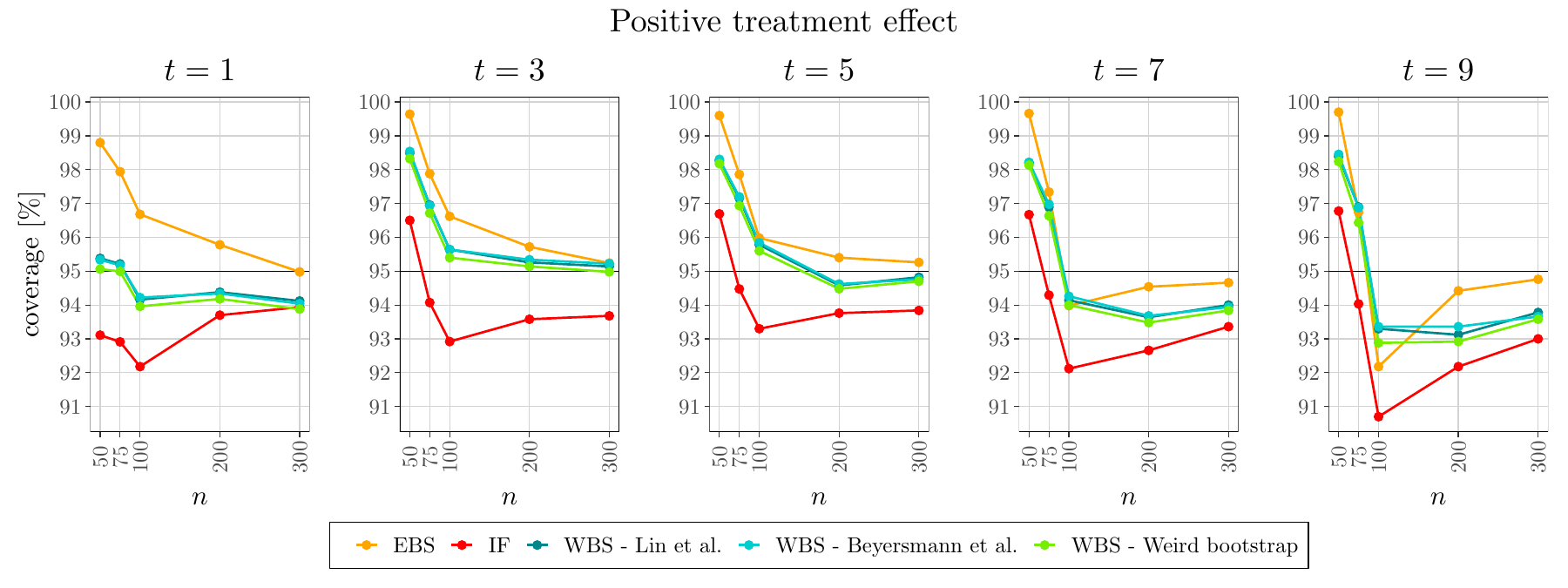}
\end{figure}

\clearpage

\subsection{High treatment probability \label{subsec:25}}
\enlargethispage{1cm}

\begin{figure}[h!tbp]
\centering \label{fig:211}
\includegraphics[width = \textwidth]{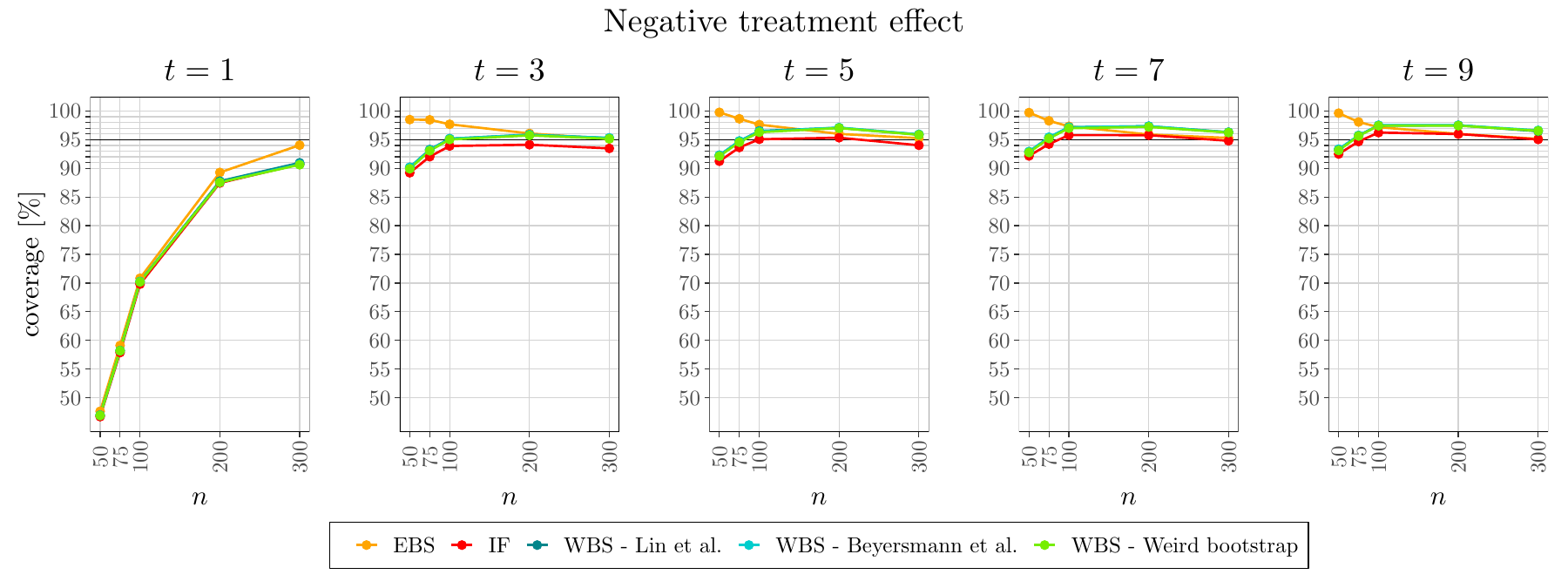}
\end{figure}

\begin{figure}[h!tbp]
\centering \label{fig:212}
\includegraphics[width = \textwidth]{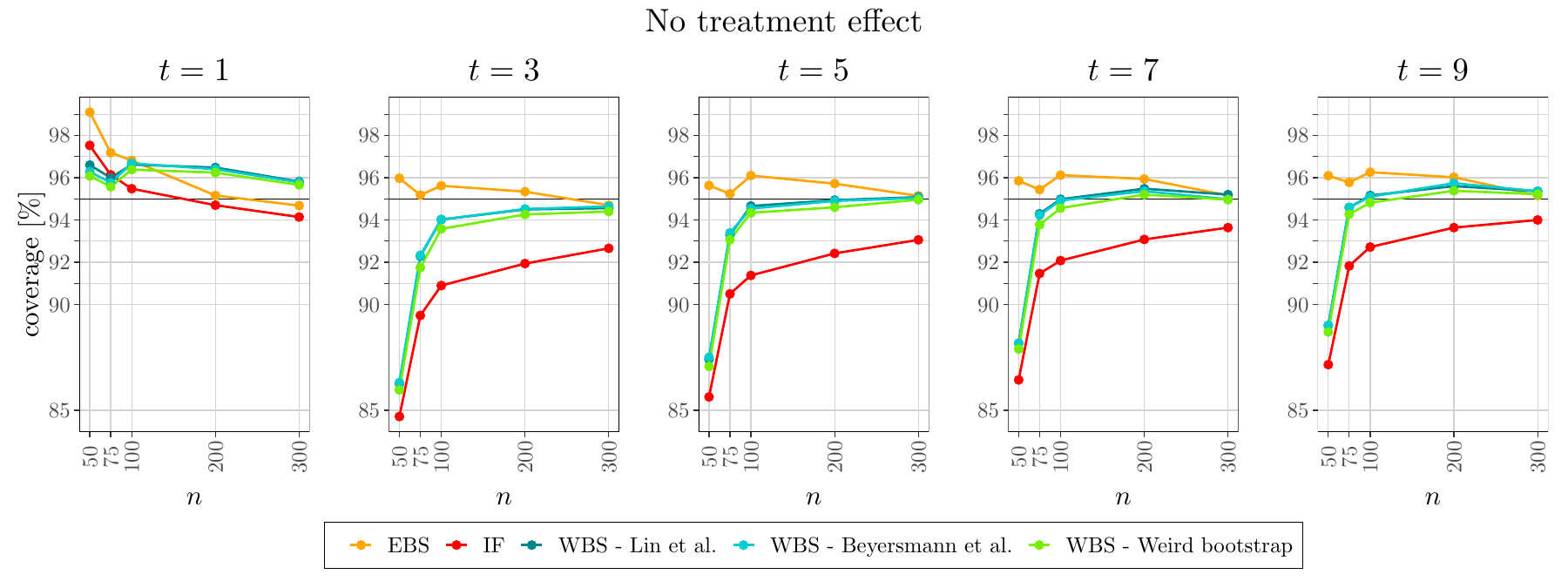}
\end{figure}

\begin{figure}[h!tbp]
\centering \label{fig:213}
\includegraphics[width = \textwidth]{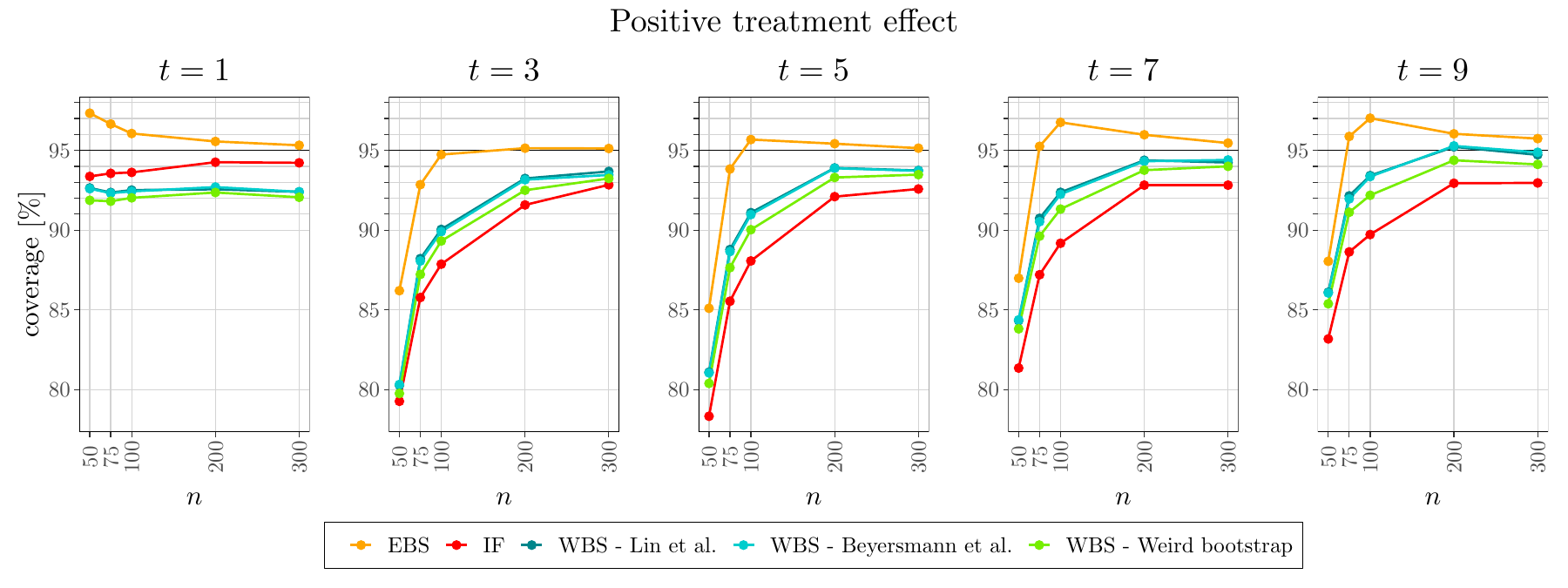}
\end{figure}

\clearpage

\subsection{Low covariance of the covariates \label{subsec:26}}
\enlargethispage{1cm}

\begin{figure}[h!tbp]
\centering \label{fig:214}
\includegraphics[width = \textwidth]{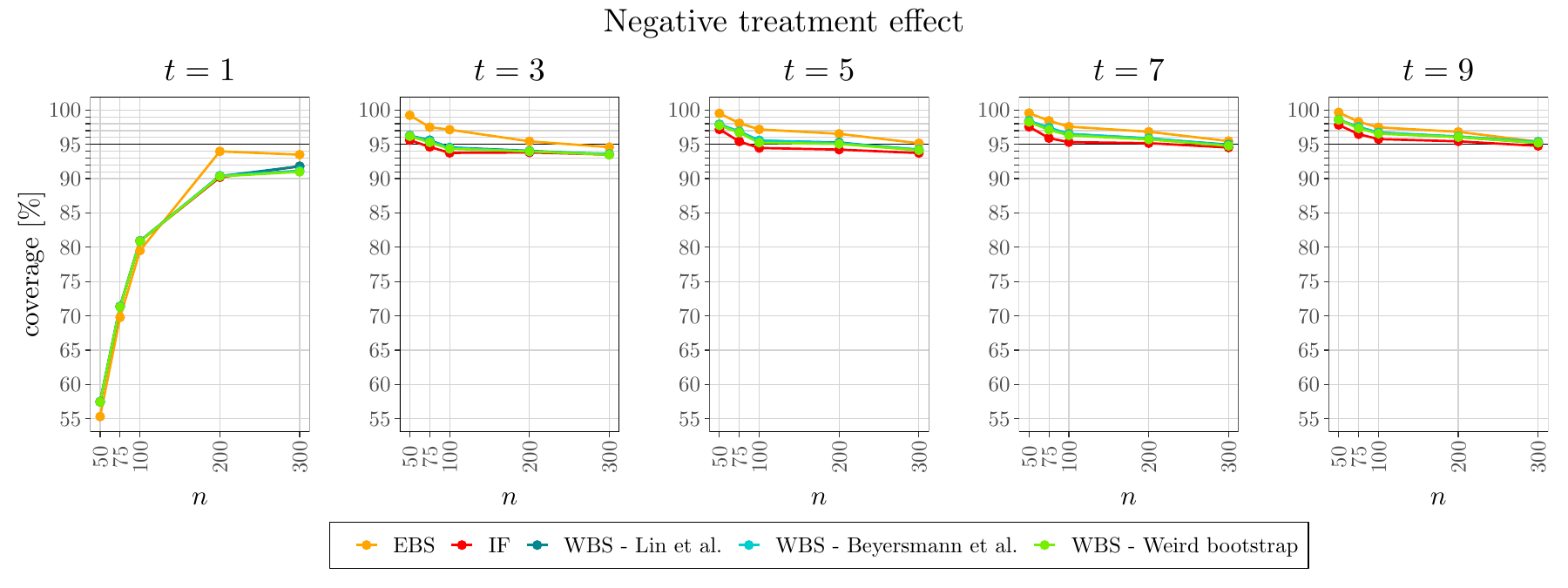}
\end{figure}

\begin{figure}[h!tbp]
\centering \label{fig:215}
\includegraphics[width = \textwidth]{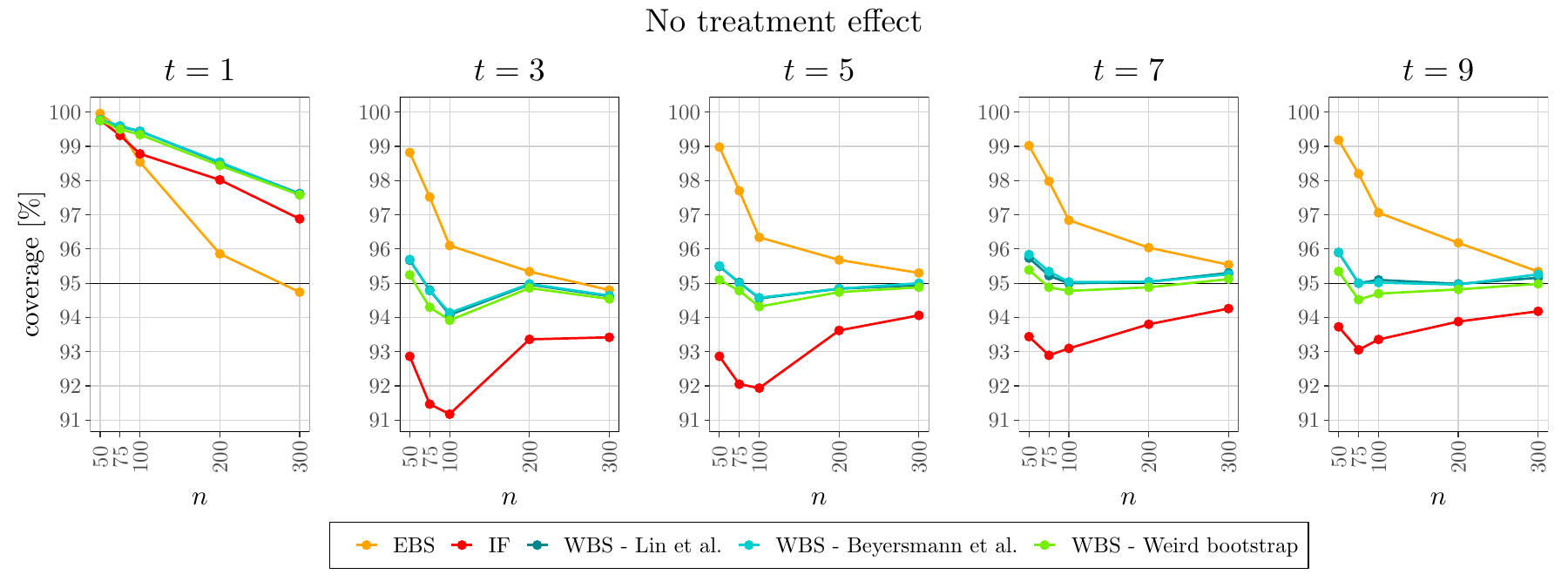}
\end{figure}

\begin{figure}[h!tbp]
\centering \label{fig:216}
\includegraphics[width = \textwidth]{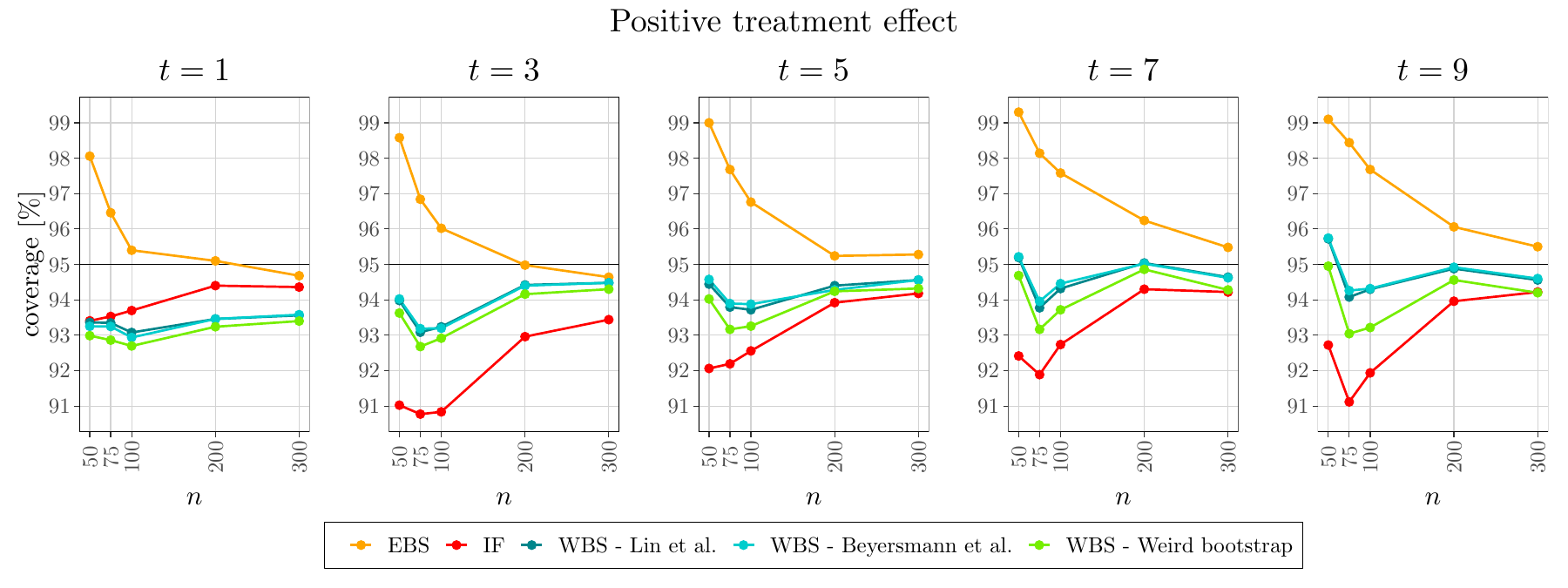}
\end{figure}

\clearpage

\subsection{High covariance of the covariates \label{subsec:27}}
\enlargethispage{1cm}

\begin{figure}[h!tbp]
\centering \label{fig:217}
\includegraphics[width = \textwidth]{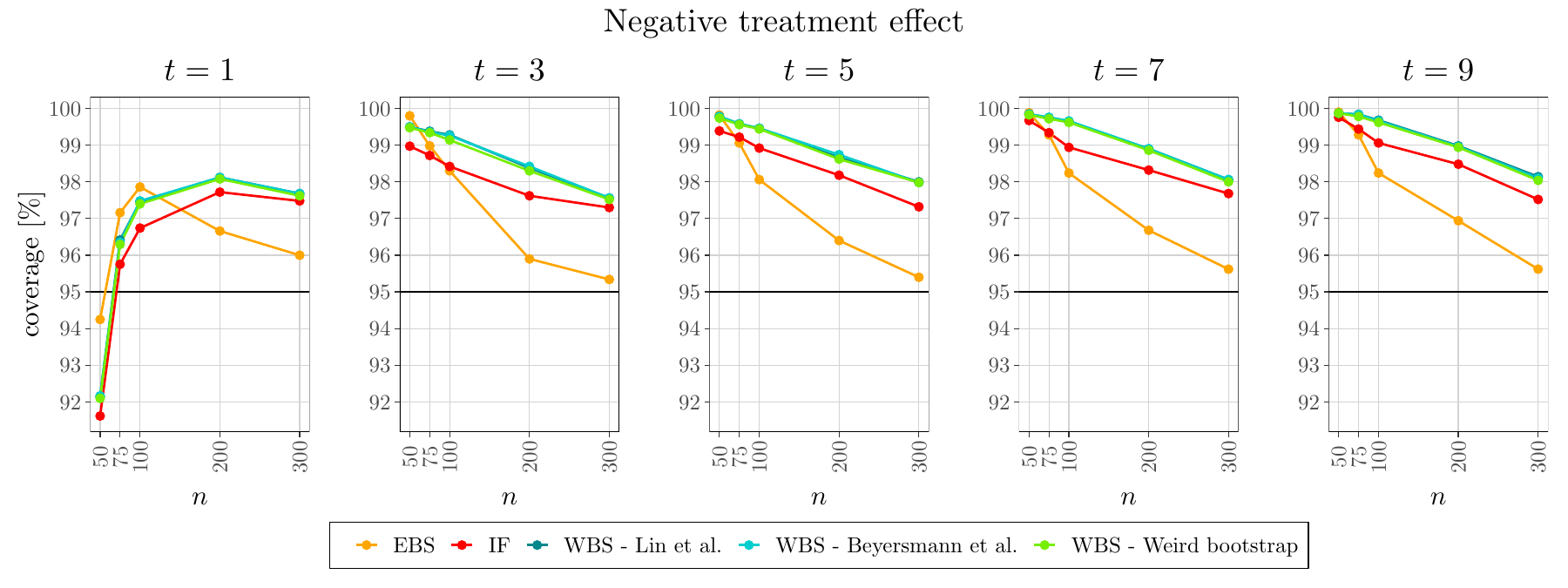}
\end{figure}

\begin{figure}[h!tbp]
\centering \label{fig:218}
\includegraphics[width = \textwidth]{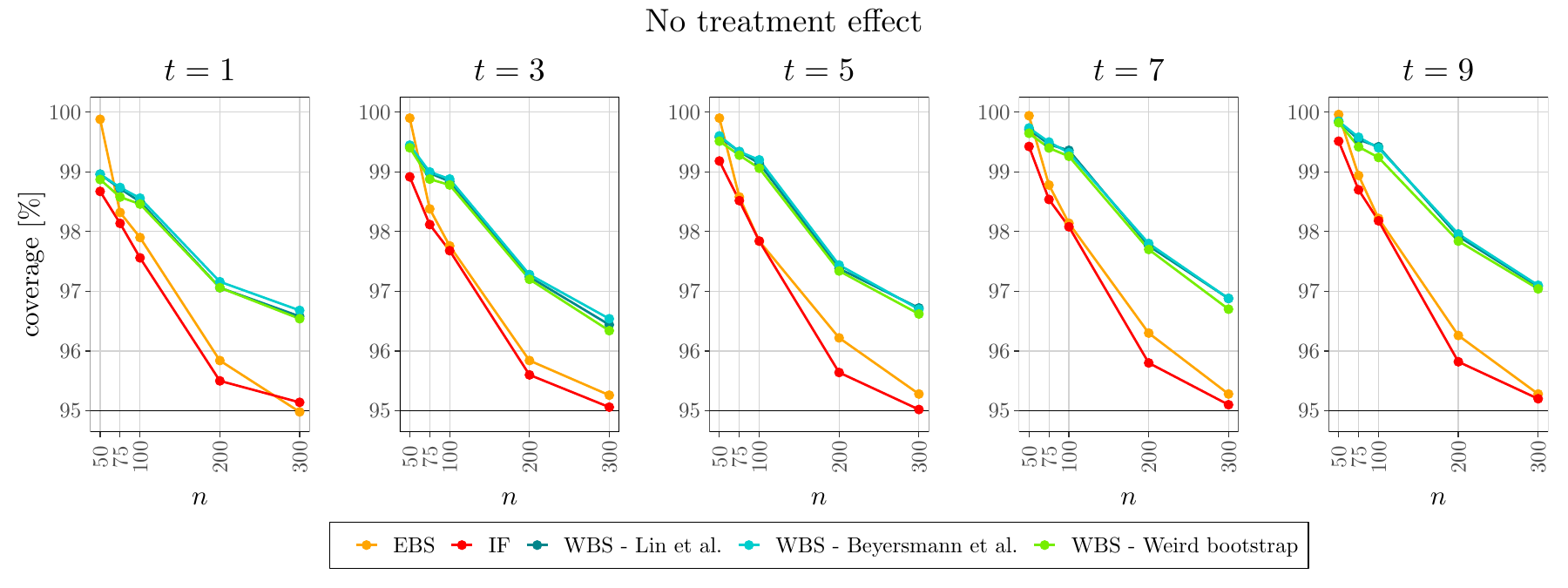}
\end{figure}

\begin{figure}[h!tbp]
\centering \label{fig:219}
\includegraphics[width = \textwidth]{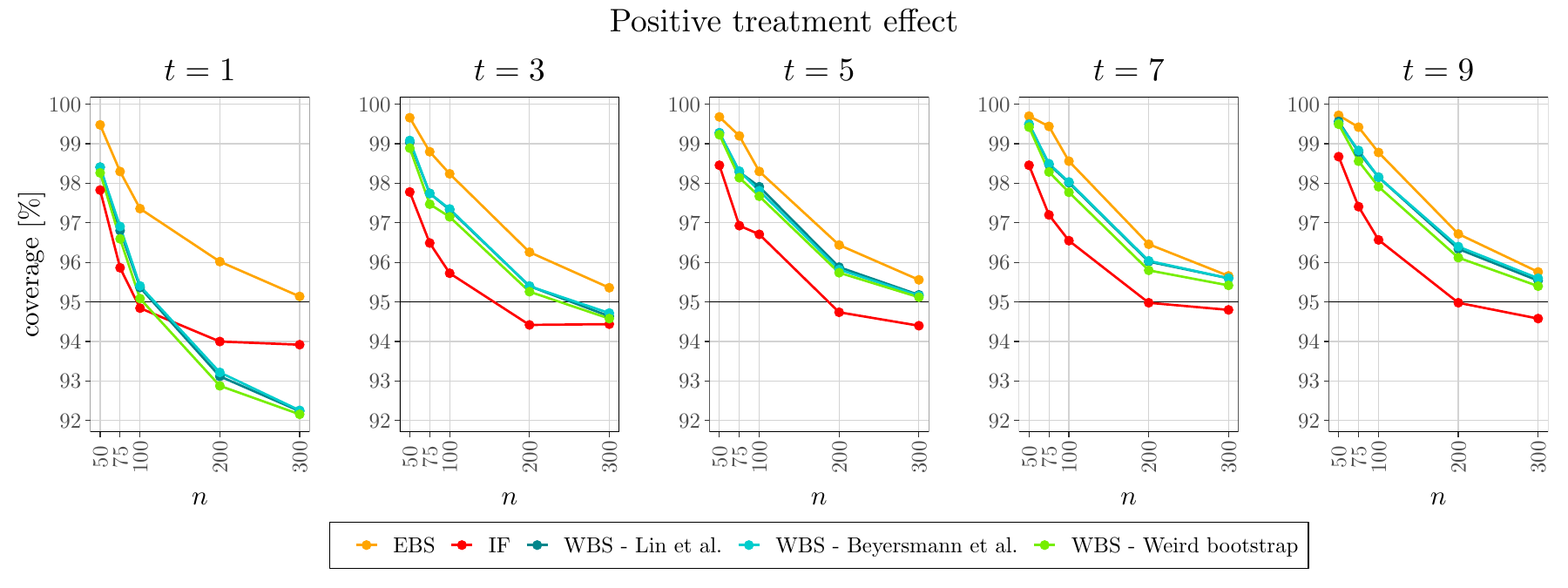}
\end{figure}

\clearpage

\subsection{Type~II censoring \label{subsec:28}}
\enlargethispage{1cm}

\begin{figure}[h!tbp]
\centering \label{fig:220}
\includegraphics[width = \textwidth]{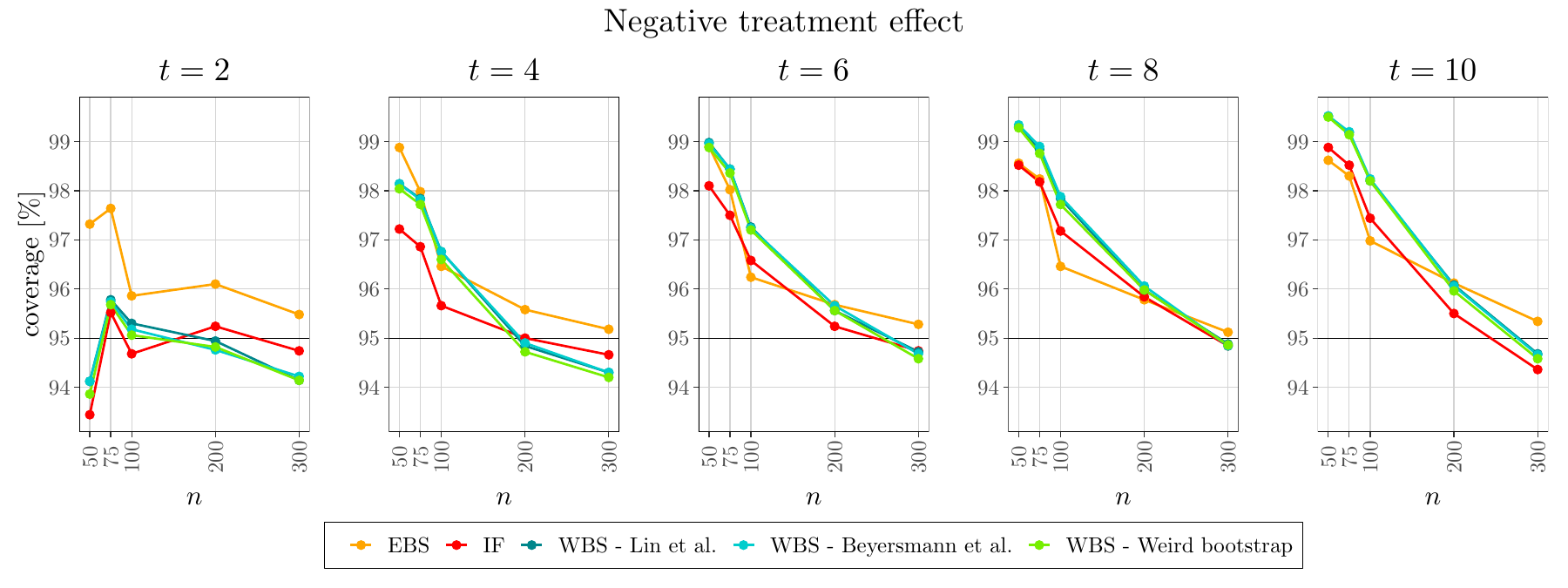}
\end{figure}

\begin{figure}[h!tbp]
\centering \label{fig:221}
\includegraphics[width = \textwidth]{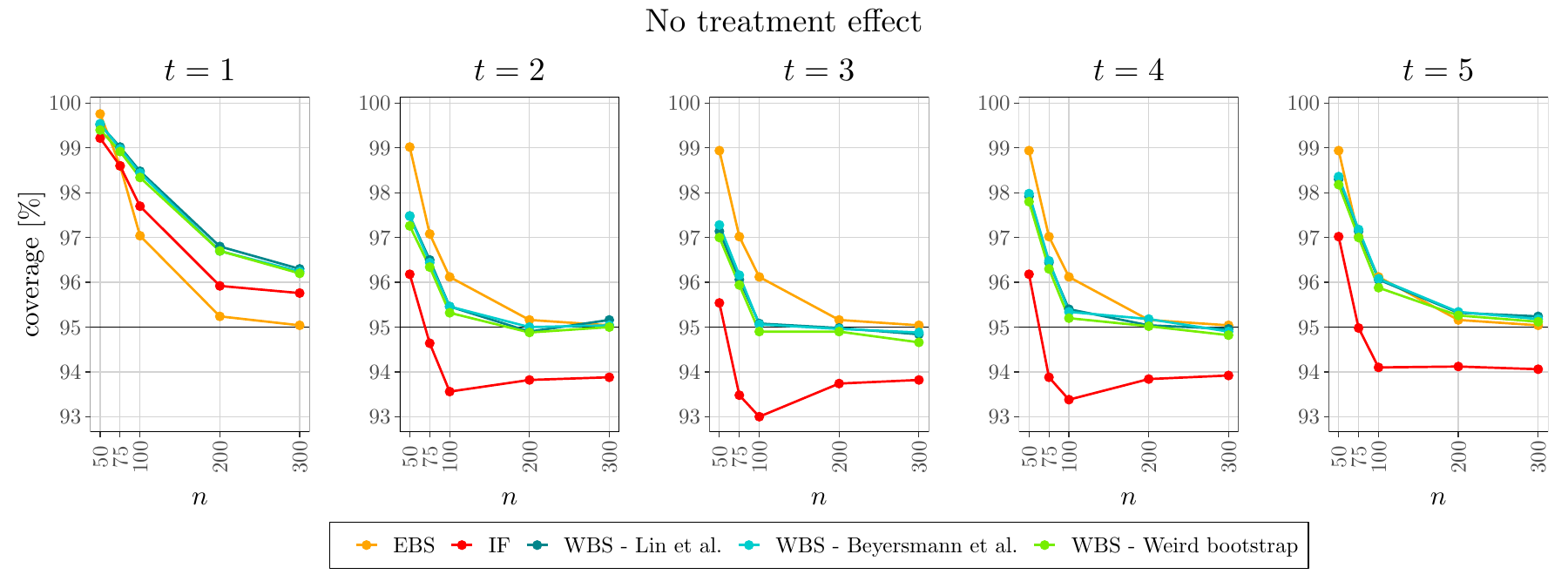}
\end{figure}

\begin{figure}[h!tbp]
\centering \label{fig:222}
\includegraphics[width = \textwidth]{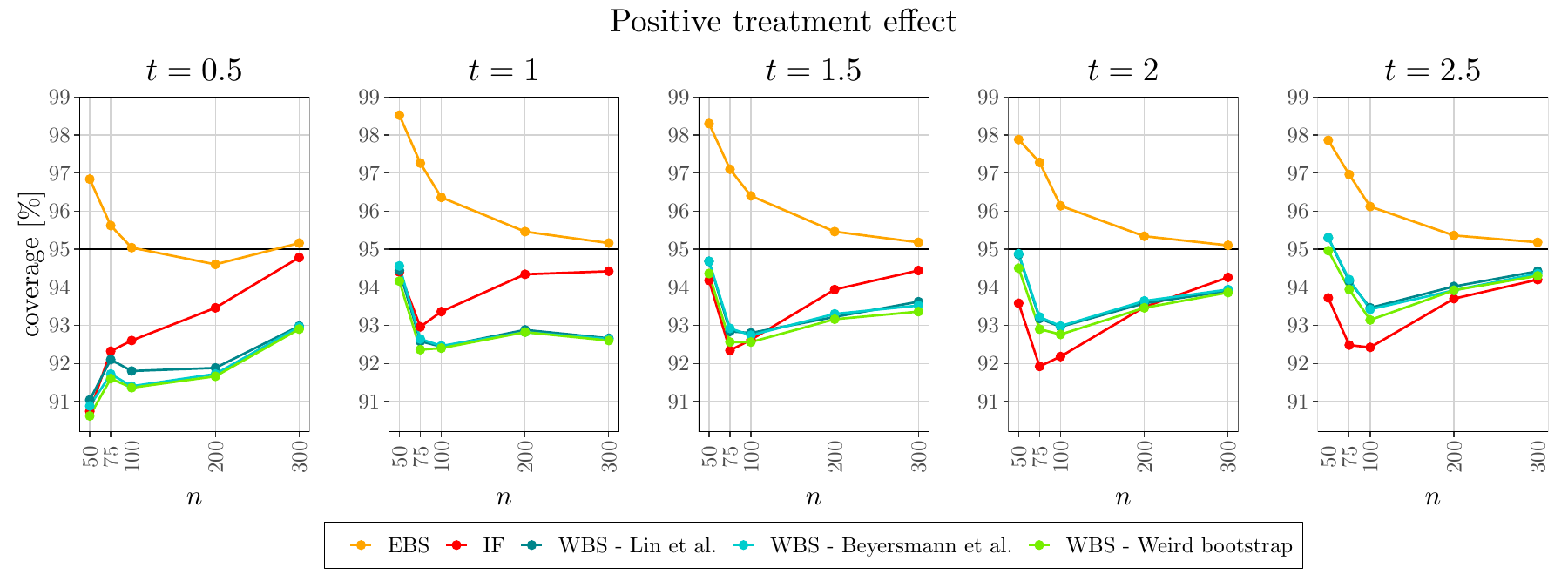}
\end{figure}

\clearpage

\section{Coverage of the confidence bands \label{sec:3}}

\subsection{No censoring \label{subsec:31}}
\enlargethispage{1cm}

\begin{figure}[h!tbp]
\centering \label{fig:31}
\includegraphics[width = 0.75\textwidth]{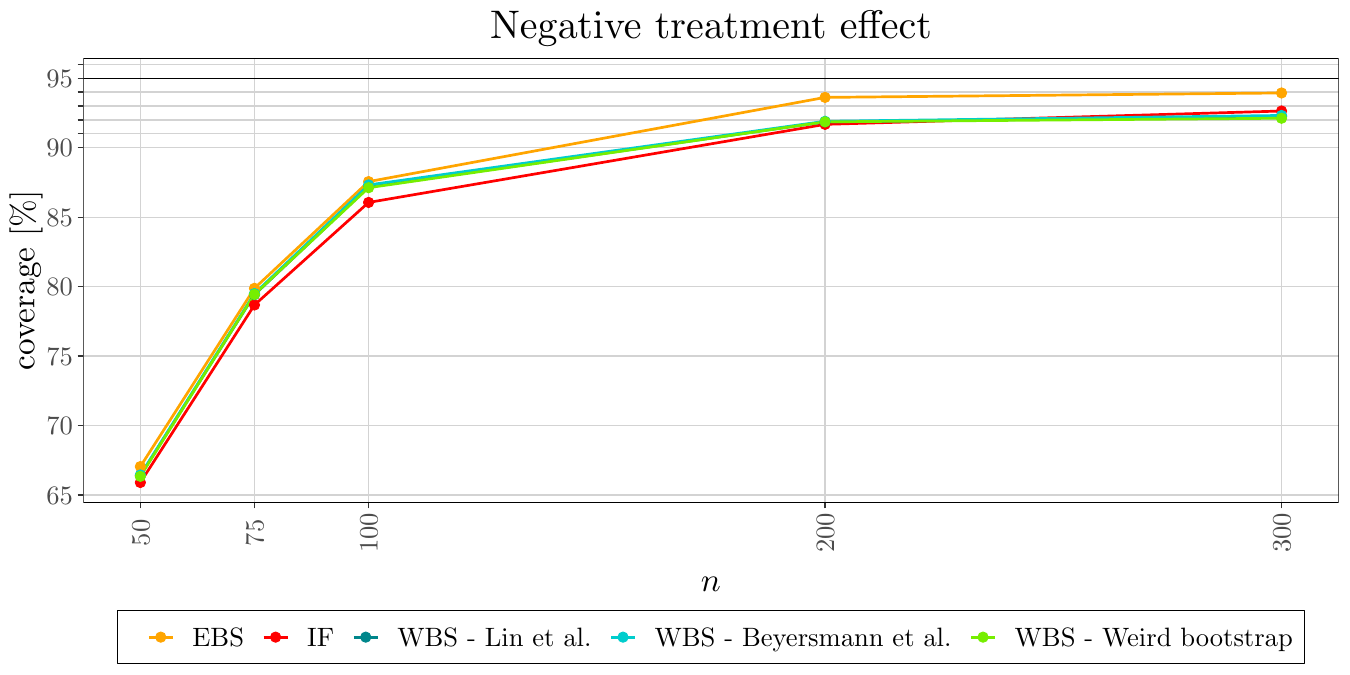}
\end{figure}

\begin{figure}[h!tbp]
\centering \label{fig:32}
\includegraphics[width = 0.75\textwidth]{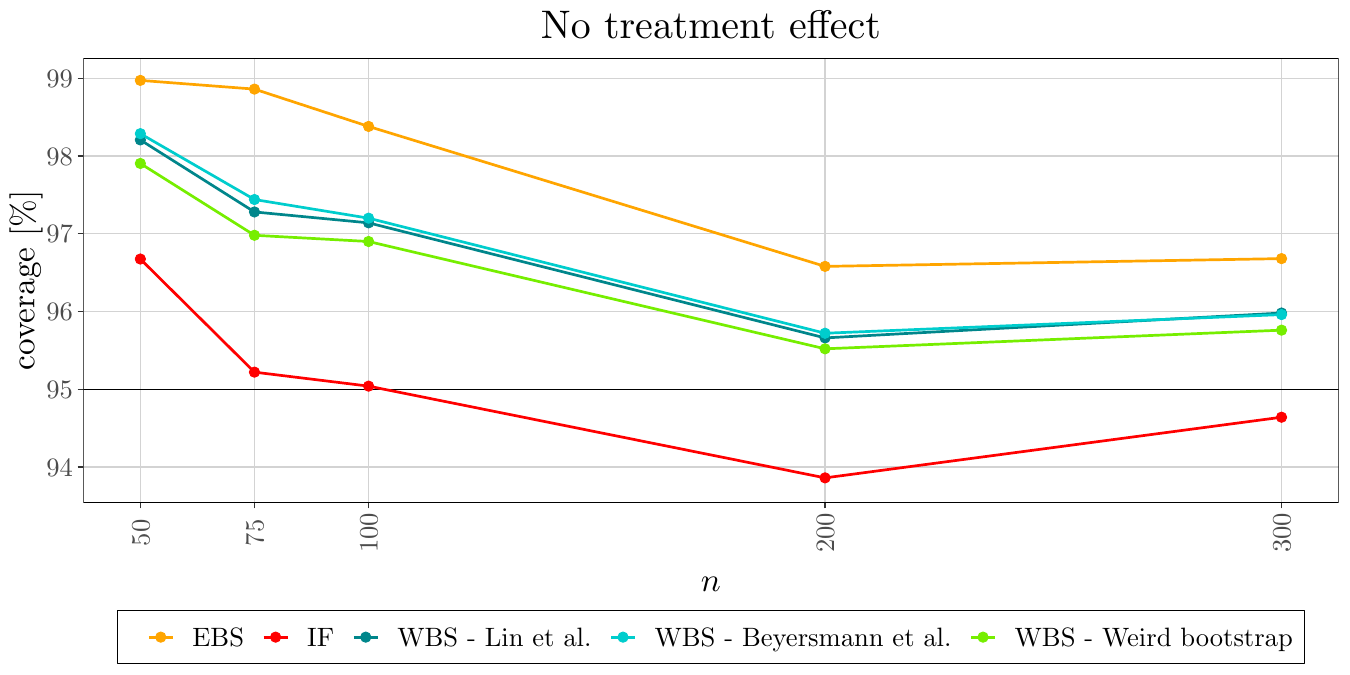}
\end{figure}

\begin{figure}[h!tbp]
\centering \label{fig:33}
\includegraphics[width = 0.75\textwidth]{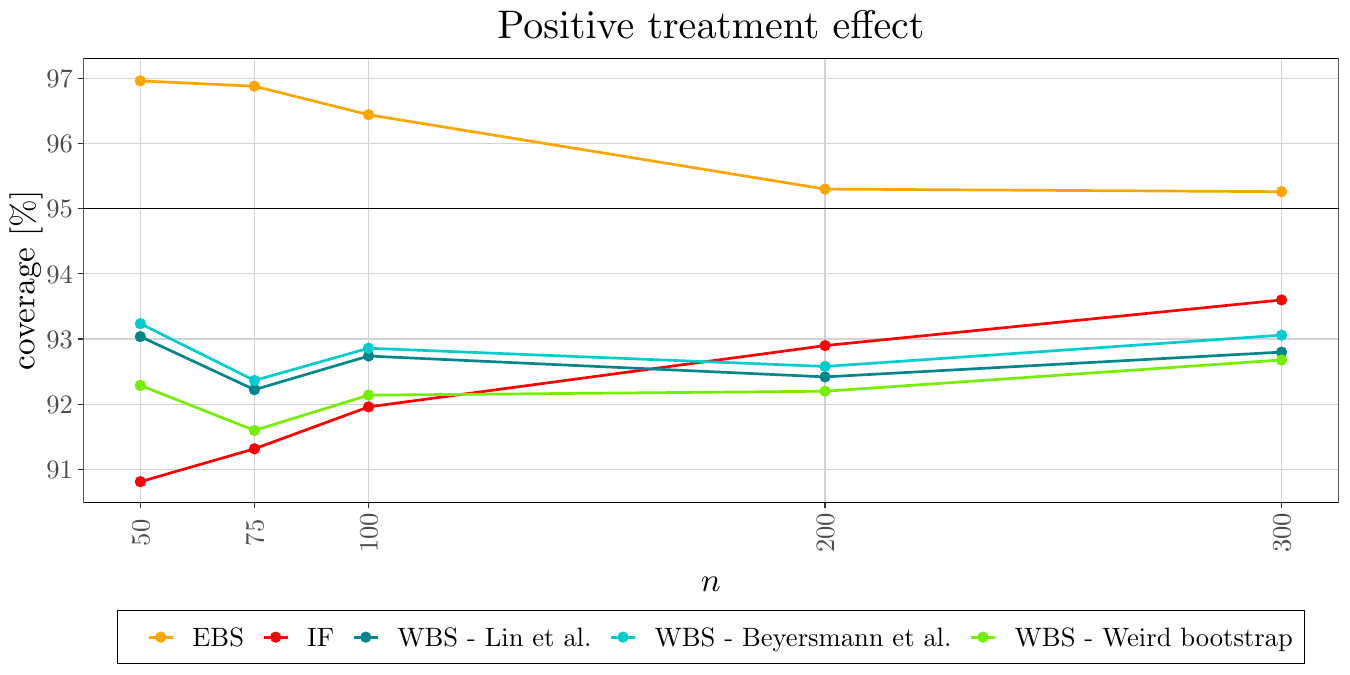}
\end{figure}

\clearpage

\subsection{Light censoring \label{subsec:32}}
\enlargethispage{1cm}

\begin{figure}[h!tbp]
\centering \label{fig:34}
\includegraphics[width = 0.75\textwidth]{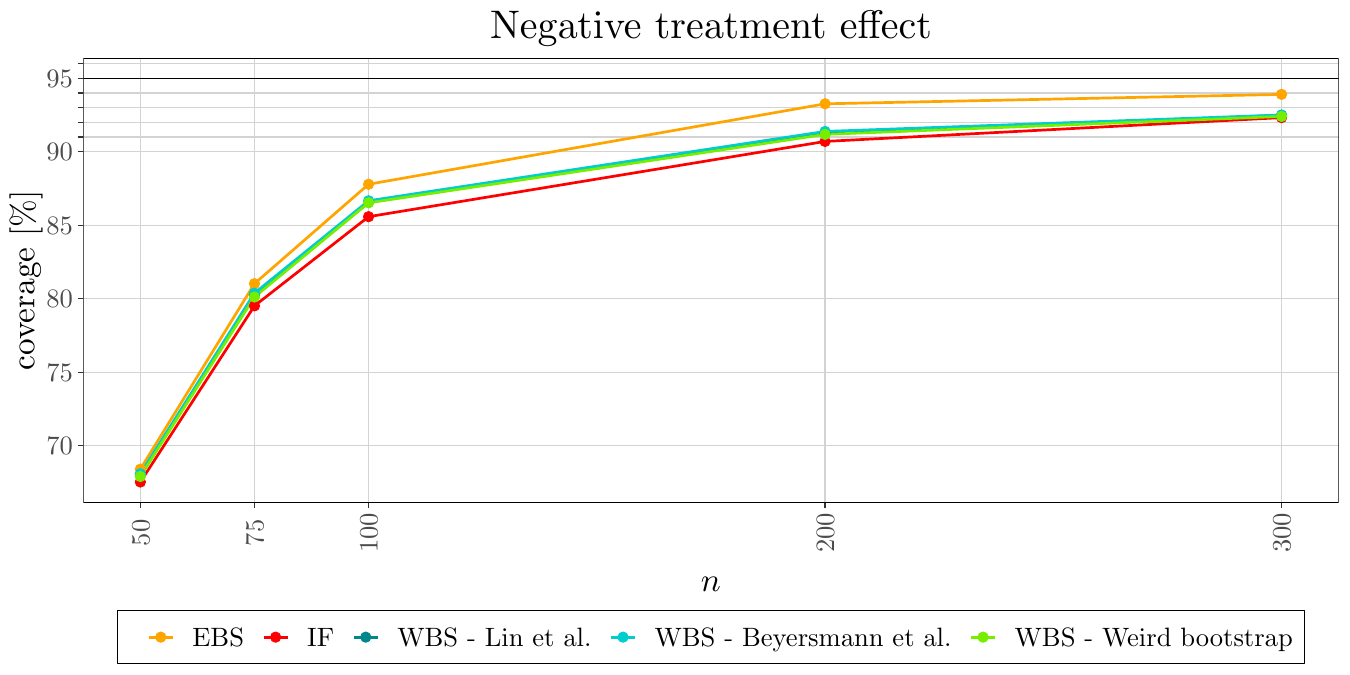}
\end{figure}

\begin{figure}[h!tbp]
\centering \label{fig:35}
\includegraphics[width = 0.75\textwidth]{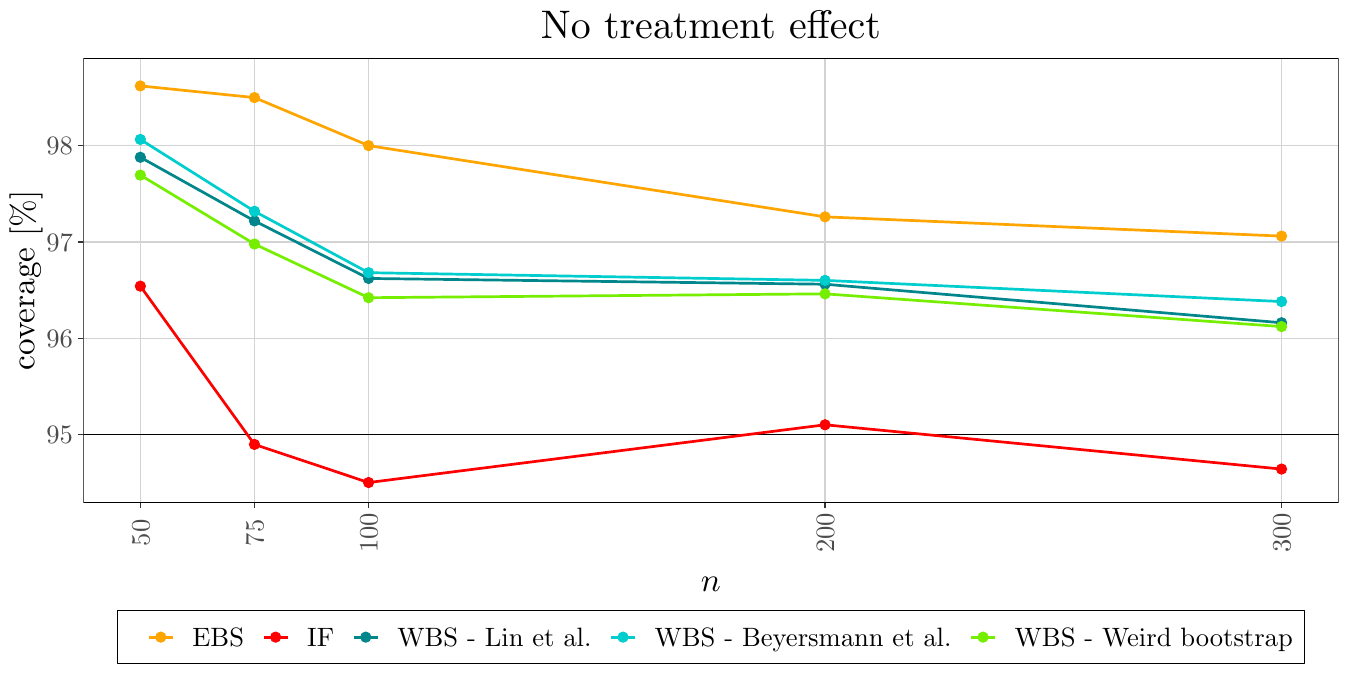}
\end{figure}

\begin{figure}[h!tbp]
\centering \label{fig:36}
\includegraphics[width = 0.75\textwidth]{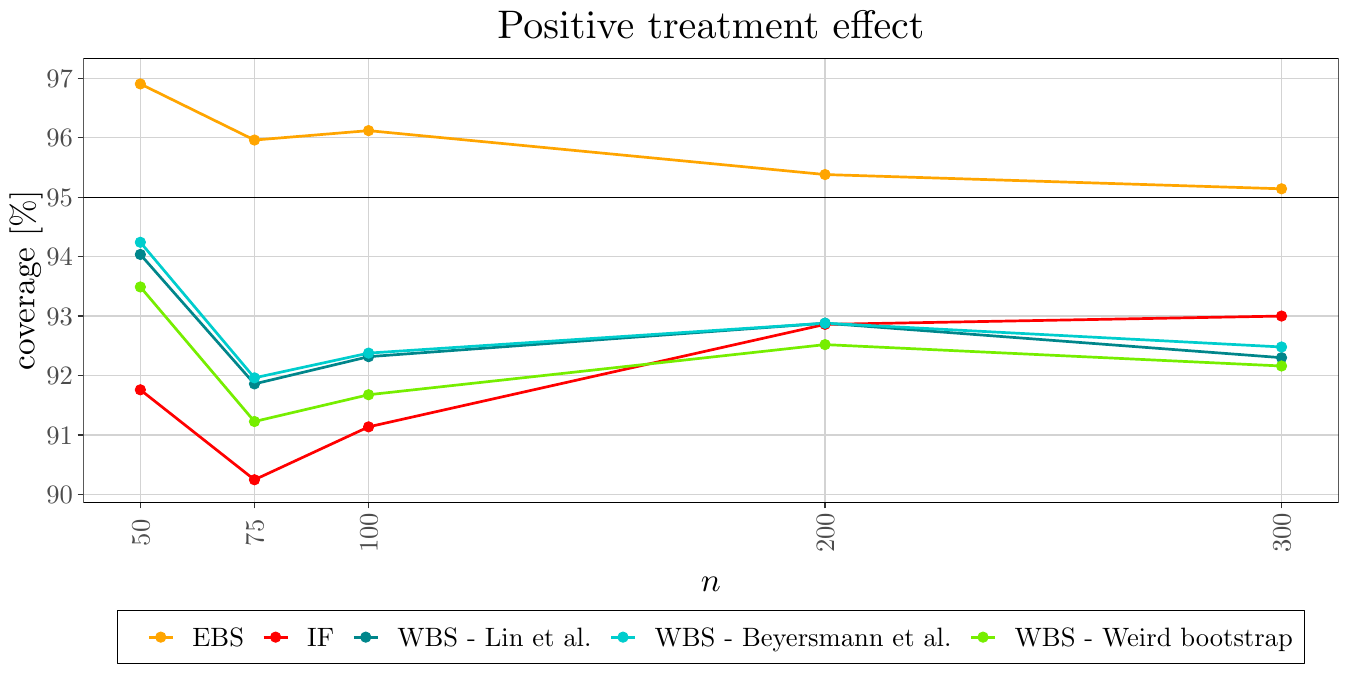}
\end{figure}
\clearpage

\subsection{Heavy censoring \label{subsec:33}}
\enlargethispage{1cm}

\begin{figure}[h!tbp]
\centering \label{fig:37}
\includegraphics[width = 0.75\textwidth]{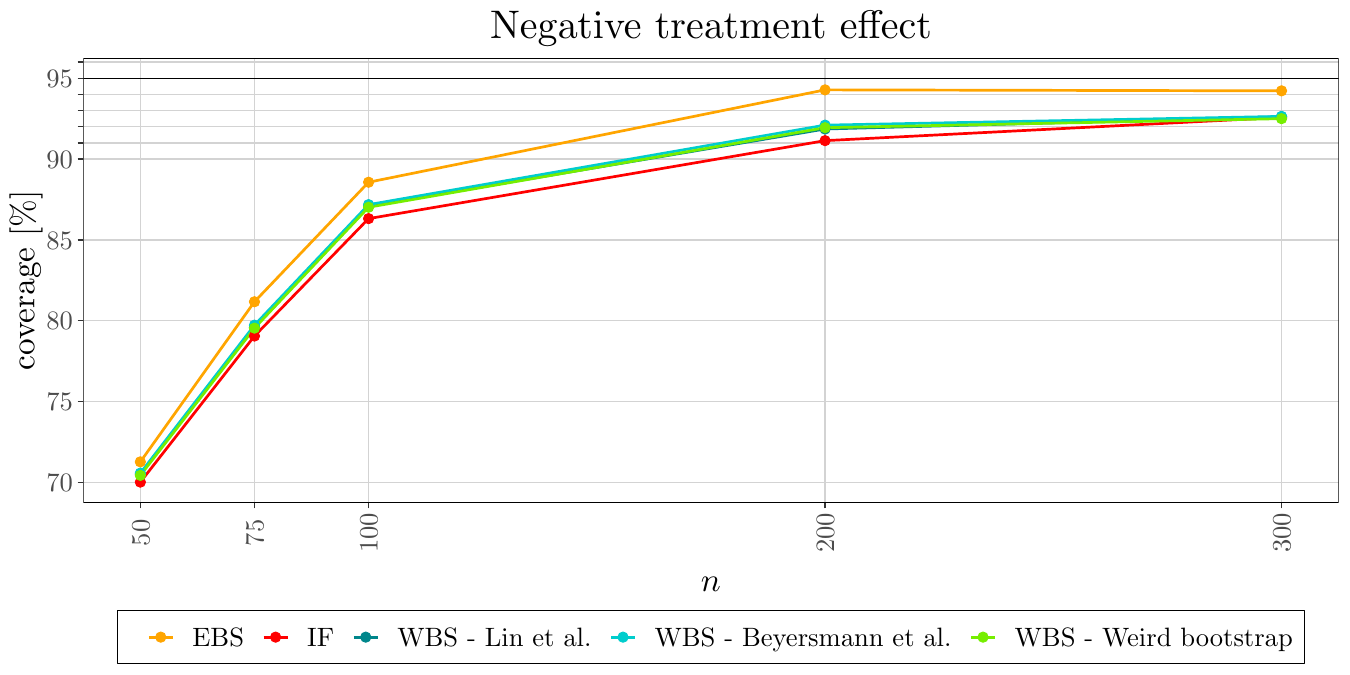}
\end{figure}

\begin{figure}[h!tbp]
\centering \label{fig:38}
\includegraphics[width = 0.75\textwidth]{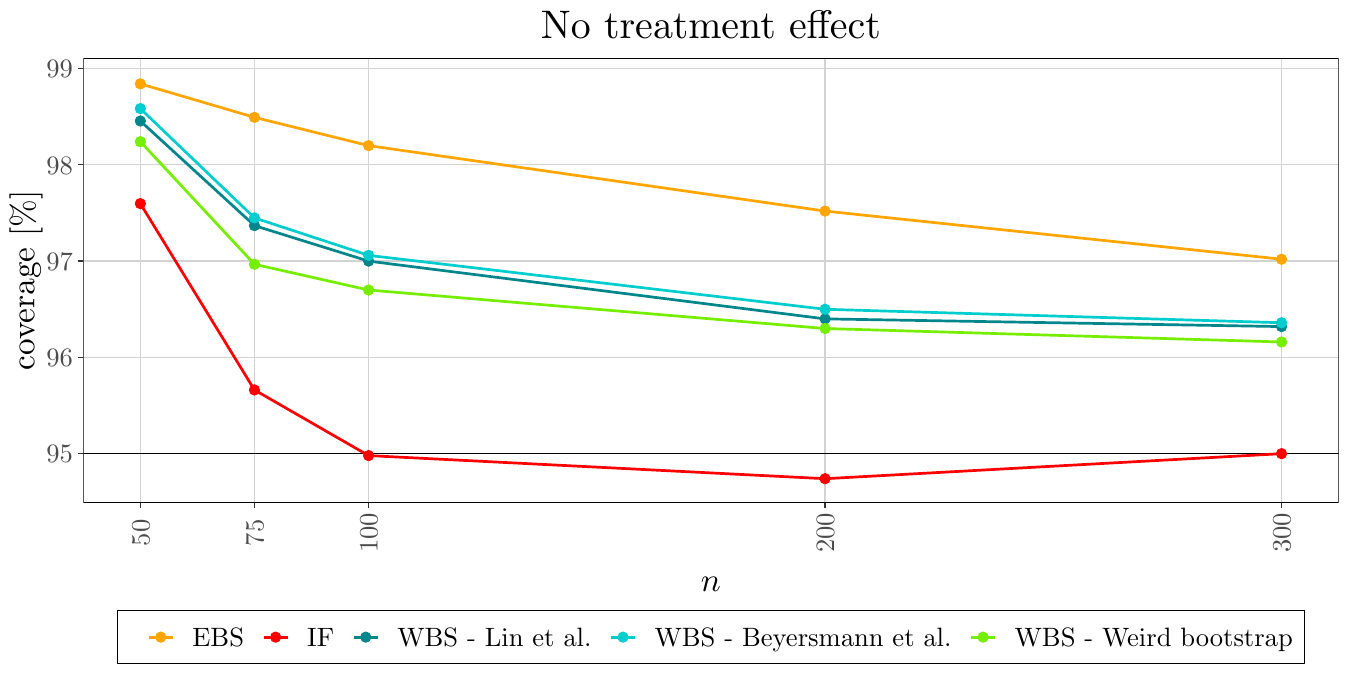}
\end{figure}

\begin{figure}[h!tbp]
\centering \label{fig:39}
\includegraphics[width = 0.75\textwidth]{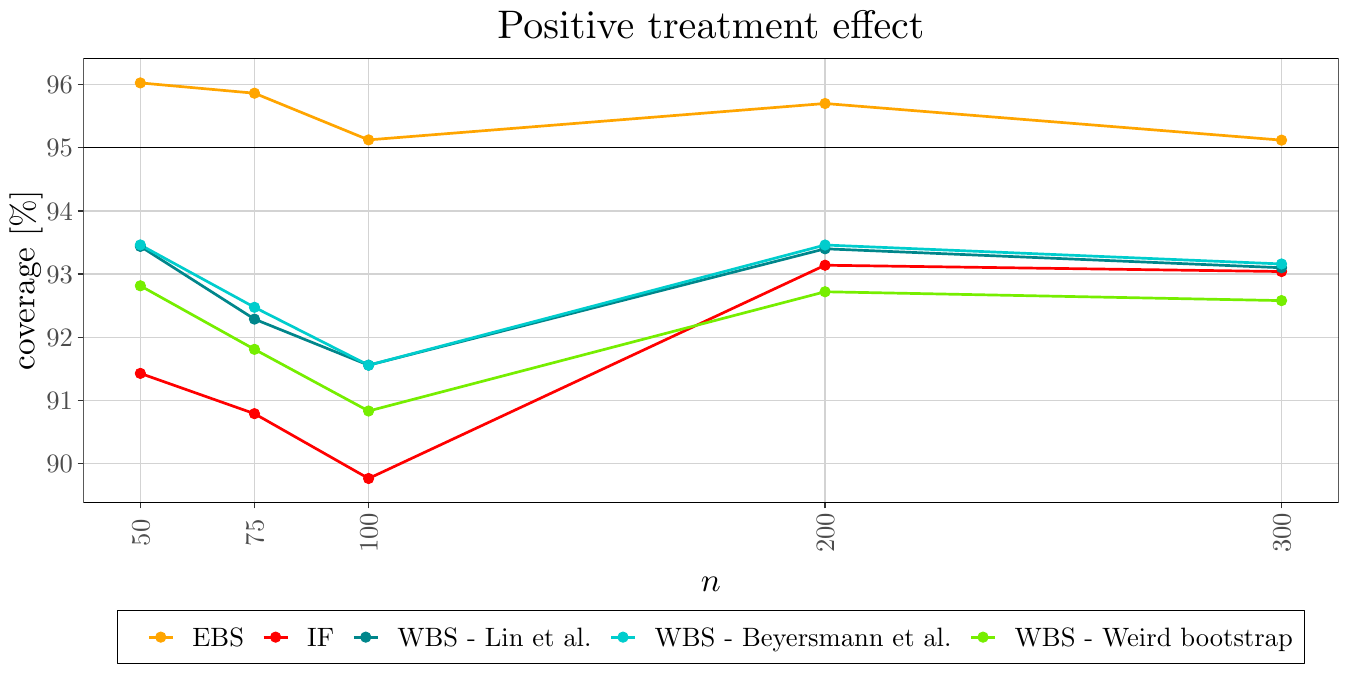}
\end{figure}

\clearpage

\subsection{Low treatment probability \label{subsec:34}}
\enlargethispage{1cm}

\begin{figure}[h!tbp]
\centering \label{fig:310}
\includegraphics[width = 0.75\textwidth]{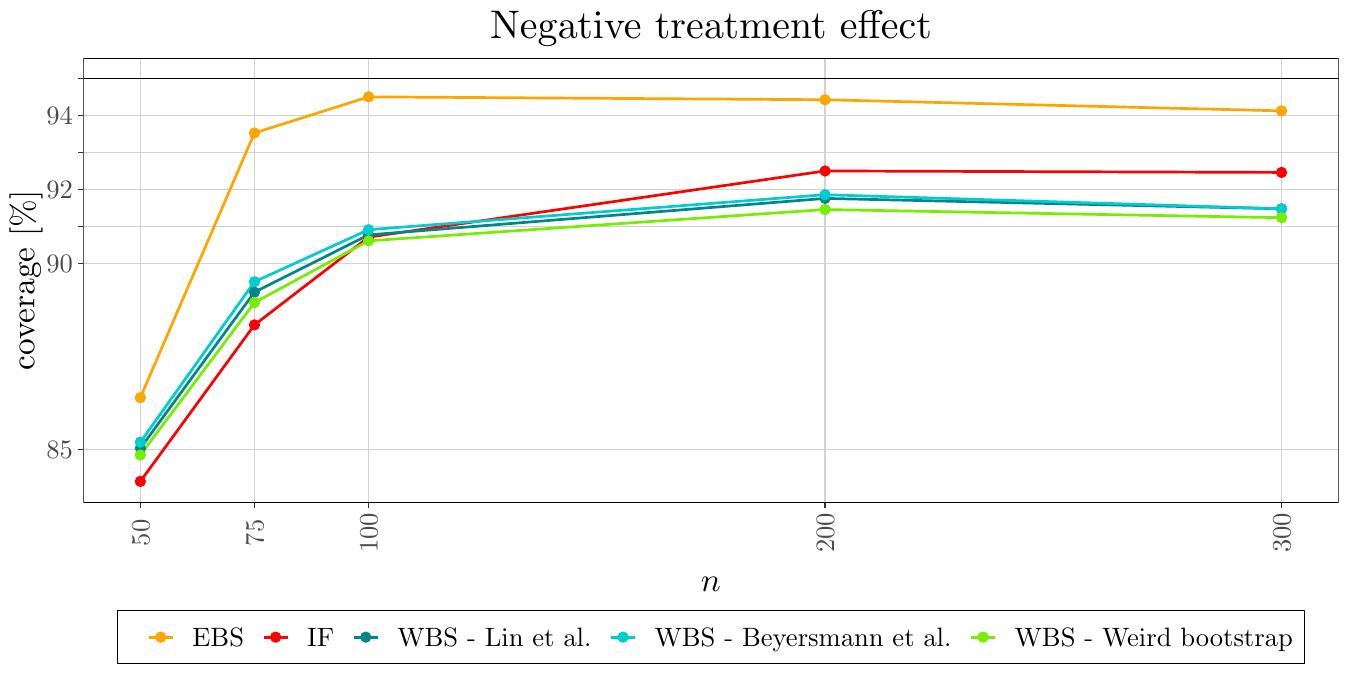}
\end{figure}

\begin{figure}[h!tbp]
\centering \label{fig:311}
\includegraphics[width = 0.75\textwidth]{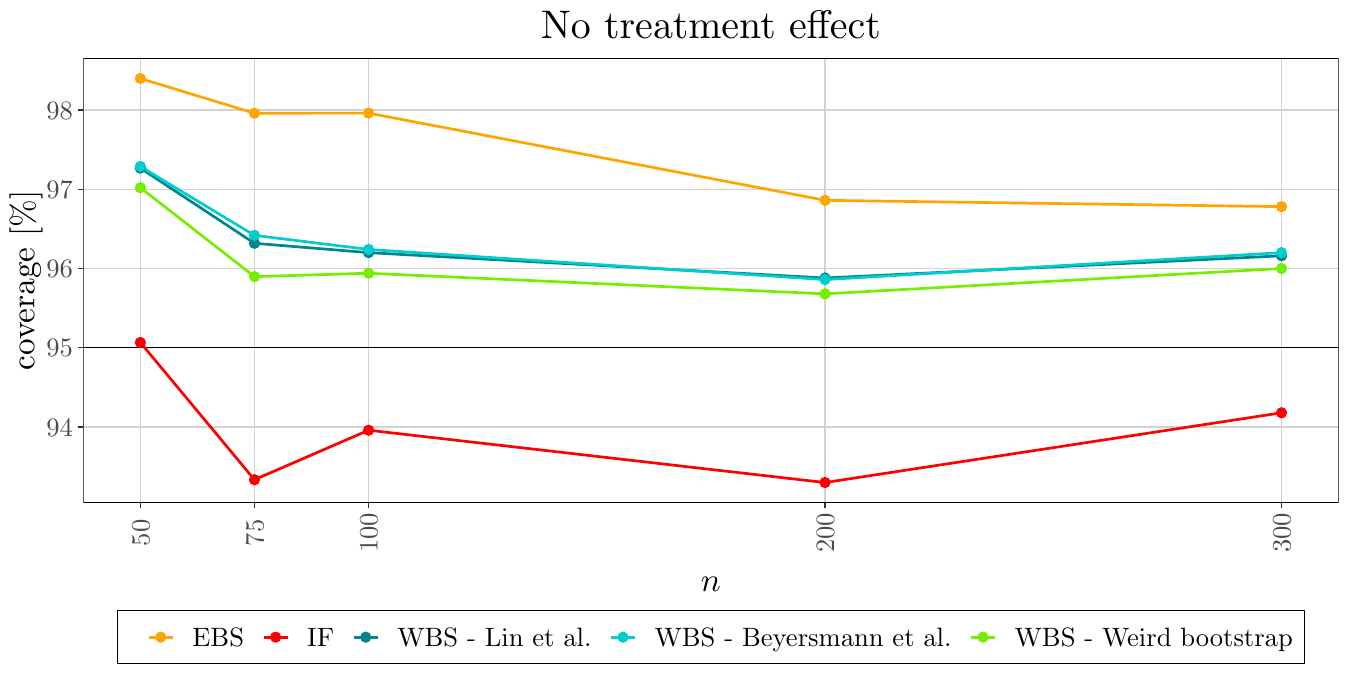}
\end{figure}

\begin{figure}[h!tbp]
\centering \label{fig:312}
\includegraphics[width = 0.75\textwidth]{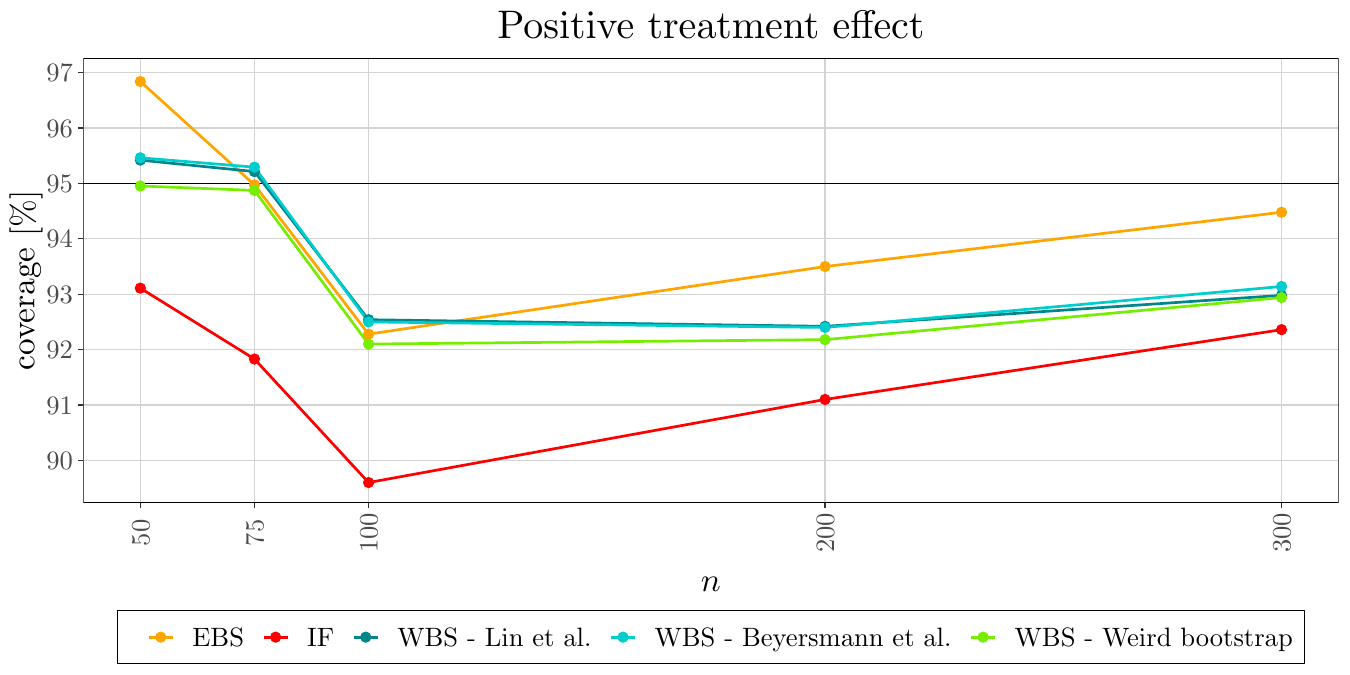}
\end{figure}

\clearpage

\subsection{High treatment probability \label{subsec:35}}

\begin{figure}[h!tbp]
\centering \label{fig:313}
\includegraphics[width = 0.75\textwidth]{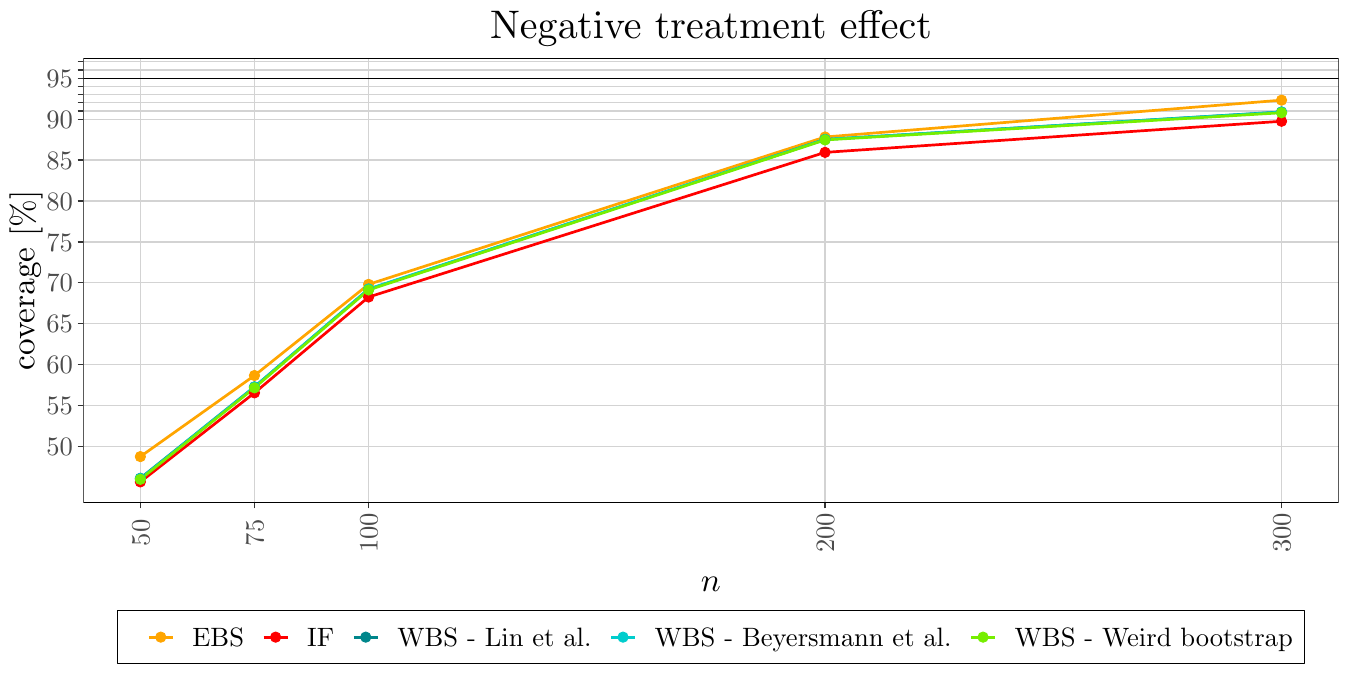}
\end{figure}

\begin{figure}[h!tbp]
\centering \label{fig:314}
\includegraphics[width = 0.75\textwidth]{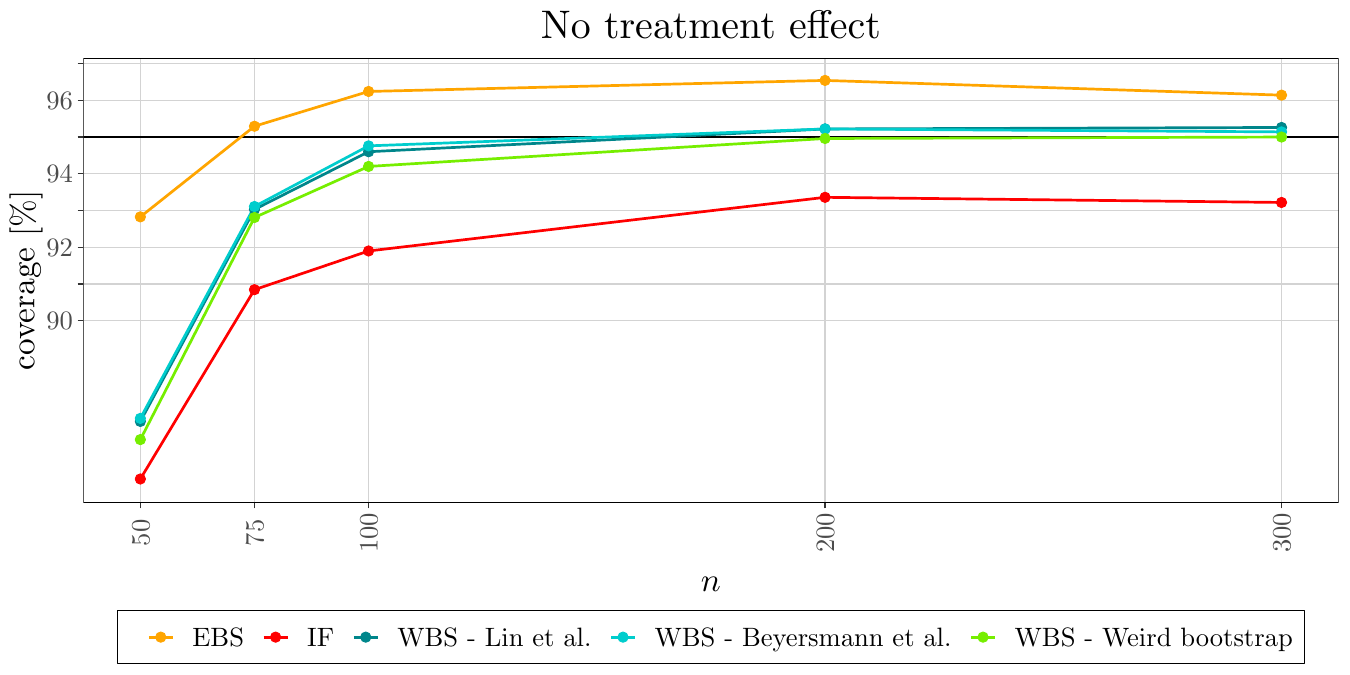}
\end{figure}

\clearpage

\subsection{Low covariance of the covariates \label{subsec:36}}
\enlargethispage{1cm}

\begin{figure}[h!tbp]
\centering \label{fig:315}
\includegraphics[width = 0.75\textwidth]{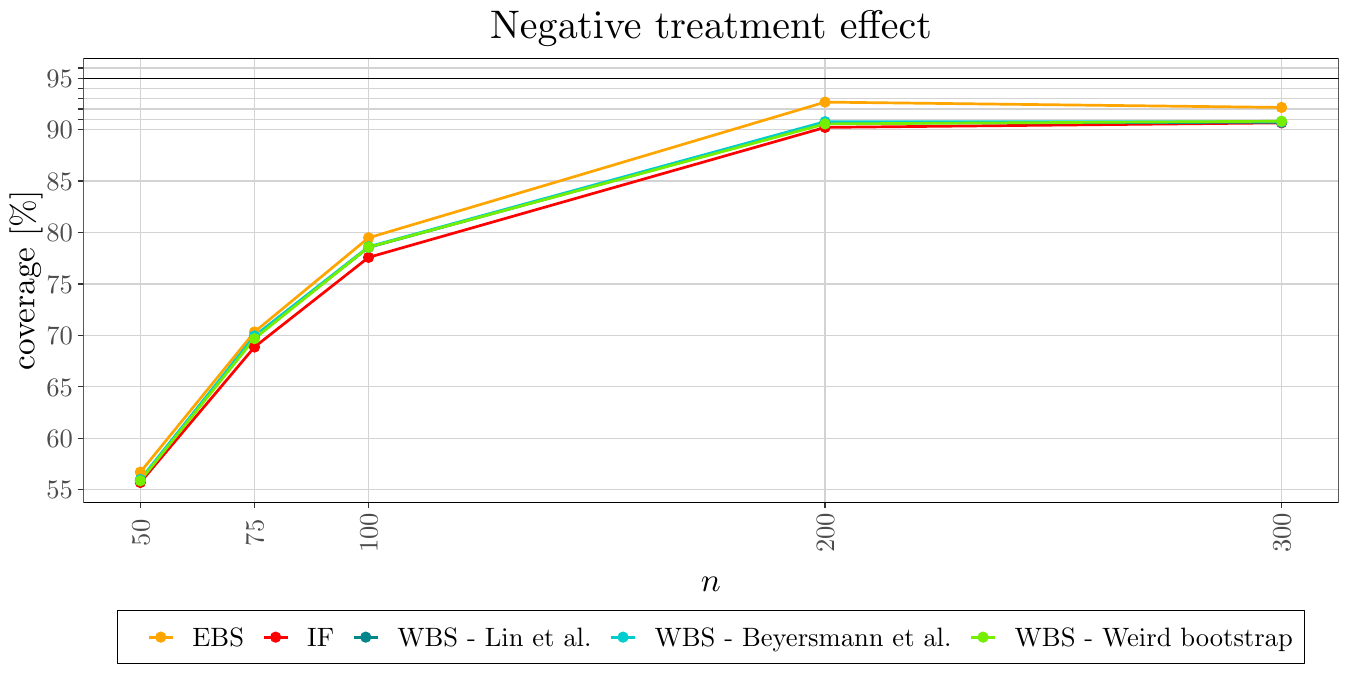}
\end{figure}

\begin{figure}[h!tbp]
\centering \label{fig:316}
\includegraphics[width = 0.75\textwidth]{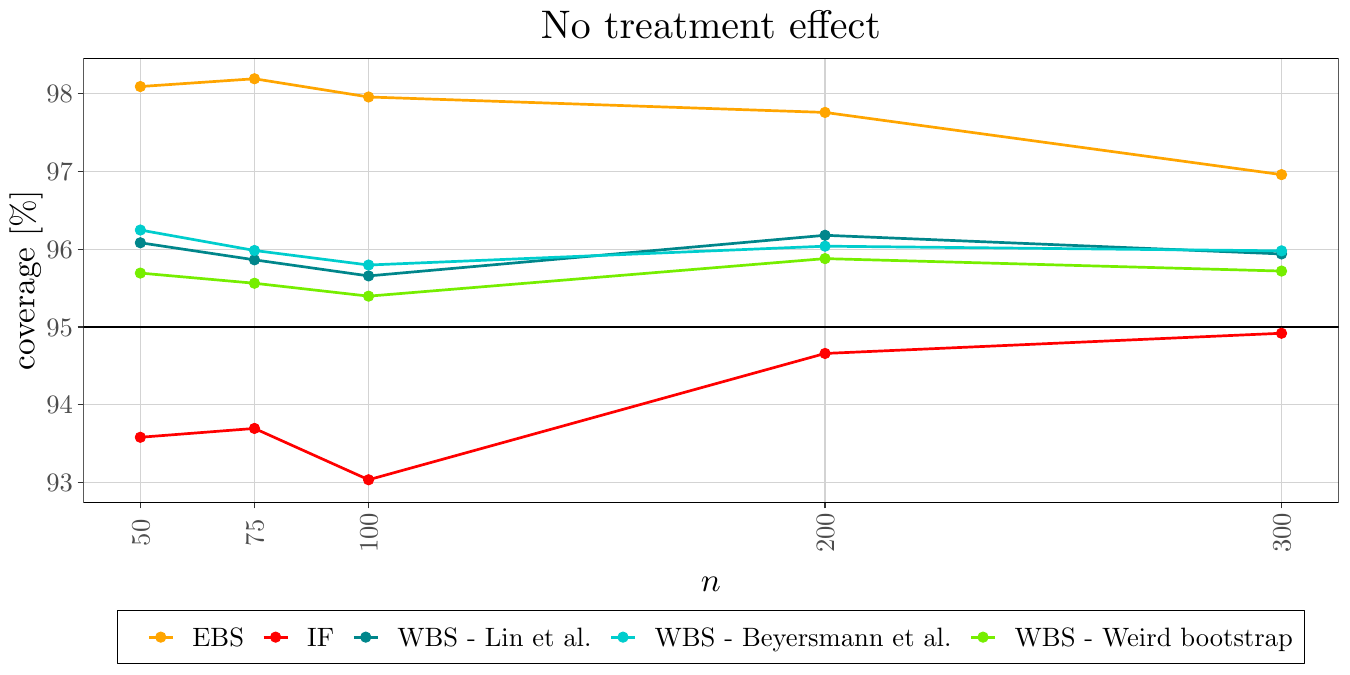}
\end{figure}

\begin{figure}[h!tbp]
\centering \label{fig:317}
\includegraphics[width = 0.75\textwidth]{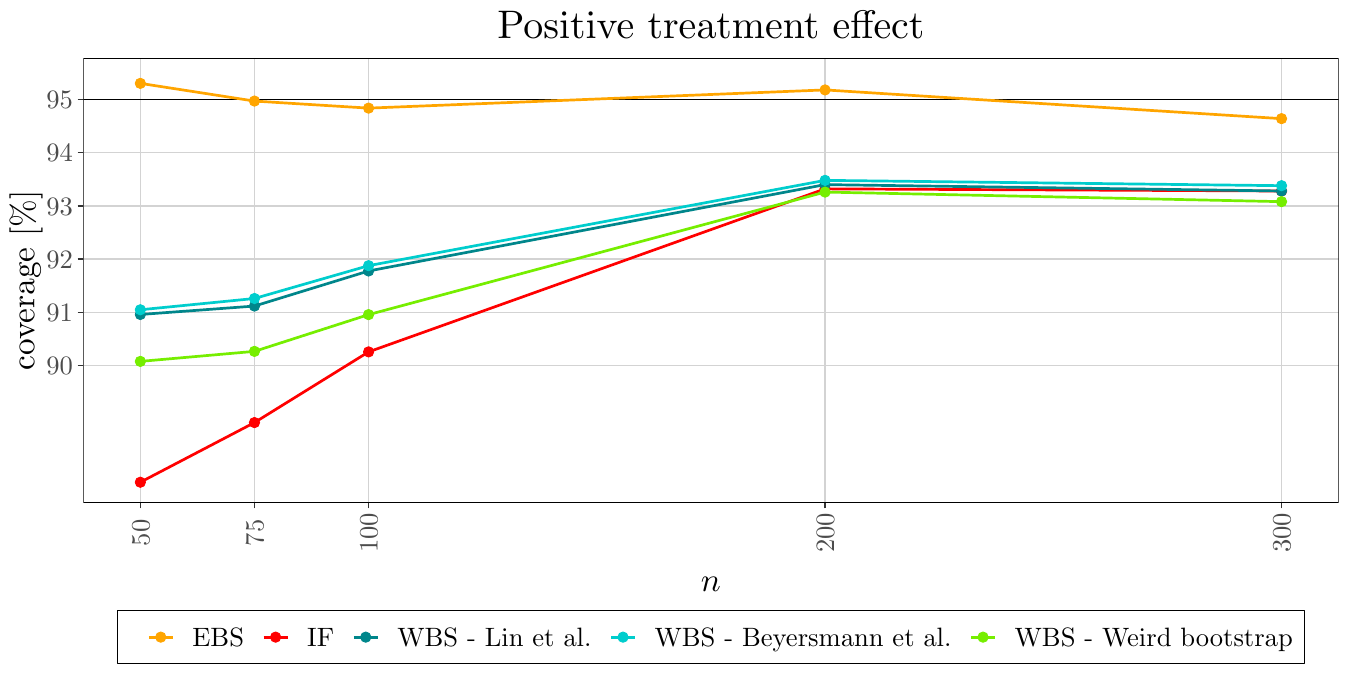}
\end{figure}

\clearpage

\subsection{High covariance of the covariates \label{subsec:37}}
\enlargethispage{1cm}

\begin{figure}[h!tbp]
\centering \label{fig:318}
\includegraphics[width = 0.75\textwidth]{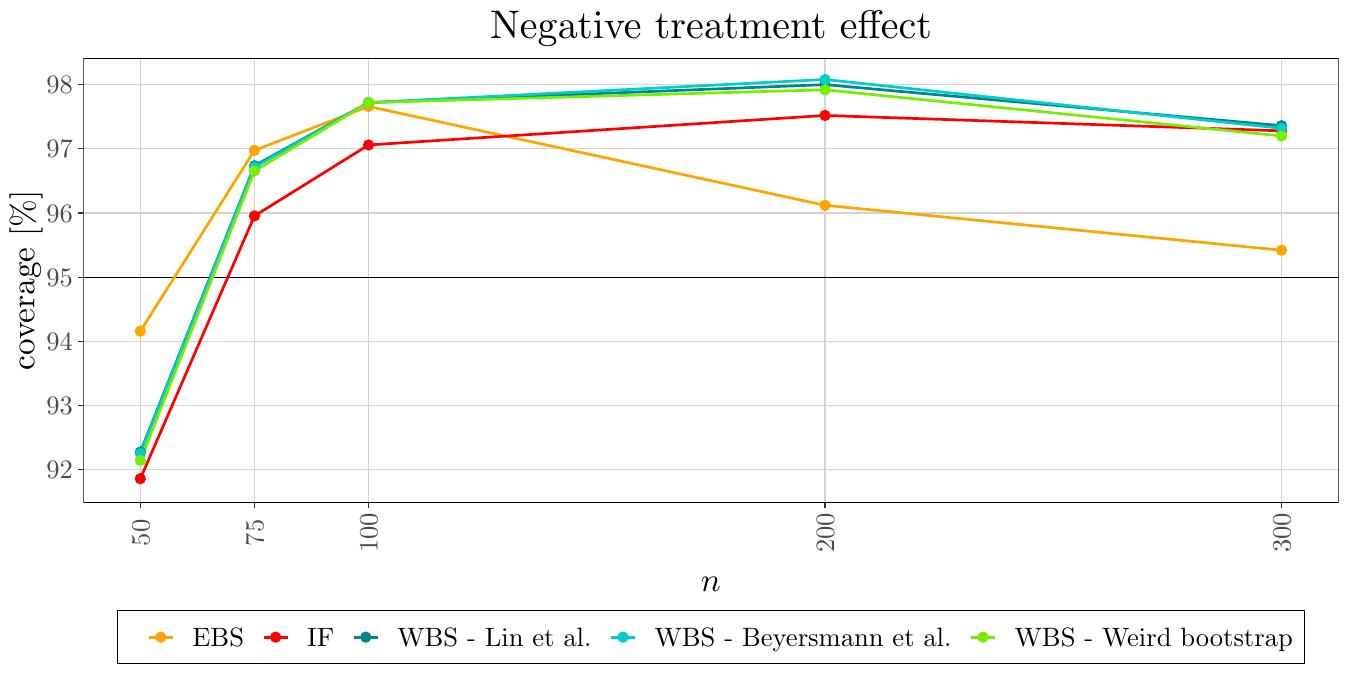}
\end{figure}

\begin{figure}[h!tbp]
\centering \label{fig:319}
\includegraphics[width = 0.75\textwidth]{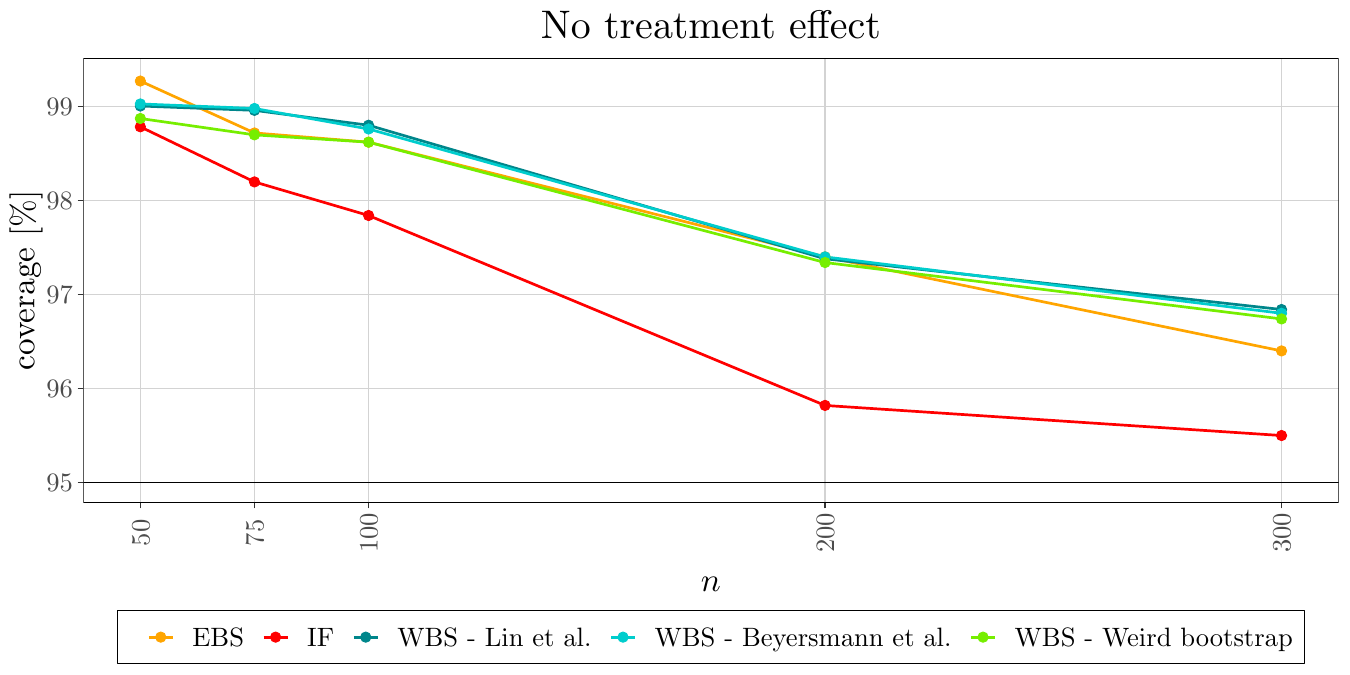}
\end{figure}

\begin{figure}[h!tbp]
\centering \label{fig:320}
\includegraphics[width = 0.75\textwidth]{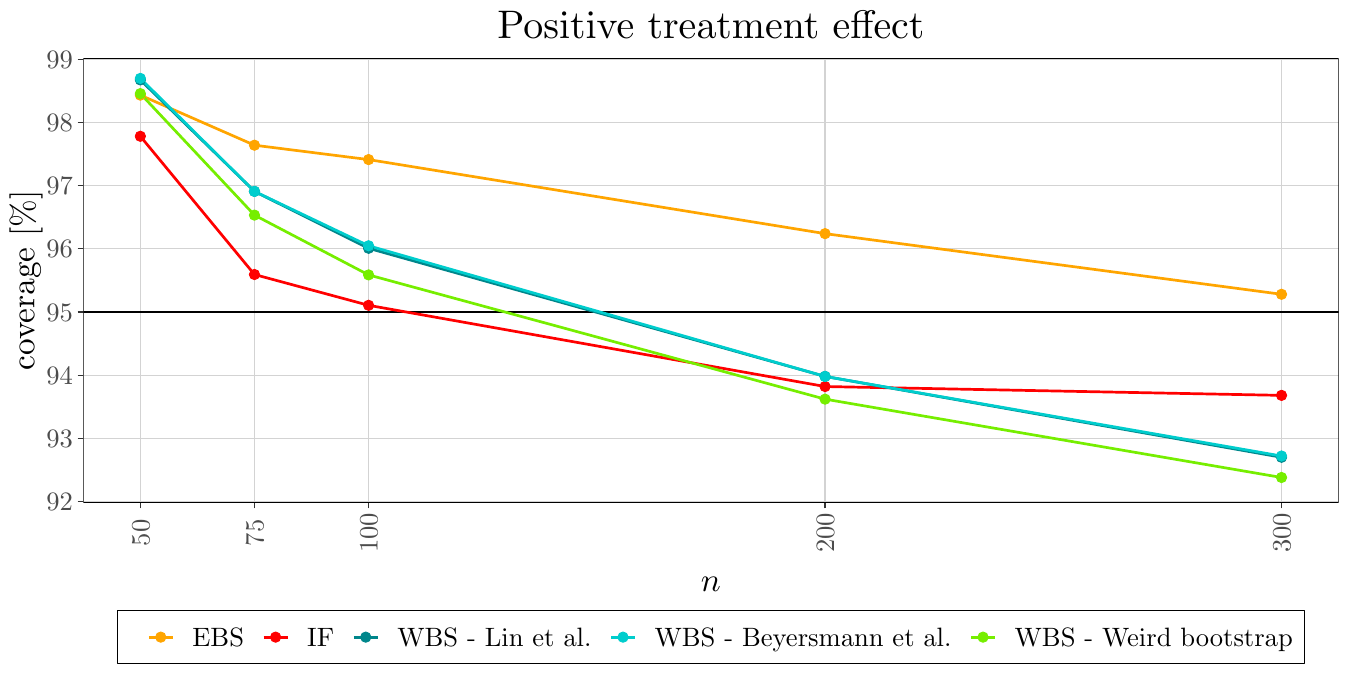}
\end{figure}

\clearpage

\subsection{Type~II censoring \label{subsec:38}}
\enlargethispage{1cm}

\begin{figure}[h!tbp]
\centering \label{fig:321}
\includegraphics[width = 0.75\textwidth]{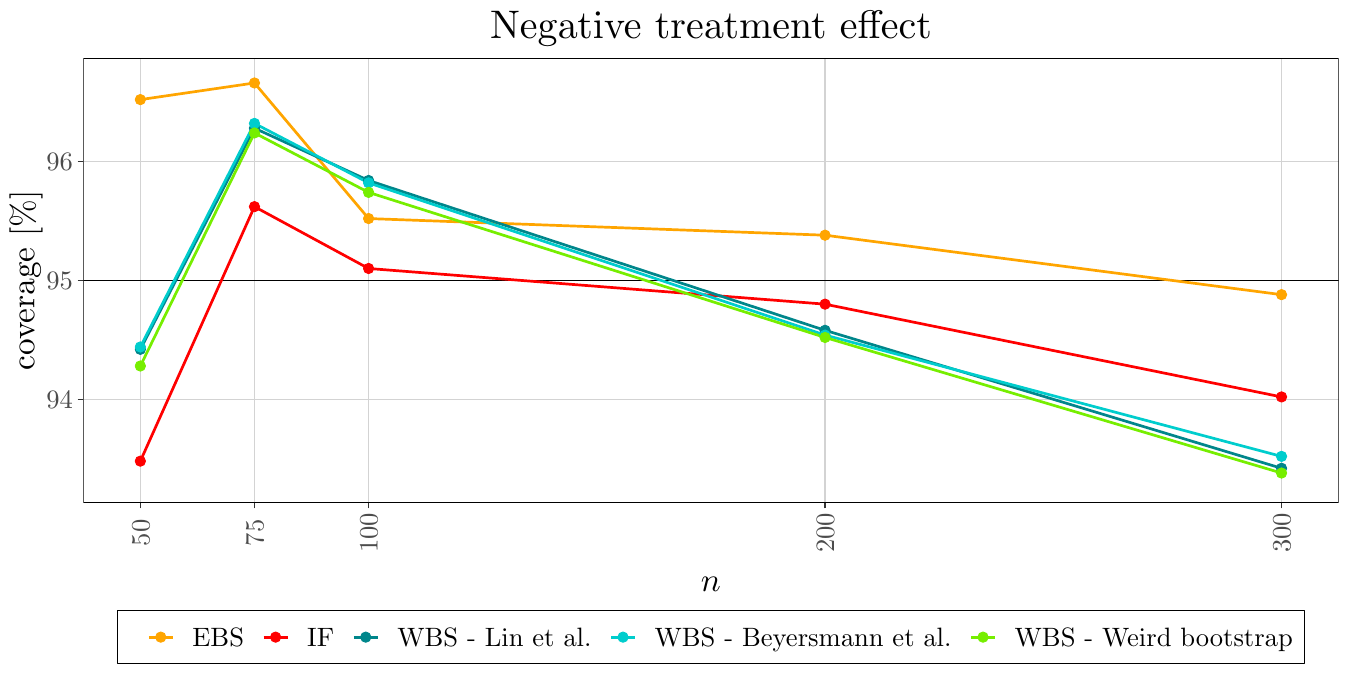}
\end{figure}

\begin{figure}[h!tbp]
\centering \label{fig:322}
\includegraphics[width = 0.75\textwidth]{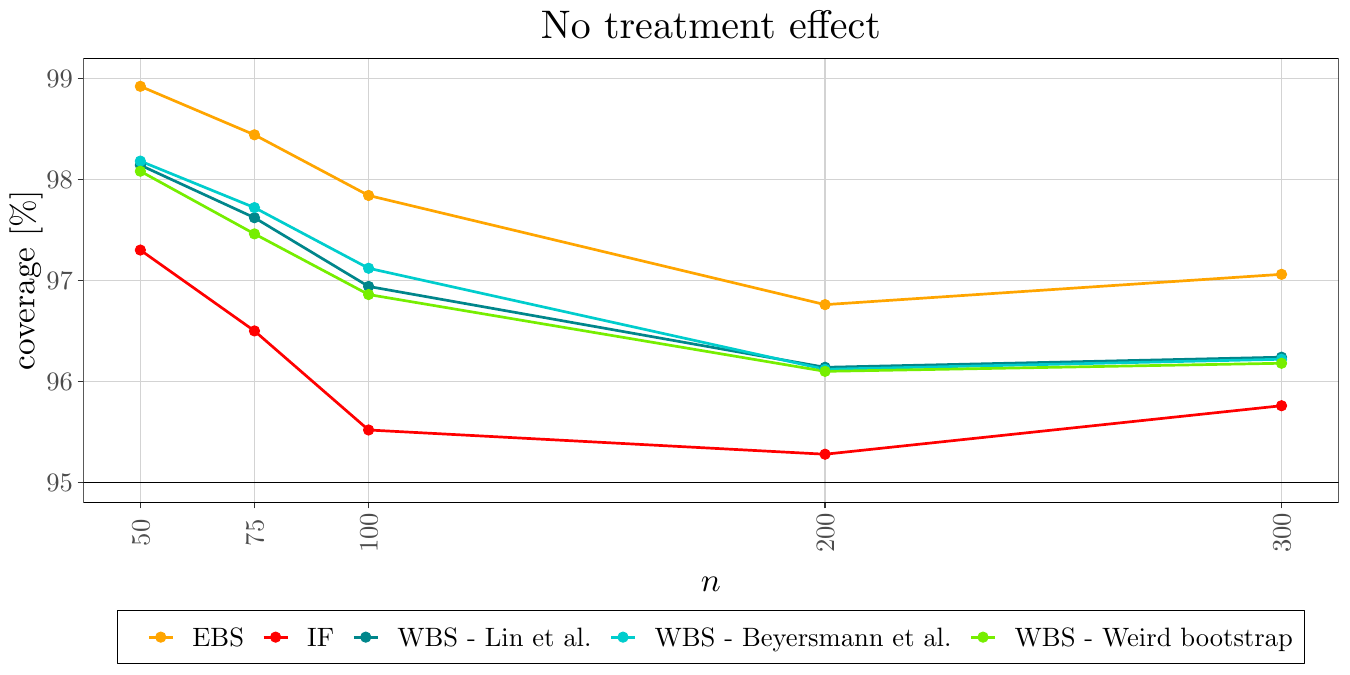}
\end{figure}

\begin{figure}[h!tbp]
\centering \label{fig:323}
\includegraphics[width = 0.75\textwidth]{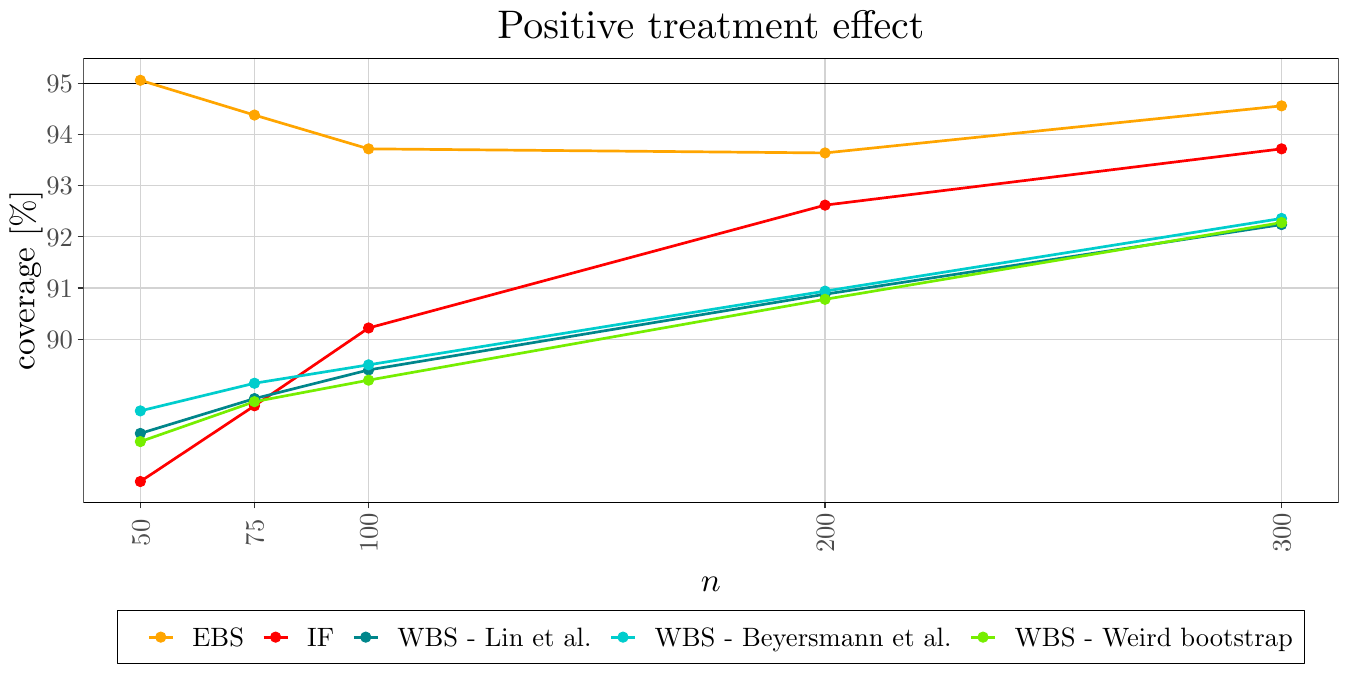}
\end{figure}

\bibliography{bib_Simulations}{}
\bibliographystyle{apalike}